\documentclass[aps,prc,twocolumn,showpacs,showkeys,preprintnumbers,amsmath,amssymb,nofootinbib,floats]{revtex4}

\newcommand{\eq}{Eq.~}

\newcommand{\refer}{Ref.~}
\newcommand{\refers}{Refs.~}
\newcommand{\figu}{Fig.~}

\newcommand{\tab}{Tab.~}
\newcommand{\sect}{Sec.~}

\newcommand{\footnoteb}[1]{$^($\protect\footnote{#1}$^)$}

\usepackage{enumerate}
\usepackage{multirow}
\usepackage{longtable}
\usepackage{epsfig}
\usepackage[latin1]{inputenc}
\usepackage[english,francais]{babel}
\usepackage{array}
\usepackage{subfigure}
\usepackage{graphics}
\begin{document}
\selectlanguage{english}
\title{New analysis method of the halo phenomenon in finite many-fermion systems\\
First applications to medium-mass atomic nuclei}
\author{V. Rotival}
\email{vincent.rotival@polytechnique.org} \affiliation{DPTA/Service
de Physique Nucl\'eaire - CEA/DAM \^Ile-de-France - BP12 - 91680
Bruy\`eres-le-Ch\^atel, France}\affiliation{National Superconducting
Cyclotron Laboratory, 1 Cyclotron Laboratory, East Lansing, MI
48824, USA}
\author{T. Duguet}
\email{thomas.duguet@cea.fr}
\affiliation{National Superconducting Cyclotron Laboratory,
1 Cyclotron Laboratory, East-Lansing, MI 48824, USA}
\affiliation{Department of Physics and Astronomy,
Michigan State University, East Lansing, MI 48824, USA}
\affiliation{CEA, Centre de Saclay, IRFU/Service de Physique Nucl{\'e}aire, F-91191 Gif-sur-Yvette, France}


\date{\today}

\begin{abstract}
A new analysis method to investigate halos in finite many-fermion systems is designed, as existing
characterization methods are proven to be incomplete/inaccurate. A decomposition of the internal wave-function of
the \mbox{$N$-body} system in terms of overlap functions allows a model-independent analysis of medium-range and asymptotic
properties of the internal one-body density. The existence of a spatially decorrelated region in the
density profile is related to the existence of three typical energy scales in the excitation spectrum of the
\mbox{$(N\!-\!1)$-body} system. A series of model-independent measures, taking the internal density as the only input, are
introduced. The new measures allow a quantification of the potential halo in terms of the average number of fermions participating to it and of its impact on the system extension. Those new "halo factors" are validated through simulations and applied to results obtained through energy density functional calculations of medium-mass nuclei. Performing spherical Hartree-Fock-Bogoliubov calculations with state-of-the-art Skyrme plus pairing functionals, a collective halo is predicted in drip-line Cr isotopes, whereas no such effect is seen in Sn isotopes.
\end{abstract}
\maketitle

\section{Introduction}
\label{sec:intro}

The study of light nuclei at the limit of stability has been possible in the last two decades thanks to the first
generations of radioactive ion beam facilities. One of the interesting phenomena observed close to the nucleon
drip-line is the formation of nuclear halos. In such systems, either the proton or the neutron density displays an
unusually extended tail due to the presence of weakly-bound nucleons~\cite{hansen87}. Since the first experimental
observation of such an exotic structure in \mbox{$^{11}$Li}~\cite{tanihata85a,tanihata85b}, other light neutron
halo systems have been identified, e.g. \mbox{$^6$He}~\cite{zhukov93},
\mbox{$^{11}$Be}~\cite{tanihata88,fukuda91,zahar93}, \mbox{$^{14}$Be}~\cite{tanihata88,thompson96},
\mbox{$^{17}$B}~\cite{tanihata88} or \mbox{$^{19}$C}~\cite{bazin95,kanungo00}. On the proton-rich side,
theoretical works demonstrated the existence of halo structures in spite of the presence of the Coulomb barrier
~\cite{zhukov95}, as was seen experimentally for \mbox{$^8$B}~\cite{minamisono92,schwab95,warner95,negoita96} and
\mbox{$^{17}$Ne}~\cite{kanungo03,jeppesen04}. Halos in excited states have been observed for
\mbox{$^{17}$F}~\cite{morlock97,ren98}, \mbox{$^{12}$B}~\cite{lin01} or \mbox{$^{13}$B}~\cite{liu01}, and several
others are predicted~\cite{chen05}. It is worth noticing that weakly-bound systems extending well beyond the classically-allowed region have also been theoretically predicted or experimentally observed for molecules (\mbox{$^3$He-$^3$He-$^{39}$K}~\cite{li06},
\mbox{$^4$He$_2$}~\cite{schollkopf96,nielsen98,grisenti00},  \mbox{$^3$He$^4$He$_2$}~\cite{bressanini02}...),
atom-positron complexes (\mbox{$e^+$Be}, \mbox{PsLi$^+$}, \mbox{PsHe$^+$}...)~\cite{mitroy05} and hypernuclei
(\mbox{$^3_\Lambda$H})~\cite{cobis97}.

The theoretical description of light halo systems is rather well under control. It usually relies on a cluster
vision where one (\mbox{$^{11}$Be}, \mbox{$^{19}$C}...) or two (\mbox{$^{11}$Li}, \mbox{$^6$He}...) loosely bound
nucleons define a low-density region surrounding a core. Assuming that core and halo degrees of freedom can be
decoupled, essentially exact solutions of the simplified many-body problem are obtained by solving the
Schr\"{o}dinger equation for two-body systems~\cite{fedorov94,nunes96a}, or Faddeev equations for three-body
ones~\cite{zhukov93,fedorov94,nunes96b,bang96}. However, the boundary between halo and non-halo nuclei is blurred
by the presence of core excitations. Indeed, the inert decoupling of the loosely bound nucleons from the core is
only an approximation. Nevertheless it has been assessed that halo systems arise when~\cite{riisager00,jensen01}
(i) the probability of nucleons to be in the forbidden region outside the classical turning point is greater than
\mbox{$50\%$}, and (ii) the cluster structure is dominant and accounts for at least \mbox{$50\%$} of the
configuration. Such conditions have been thoroughly studied~\cite{fedorov93,jensen00} and found to be fulfilled
when (a) the separation energy of the nucleus is very small, of the order of \mbox{$2$~MeV$/A^{2/3}$}, (b) the
loosely bound nucleons occupy low angular-momentum states (\mbox{$\ell=0$} or \mbox{$\ell=1$}) for two-body
clusters, or low hyperangular momentum states (\mbox{$K=0$} or \mbox{$K=1$}) for three-body ones in order to limit the
effect of the centrifugal barrier that prevents nucleons from spreading out~\cite{fedorov94b}, and (c) the charge
of the core remains small for proton halos. The latter requirement might be weakened because of a potential
Coulomb-induced rearrangement of the single-particle states~\cite{liang07}.

Going to heavier nuclei, few-body techniques face theoretical and computational limits because of the large
number of degrees of freedom involved. Single-reference energy density functional (SR-EDF) methods under the form
of self-consistent Hartree-Fock-Bogoliubov (HFB) calculations become appropriate~\cite{ring80a,bender03b}. The
EDF, either non-relativistic (Skyrme~\cite{skyrme56,vautherin72a} or Gogny~\cite{decharge80a}) or relativistic
~\cite{bouyssy87,reinhard89,gambhir90,ring93,todd03}, constitutes the only phenomenological input to the method.
Phenomenological functionals have now reached an accuracy suitable for comparison of various observables with
experimental data over the known part of the nuclear chart~\cite{samyn02,goriely02,goriely03,samyn04}. However,
properties of current EDFs are not yet under control in extreme conditions, where low-density configurations,
isospin or surface effects come strongly into play. Thus, the capacity of existing functionals to predict
properties of exotic nuclei, such as their limits of stability, remains rather weak~\cite{doba02b}. In that
respect, the input from the coming generation of radioactive beam facilities
(FAIR at GSI, RIBF at RIKEN, REX-ISOLDE at CERN, SPIRAL2 at GANIL...)
will help to further constrain models and to design a universal EDF.

Halo structures may contribute significantly to such a quest as they emphasize low-density configurations and
surface/finite-size effects. Their study in medium-mass nuclei might provide relevant information regarding
isovector density dependencies and gradient/finite-size corrections in the energy functional. In particular, the pairing strength
in low density regimes and the evolution of shell structures towards the limit of stability might be
constrained. However, two questions arise as we discuss potential medium-mass halos. Indeed, medium-mass nuclei
are (i) large enough that the cluster picture at play in light nuclei needs to be revisited, in such a way that
our understanding of the halo phenomenon might change significantly, and (ii) light enough that explicit
correlations associated with symmetry restorations and other large amplitude motions are important and may impact
halo properties. Including such correlations require to perform multi-reference (MR) EDF
calculations based on projection techniques and on the generator coordinate
method (GCM)~\cite{bender03c,duguet03c,egido04}.

The first part of the present work, is dedicated to introducing a new method to
identify and characterize halo structures in finite many-fermion systems. Although we only apply the method to even-even, spherical, medium-mass nuclei in the present paper, its range of applicability is wider~\cite{rotival08b}. Regarding nuclei, extensions of the method to odd and deformed
systems can be envisioned. The charge restriction for proton halos identified in light nuclei is such that we do
not expect proton halos in medium-mass systems. As a result, the present work focuses on exotic structures at the
 neutron drip-line.

The article is organized as follows. \sect\ref{sec:std} provides a brief overview of the features that are crucial
to the formation of halos. In \sect\ref{sec:analysisold}, the limitations of existing tools used to characterize
skins and halos, such as the Helm model~\cite{mizutori00}, are highlighted. A new method to properly identify and
characterize halo features of weakly-bound systems in a model-independent fashion is introduced in
\sect\ref{sec:newcrit}. We validate the method using a selection of toy models before applying it to the results
of self-consistent spherical HFB calculations of Cr and Sn isotopes in \sect\ref{sec:1stresults}. The
latter section is also devoted to a critical discussion of our results. Our conclusions are given in
\sect\ref{sec:concl}.

\section{Basic features of halo systems}
\label{sec:std}

The goal of the present section is to outline some of the elements that are crucial to the formation of halos.
This will serve as an introduction to the more quantitative discussion proposed later on as we develop our new
analysis method. For convenience, the discussion is conducted within the EDF framework whose basic aspects are
briefly recalled at first. Note however that the features discussed are not specific to a particular many-body
method or approximation but constitute generic aspects of halos. For simplicity, spin and isospin indices are dropped in the present section.

\subsection{Elements of the nuclear EDF method}
\label{sec:EDF}

The nuclear EDF approach is the microscopic tool of choice to study medium-mass and heavy nuclei in a systematic
manner~\cite{bender03b}. We consider a Single-Reference EDF formalism. In such an implementation, the energy is
postulated under the form of a functional ${\cal E}[\rho,\kappa,\kappa^{\ast}]$ of the (local or non-local)
density $\rho$ and pairing tensor $\kappa$. The density matrix and the pairing tensor are further represented
through a {\it reference state} $| \Phi \rangle$

\begin{eqnarray}
\rho_{ij} & \equiv & \frac{\langle \Phi | c^{\dagger}_{j} c_{i}
            | \Phi \rangle}
           {\langle \Phi  | \Phi  \rangle} \, \, \, ,
  \label{intrrho}    \\
\kappa_{ij} & \equiv & \frac{\langle \Phi | c_{j} c_{i}
            | \Phi \rangle}
           {\langle \Phi | \Phi \rangle} \, \, \, ,
  \label{intrkappa}
\end{eqnarray}
which takes the form of a quasiparticle vacuum and which reduces to a standard Slater determinant if no explicit $\kappa$ dependence of the EDF (${\cal E}$) is considered. Such a product state reads

\begin{equation}
| \Phi \rangle = {\cal C} \, \prod_\nu \beta_\nu | 0 \rangle \, \, \, , \label{initialstates}
\end{equation}
where ${\cal C}$ is a complex normalization number whereas the quasiparticle operators $(\beta^{\dagger}_\nu,
\beta_\nu)$ are obtained through the Bogoliubov transformation $(U,V)$ of the creation and annihilation operators
$(c^{\dagger}_i, c_i)$ defining an arbitrary single-particle basis

\begin{eqnarray}
\label{eq:bogo} \beta^{\dagger}_\nu & \equiv & \sum_i  U_{i \nu} \, c^{\dagger}_i + V_{i \nu} \, c_i  \, \, \, .
\end{eqnarray}

The equations of motion, the so-called HFB equations, are obtained by minimizing the energy ${\cal
E}[\rho,\kappa,\kappa^{\ast}]$ with respect to the degrees of freedom \mbox{$(\rho_{ij},\rho^*_{ij},\kappa_{ij},\kappa^*_{ij})_{i\le
j}$}, under the constraint that the neutron and proton numbers are fixed on the average in the reference state $|
\Phi \rangle$. This leads to solving the eigenvalue problem

\begin{equation}
\left(
\begin{array}{cc}
h-\lambda &\Delta\\
-\Delta^{*}&-(h^*-\lambda)
\end{array}
\right)\left(\begin{array}{c} U\\
V\end{array}\right)_\nu=E_\nu\left(\begin{array}{c} U\\ V\end{array}\right)_\nu \, \, \, ,
\label{eq:hfb_equation_2N_space}
\end{equation}
where the one-body field $h$ and the pairing field $\Delta$ are defined as

\begin{equation}
h_{ij}\equiv\frac{\partial \mathcal{E}}{\partial \rho_{ji}} \, \, \, , \, \, \,  \Delta_{ij}\equiv\frac{\partial
\mathcal{E}}{\partial \kappa^*_{ij}} \, \, \, ,
\end{equation}
\mbox{$\lambda<0$} being the chemical potential. Solutions of \eq(\ref{eq:hfb_equation_2N_space}) are the
quasiparticle eigenstates \mbox{$(U,V)_\nu$} whose occupations $N_\nu$
are defined through the norm of the lower components $V_{\nu}$

\begin{equation}
N_{\nu}\equiv\sum_{k}\left| V_{\nu k}\right|^2=\int \left|V_{\nu}(\vec{r}\,)\right|^2\,d\vec{r} \, \, \, .
\label{eq:def_qp_occ}
\end{equation}

In order to analyze the properties of the many-body system, it is convenient to introduce the {\it canonical
basis}\footnoteb{The canonical basis is the name given to the {\it natural basis} in the context of HFB
calculations.} \mbox{$\{|\phi_i\rangle\}$}~\cite{ring80a,tajima92a}. In this basis, individual states can be
grouped in conjugated pairs \mbox{$(i,\bar{\imath})$}. The one-body density ${\rho}$ is diagonal whereas the
pairing tensor ${\kappa}$ takes its canonical form

\begin{eqnarray}
\rho_{ij}&\equiv&{v_i}^2 \, \delta_{ij}\,,\\
\kappa_{ij}&\equiv&u_{{i}} \, v_{{i}} \, \delta_{\bar{\imath}j}\,,
\end{eqnarray}
where \mbox{$u_{i}=u_{\bar{\imath}}>0$} and \mbox{$v_{i}=-v_{\bar{\imath}}$} play the role of BCS-like
coefficients; $v_i^2$ being the canonical occupation number. Even though the EDF method is not an independent
particle theory, it is convenient to use the canonical basis for analysis purposes as it provides the most intuitive
single-particle picture and allows one to define individual "energies" and "pairing gaps" through
\begin{eqnarray}
e_i&\equiv&h_{ii} \, \, \, , \\
\Delta_i&\equiv&\Delta_{i\bar{\imath}} \, \, \, .
\end{eqnarray}

\subsection{Importance of low angular-momentum orbits}
\label{sec:lowlmanybody}

We first discuss the impact of low-angular momentum orbitals\footnoteb{Although the notion of {\it orbital} often
refers to an independent-particle picture or a Hartree-Fock approximation, it is important to note that the EDF
method includes correlations beyond such approximations. In fact, and as discussed in \sect\ref{sec:newcrit}, the
notion of {\it orbital} should rather be replaced by the one of {\it overlap function} in the present
discussion.} on the density profile of halo nuclei. To do so, we first use the realization of the
EDF method in which the reference state is taken as a Slater determinant. This corresponds to eliminating the
dependence of the EDF on anomalous densities and thus the {\it explicit} treatment of pairing correlations.
It is important to stress that, at least in principle, this does not mean that the effect of
superfluidity could not be accounted for in such a realization of the EDF method. It would, however, certainly
require the design of more involve energy functionals ${\cal E}[\rho]$ that those used traditionally; i.e. Skyrme~\cite{vautherin72a} and
Gogny~\cite{decharge80a} EDFs.

Within such a realization of the EDF method, the HFB equations reduce to a standard one-body eigenvalue problem
that provides the orbitals $\varphi_{\nu}(\vec{r}\,)$ from which the auxiliary Slater determinant $| \Phi \rangle$ is
constructed. Such a basis coincides in this case with both the canonical basis and the quasiparticle basis introduced in
\sect\ref{sec:EDF}. Restricting the description to spherical systems, considering for simplicity a
multiplicative local potential $U(r)$ and forgetting about the spin degree of freedom, it can be
proven~\cite{riisager92} that the density $\rho(r)$
behaves asymptotically as \mbox{$e^{-2\kappa_0\,r}/(\kappa_0\,r)^2$}, where the
decay constant $\kappa_0=\sqrt{-2m\epsilon_0/\hbar^2}$ is related to the eigenenergy $\epsilon_0$ of the least bound occupied orbital in the
reference Slater determinant. As the density used in the SR-EDF method is meant to reproduce the internal
local density (see Appendix~\ref{app:intrinsic}), an analogue of Koopmans' theorem~\cite{koopmans34} holds, that is $\epsilon_0$ is equal to minus
the one-nucleon separation energy \mbox{$S_n=E_0^{N-1}-E_0^{N}$}, where $E_0^{N}$ is the ground state internal
energy of the \mbox{$N$-body} system. As a result, long density tails arise for weakly-bound systems; i.e. in the
limit \mbox{$S_n=|\epsilon_0|\rightarrow0$}.

A more quantitative characterization of the density is provided by its radial moments \mbox{$\langle r^n\rangle$}.
Such moments are of special interest in the case of halo systems. At long distances, the dominant contribution to
\mbox{$\langle r^n \rangle$} comes from $\varphi_0$. In the
limit of weak binding \mbox{$\epsilon_0\rightarrow0$}, the individual moment
\mbox{$\langle r^n\rangle_0$} (i) diverges as
\mbox{$\displaystyle\epsilon_0^{\frac{2\ell-1-n}{2}}$} for \mbox{$n>2\ell-1$}, (ii)
diverges as \mbox{$\displaystyle \ln(\epsilon_0)$} for \mbox{$n=2\ell-1$}, or (iii)
remains finite for \mbox{$n<2\ell-1$}~\cite{riisager92}. In particular, one
finds that the wave function
normalization \mbox{$\langle r^0\rangle_0$} diverges for $s$ waves,
whereas the second moment \mbox{$\langle r^2\rangle_0$} diverges for both
$s$ and $p$ waves. As a result, the root-mean-square (r.m.s.) radius, defined as

\begin{equation}
R_{\mathrm{r.m.s.}}\equiv\sqrt{\frac{\langle r^2\rangle}{\langle r^0\rangle}} \, \, \, ,
\end{equation}
diverges as \mbox{$\epsilon_{0}\rightarrow0$} if $\varphi_0$ corresponds to a $s$ or a $p$ wave. It diverges as
\mbox{$\epsilon_0^{-\frac{1}{2}}$} for a $s$ wave and as $\epsilon_0^{-\frac{1}{4}}$ for a $p$ wave.
The centrifugal barrier confines wave functions with higher orbital-angular momenta, in such a way that $R_{\mathrm{r.m.s.}}$
remains finite as \mbox{$\epsilon_0\rightarrow0$} if $\varphi_0$ has an angular momentum \mbox{$\ell\ge2$}.
Equivalent arguments are found in the case of three-body systems~\cite{fedorov94b}.

According to the above analysis, only low-lying $s$ or $p$ waves near the threshold are able to extend
significantly outside the classically forbidden region. The consequences of such patterns are that (i)
one usually focuses on the evolution of the neutron r.m.s radius as a function of neutron number, looking for a sudden increase as a signature of the building of a halo, (ii) the presence and occupation of low-lying $s$ or $p$ waves are often seen as a
prerequisite for the formation of neutron halos, (iii) orbitals with $\ell \ge 2$ are not believed to contribute to
halos. However, it is important to notice that \mbox{$\langle r^2 \rangle$} is only the leading moment in the
representation of the density. The complete expansion of $\rho(r)$ involves moments of higher orders which probe
the nuclear density at increasing distances. Even if those higher-order moments weight usually little in the
expansion, one cannot rule out \mbox{$\ell\ge$2-type} halo structures, as \mbox{$\langle r^n\rangle_{0}$} with $n
\ge 2$ diverges in the limit \mbox{$\epsilon_0\rightarrow0$} for such angular momenta: \mbox{$\langle r^4 \rangle$}
diverges for \mbox{$\ell=0,1,2$}, \mbox{$\langle r^6 \rangle$} diverges for \mbox{$\ell=0,1,2,3$}... and so
on~\cite{jensen04}.

\subsection{Role of pairing correlations}
\label{sec:introHFB}

Theoretical investigations of nuclei far from stability, either within
non-relativistic~\cite{doba84a,doba96,terasaki96} or relativistic~\cite{toki91,sharma93,suagahara94} EDF
frameworks, have pointed out the importance of pairing correlations. This makes the implementation of the SR-EDF
method in terms of a quasiparticle vacuum more successful in practice than the one based on a reference
Slater determinant.

The explicit treatment of pairing correlations through dependencies of the nuclear EDF on the anomalous density
changes qualitatively the density profile in loosely bound systems. By studying the asymptotic form of the
quasiparticle wave-functions solution of \eq(\ref{eq:hfb_equation_2N_space}), it is easy to show that the decay
constant $\kappa_0$ at play is now \mbox{$\kappa_0=\sqrt{-2m\epsilon_0/\hbar^2}$}, where
\mbox{$|\epsilon_{0}|\equiv E_0-\lambda$} and \mbox{$E_0\equiv{\rm min}_\nu[E_\nu]$} is the lowest quasiparticle energy
solution of \eq(\ref{eq:hfb_equation_2N_space}). Considering the most extreme case of a canonical state lying at
the Fermi level at the drip-line (\mbox{$e_0\approx\lambda\approx 0$}), one sees that
\mbox{$|\epsilon_{0}|\approx E_{0}\approx\Delta_0\ge0$}. Therefore, everything else being equal, paired densities decrease faster than
unpaired ones at long distances. Because the decay constant does not go to zero as \mbox{$e_0\approx\lambda\approx
0$}, the second moment of the density cannot diverge, whatever the angular momentum of the least bound
quasiparticle. In other words, the effect of pairing correlations is to induce a generic \textit{anti-halo
effect} by localizing the density~\cite{bennaceur99,bennaceur00}.

Two additional effects may however blur such a picture. First, recent HFB calculations performed in terms of a
fixed one-body Wood-Saxon potential have shown that such a pairing anti-halo effect could be ineffective under
extreme conditions~\cite{hamamoto03,hamamoto04}. Indeed, very weakly bound $s_{1/2}$ states (bound by a few keVs)
tend to decouple from the pairing field because of their abnormal extension. As a consequence, \mbox{$E_0={\rm
min}_\nu[E_\nu]$} tends towards zero again as \mbox{$e_0\approx\lambda\approx 0$} and the r.m.s. radius of such an
unpaired orbital may diverge, contributing strongly to the formation of a halo. Although this possibility is to
be considered in principle, the depicted situation of a $\ell=0$ orbit bound by a few keVs right at the drip-line
is rather improbable and would be highly accidental in realistic nuclei.
Second, the pair scattering distributes particles over several canonical orbitals located around the Fermi level. As
compared to the implementation of the EDF based on a Slater determinant, this might lead to the promotion of
particles from low/high angular-momentum states to high/low angular momentum orbitals~\cite{grasso06}. Depending on
the situation, this will favor or inhibit the formation of halos. As opposed to the anti-halo effect discussed
above, the way this process impacts halos depends on the system and on the particular distribution of orbitals
around the Fermi energy at the drip-line.

\section{Existing investigations and analysis methods}
\label{sec:analysisold}

Halo properties of medium-mass drip-line nuclei have been studied for various isotopic chains using relativistic
or non-relativistic EDF
methods~\cite{meng98,nerlo00,mizutori00,im00,sandulescu03,geng04,kaushik05,grasso06,terasaki06}. Owing to the
discussion provided above, the evolution of the r.m.s radii along isotopic chains is often used to characterize
halos in a qualitative manner. One needs however more quantitative characterizations of the halo itself. For
example, the concept of {\it giant halo} was recently introduced on the basis of summing up the occupations of
low-lying orbitals with large r.m.s. radii~\cite{meng98}. Such halo structures, supposedly composed of six to
eight neutrons, have been characterized through relativistic and non-relativistic
methods~\cite{sandulescu03,geng04,kaushik05,grasso06,terasaki06}, mainly for Zr and Ca isotopes, and were related
to the presence of \mbox{$\ell=1$} states close to the Fermi level at the drip-line. Finding giant halos in medium-mass nuclei is intuitively
surprising. Indeed, spatially decorrelated neutrons seem less likely to appear as the mass of the system increases
and their behavior tends to become more collective. We will come back to this point.

The present section is devoted to discussing observables and analysis tools that are usually used to identify
and quantify halo signatures in nuclear systems. The purpose is to introduce generic features which turn out to be
useful later on and, above all to demonstrate the limitations of existing analysis tools.

Chromium and tin isotopic chains are chosen as testing cases throughout this work. Calculations
are performed using the non-relativistic HFB spherical code \verb1HFBRAD1~\cite{bennaceur05a}.
In \verb1HFBRAD1, the space is discretized within a sphere using vanishing boundary conditions for the wave
functions (Dirichlet conditions). Convergence of the calculations as a function of numerical parameters
has been checked for all results presented here. The Skyrme SLy4
functional~\cite{chabanat97,chabanat98} is employed in the particle-hole channel. The particle-particle effective
vertex is a density-dependent delta interaction (DDDI) corresponding to a "mixed-type" pairing. Its
density-dependent form factor is a compromise between a pairing which is constant over the nucleus volume
("volume-type"), and one which is peaked at the nucleus surface
("surface-type")~\cite{tondeur79,krieger90,bertsch,tajima93,terasaki95}. To avoid the ultra-violet divergence associated with the local
nature of the pairing functional, a phenomenological regularization scheme corresponding to a smooth cutoff at
$60$~MeV in the single-particle equivalent spectrum is used~\cite{doba96}. Such a pairing scheme is
referred to as REG-M.

The HFB problem is solved self-consistently. Thus, the shape of the central potential cannot be manually adjusted to
reduce the binding energy of weakly-bound orbitals and halo candidates can only be identified
a posteriori.

\subsection{First characterizations}
\label{sec:test}

\subsubsection{Chromium isotopes}
\label{sec:test_Cr}

\begin{figure}[hptb]
\includegraphics[keepaspectratio, angle = -90, width = \columnwidth]{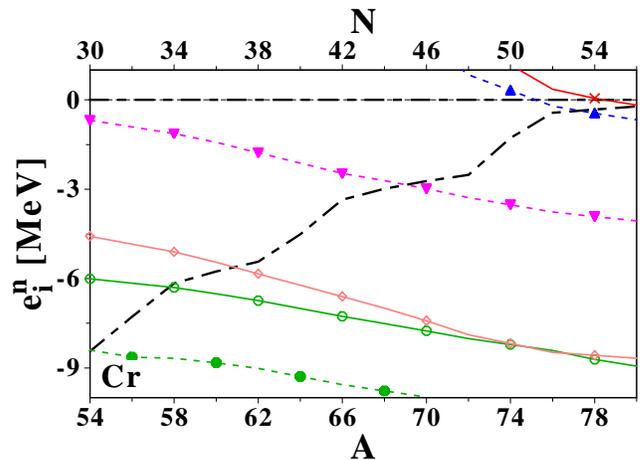}
\caption{ \label{fig:Cr_cano_spectrum} (Color Online) Neutron canonical energies $e^n_i$ along the Cr isotopic chain, obtained
through spherical HFB calculations with the \{SLy4+REG-M\} functional. Conventions used in all the figures of the
article are given in \figu\ref{fig:ref}.}
\end{figure}

Among all medium-mass nuclei predicted to be spherical~\cite{hilaire07,doba04b}, Chromium isotopes (\mbox{$Z=24$}) located at the neutron drip-line are good halo candidates. In \figu\ref{fig:Cr_cano_spectrum}, neutron canonical energies $e^n_i$ in the vicinity of the positive energy threshold are plotted along the Chromium chain, \mbox{$^{80}$Cr} being the predicted
drip-line nucleus. The presence of low-lying $3s_{1/2}$ and $2d_{5/2}$ orbitals at the drip-line
provides ideal conditions for the formation of halo structures.\\

\begin{figure}[hptb]
\includegraphics[keepaspectratio, angle = -90, width = 0.8\columnwidth]{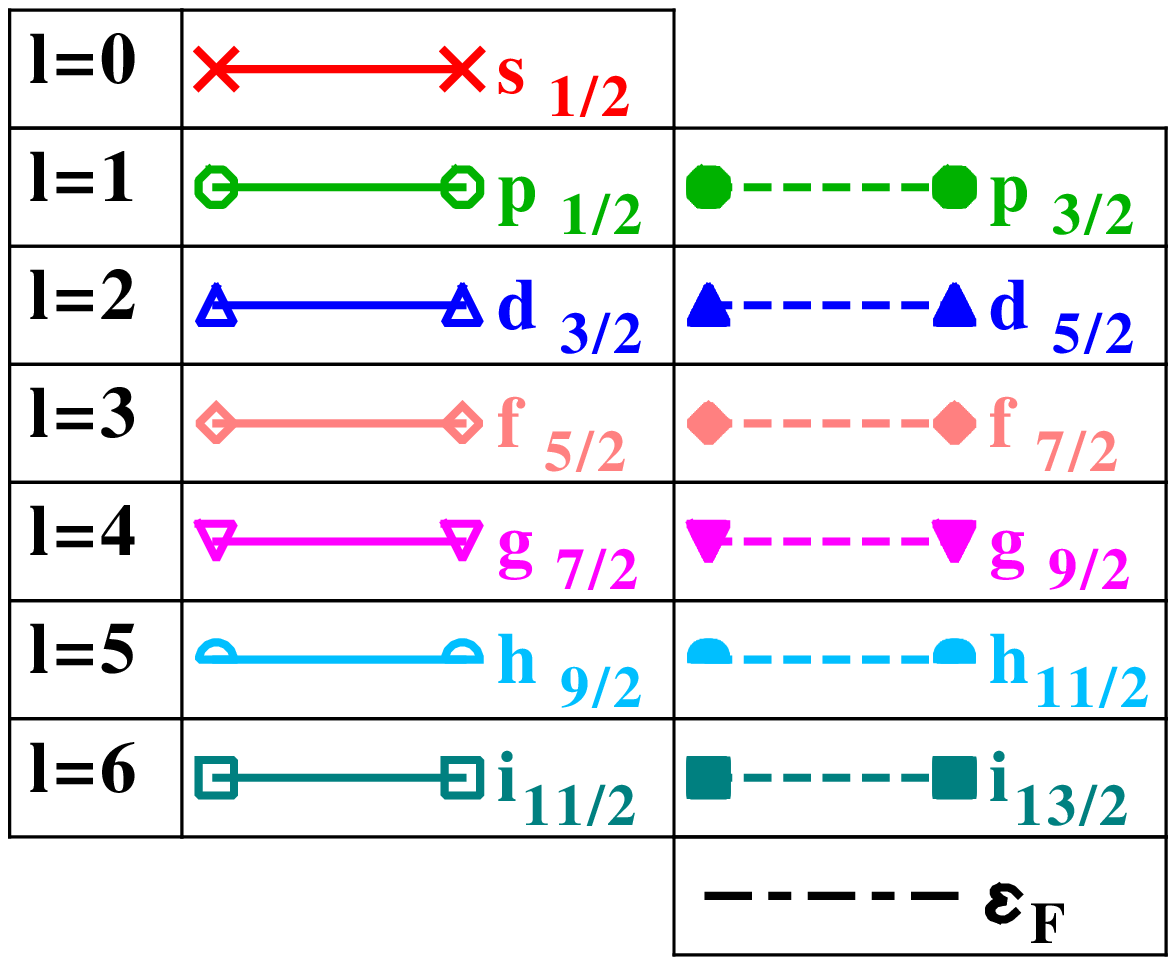}
\caption{\label{fig:ref} (Color Online) Conventions used in all the figures for the labeling of individual states and of the
chemical potential.}
\end{figure}
\begin{figure}[hptb]
\includegraphics[keepaspectratio,angle = -90,  width = \columnwidth]{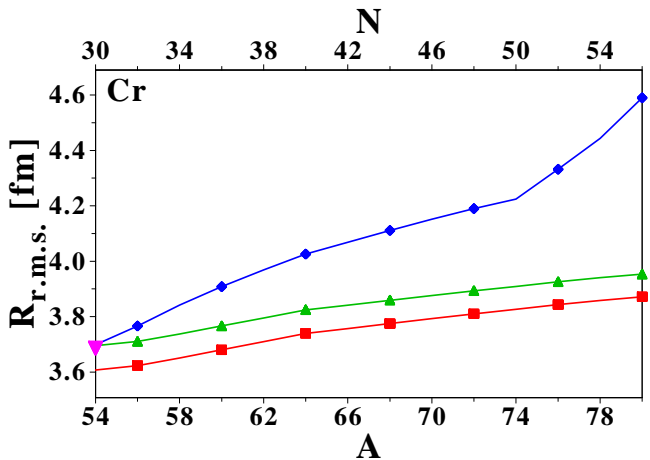}
\caption{ \label{fig:Cr_rms} (Color Online) Same as \figu\ref{fig:Cr_cano_spectrum} for proton ($\blacksquare$),
neutron ($\blacklozenge$) and charge ($\blacktriangle$) r.m.s.
radii. Experimental values for charge r.m.s. radii are indicated
when available ($\blacktriangledown$), along with experimental error bars~\protect\cite{angeli04}.}
\end{figure}
\begin{figure}[hptb]
\includegraphics[keepaspectratio, angle = -90, width = \columnwidth]{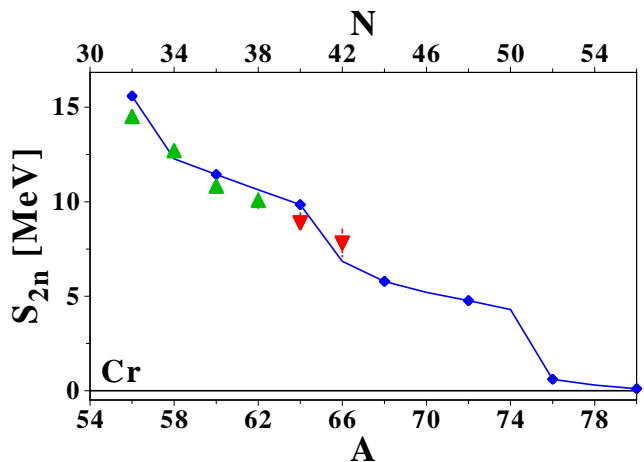}
\caption{ \label{fig:Cr_s2n} (Color Online) Same as \figu\ref{fig:Cr_cano_spectrum} for two-neutron separation
energies $S_{2n}$ ($\blacklozenge$). Experimental values are
indicated when available~\protect\cite{audi2003} ($\blacktriangle$ when both
masses are known, $\blacktriangledown$ when at least one comes from
mass extrapolation), along with experimental error bars.}
\end{figure}

As discussed in \sect\ref{sec:lowlmanybody}, the abnormal extension of the one-body neutron density is usually
characterized through the evolution of the neutron r.m.s. radius as one approaches the drip-line, as
presented in \figu\ref{fig:Cr_rms}. A significant kink in the neutron r.m.s. radius is seen at the \mbox{$N=50$}
shell closure. Such a kink is usually interpreted as a signature of the emergence of a neutron
halo~\cite{meng98,mizutori00}. However, this could equally be due to a simple shell effect. Indeed, as the
\mbox{$N=50$} gap is crossed, the two-neutron separation energy $S_{2n}$ drops, as seen in \figu\ref{fig:Cr_s2n}.
As a result, the decay constant $\kappa_0$ of the one-body density is largely reduced. However, a genuine halo
phenomenon relates more specifically to the presence of nucleons which are spatially decorrelated from a core.
Even though the case of drip-line Cr isotopes seems favorable, as the $S_{2n}$ drops to almost zero at $N=50$, the
occurrence of a halo cannot be thoroughly addressed by only looking at the evolution of the neutron
r.m.s. radius.


\subsubsection{Tin isotopes}
\label{sec:test_Sn}

\begin{figure}[hptb]
\includegraphics[keepaspectratio,angle = -90,  width = \columnwidth]{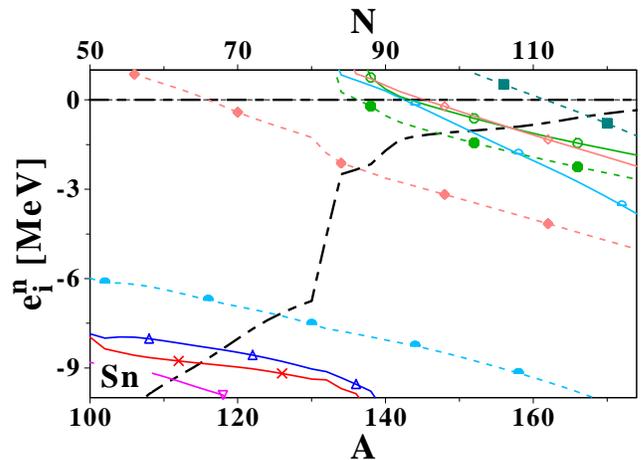}
\caption{ \label{fig:Sn_cano_spectrum} (Color Online) Same as \figu\ref{fig:Cr_cano_spectrum} for Sn isotopes.}
\end{figure}

Sn isotopes (\mbox{$Z=50$}) are considered as a milestone for EDF methods and are rather easy to produce in radioactive beam experiments because of their magic proton number. In particular, the fact that it is a long
isotopic chain is convenient for systematic studies. At the neutron drip-line, which corresponds to
\mbox{$^{174}$Sn} for the \{SLy4+REG-M\} parameter set, the least-bound orbitals are mostly odd-parity states. Among them, $3p_{3/2}$ and $3p_{1/2}$ states might contribute significantly to the formation of a halo (\figu\ref{fig:Sn_cano_spectrum}).

\begin{figure}[hptb]
\includegraphics[keepaspectratio, angle = -90, width = \columnwidth]{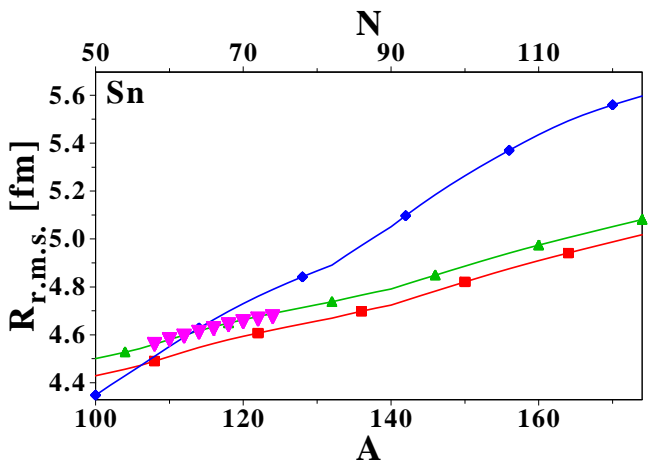}
\caption{ \label{fig:Sn_rms} (Color Online) Same as \figu\ref{fig:Cr_rms} for
Sn isotopes.}
\end{figure}

However, whereas those \mbox{$\ell=1$} states are relatively well bound, the least bound orbital is the
$1i_{13/2}$ \mbox{($\ell=6$)} intruder state which is strongly affected by the confining centrifugal barrier.
Nevertheless, the neutron r.m.s. radius (\figu\ref{fig:Sn_rms}) exhibits a weak kink at the \mbox{$N=82$} shell
closure, which has been interpreted as a halo signature~\cite{mizutori00}.

\begin{figure}[hptb]
\includegraphics[keepaspectratio, angle = -90, width = \columnwidth]{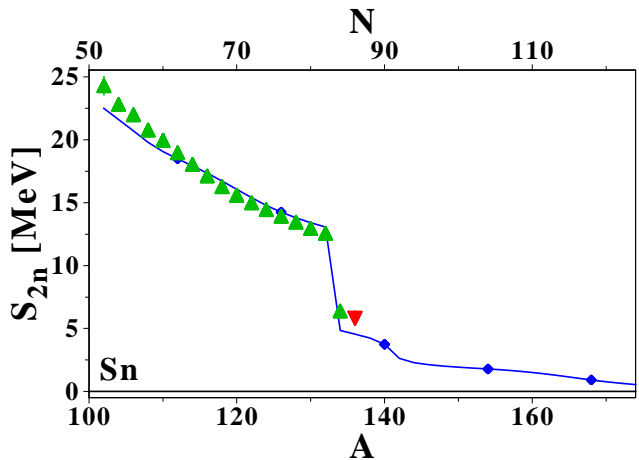}
\caption{ \label{fig:Sn_s2n} (Color Online) Same as \figu\ref{fig:Cr_s2n}
for Sn isotopes.}
\end{figure}

As pointed out previously, an analysis based only on r.m.s. radii is incomplete and can even be misleading. Indeed,
although the shell effect at the \mbox{$N=82$} magic number generates a sudden decrease of the $S_{2n}$, the
latter does not drop to zero, as seen in \figu\ref{fig:Sn_s2n}. A direct connection between the kink
of the r.m.s. radius and the formation of a neutron halo is thus dubious. This point will be further discussed
below.

In any case, the analysis based on neutron r.m.s. radii is useful but insufficient to characterize halo in a
manner that allows the extraction of information useful to nuclear structure and theoretical
models. The characterization of halos through the definition of the neutron matter thickness and the one-neutron
region thickness is possible~\cite{im00} but remains arbitrary and correlated to a one-neutron halo hypothesis.
Another way is to extract so-called "halo factors" from the individual spectrum through antiproton
annihilation probing the nuclear density extension~\cite{lubinski98,nerlo00}. However, such tools do not allow the
extraction of quantitative properties, such as the actual number of nucleons participating in the halo. They also
define the halo as the region where the neutron density dominates the proton one, which is an admixture of the
neutron skin and the (potential) halo.

\subsection{The Helm model}
\label{sec:helm_intro}

\subsubsection{Introduction}
\label{sec:helm_overview}

The Helm model has recently been exploited to remedy to the lack of quantitative measure of halo existence and properties~\cite{mizutori00}.
Originally, the purpose of the Helm model \refers\cite{helm56,rosen57,raphael70} was to fit experimental charge densities, using a few-parameter anzatz, in view of analyzing electron scattering data. The normalized nuclear charge density is approximated by the convolution of a sharp-sphere density of radius $R_0$ defining the nuclear extension and of a gaussian of width $\sigma$ describing the surface
thickness. The r.m.s. radius of the Helm density solely depends on $R_0$ and $\sigma$ and reads as

\begin{equation}
R_{\mathrm{r.m.s.}}^H=\sqrt{\frac{\int \rho_H(r)\,r^4\,dr}{\int
\rho_H(r)\,r^2\,dr}}=\sqrt{\frac{3}{5}\left(R_0^2+5\sigma^2\right)}\,.
\end{equation}

This model has been used to study neutron skins and halos in medium-mass nuclei close to the neutron
drip-line~\cite{mizutori00}. Proton and neutron densities were defined as a superposition of a core density
$\rho^q_{\mathrm{core}}$ plus a tail density $\rho^q_{\mathrm{tail}}$ describing, when appropriate, the halo. The idea was to
reproduce the core part $\rho^q_{\mathrm{core}}$ using the Helm anzatz $\rho^q_H$, normalized to
the nucleon number $N^q$ (\mbox{$N^q=N$} or $Z$). Thus, the two free parameters \mbox{$(R_0^q, \sigma^q)$} were adjusted
on the high momentum part of the realistic form factor

\begin{equation}
F^q(k)=4\pi\int\rho^q(r)\,r^2\,j_0(k\,r)\,dr\,,
\end{equation}
where $\rho^q(r)$ is the density coming out of the many-body calculations. It was suggested in
\refer\cite{mizutori00} to evaluate (i) $R^q_0$ through the first zero $k_1^q$ of the realistic form factor:
\mbox{$R^q_0=z_{1}^1/k_1^q$}, where $z_{1}^1$ is the first zero of the Bessel function $j_1$ \mbox{($z_1^1\approx
4.49341$)}, and (ii) $\sigma^q$ by comparing the model and realistic form factors at their first extremum $k^q_M$
(a minimum in the present case). Then, the following radii are defined (i) \mbox{$ R_{\mathrm{geom}}(q)=\sqrt{5/3}\,R_{\mathrm{r.m.s.}}(q)$} ({geometric} radius) for
realistic densities, and (ii) \mbox{$ R_{\mathrm{Helm}}(q)=\sqrt{5/3}\,R_{\mathrm{r.m.s.}}^H(q)=\sqrt{{R^q_{0}}^2+5\,{\sigma^q}^2}$}
(Helm radius) for model densities.

Adjusting the Helm parameters to the high momentum part of the realistic form factor was meant to make the fitting
procedure as independent of the asymptotic tail of $\rho^q(r)$ as possible. Constructed in this way, $R_{\mathrm{Helm}}(n)$
should not incorporate the growth of $R_{\mathrm{geom}}(n)$ when the neutron separation energy drops to zero and the
spatial extension of weakly-bound neutrons increases dramatically. In addition, it was checked that the difference
between $R_{\mathrm{Helm}}(p)$ and $R_{\mathrm{geom}}(p)$ was negligible near the neutron drip-line. From these observations, the
{neutron skin} and {neutron halo} contributions to the geometric radius were defined as\footnoteb{Similar
definitions could be applied to nuclei close to the proton drip-line, where a proton halo is expected instead of a
neutron one.}

\begin{equation}
\left\{
\begin{array}{l}
\displaystyle\Delta R_{\mathrm{skin}}\equiv R_{\mathrm{Helm}}(n)-R_{\mathrm{Helm}}(p) \, \, \, , \\\\
\displaystyle\Delta R_{\mathrm{halo}}\equiv R_{\mathrm{geom}}(n)-R_{\mathrm{Helm}}(n) \, \, \, .
\end{array}
\right.\label{eq:helm_halo_crit}
\end{equation}

\subsubsection{Limitations of the Helm model}

\label{sec:helm_results}
\begin{figure}[hptb]
\begin{minipage}{\columnwidth}
\includegraphics[keepaspectratio, angle = -90, width = \columnwidth]{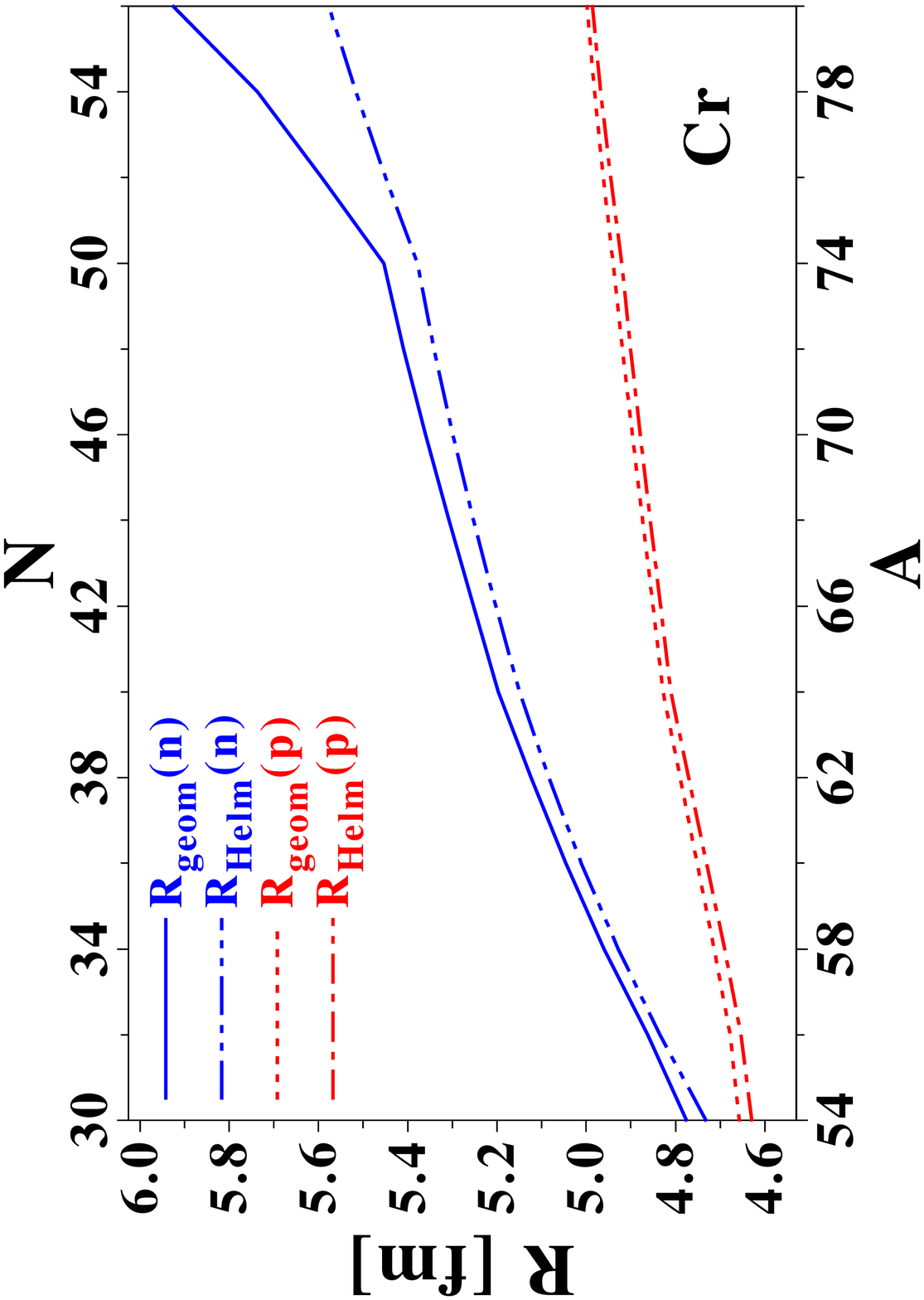}
%\\
\includegraphics[keepaspectratio, angle = -90, width = \columnwidth]{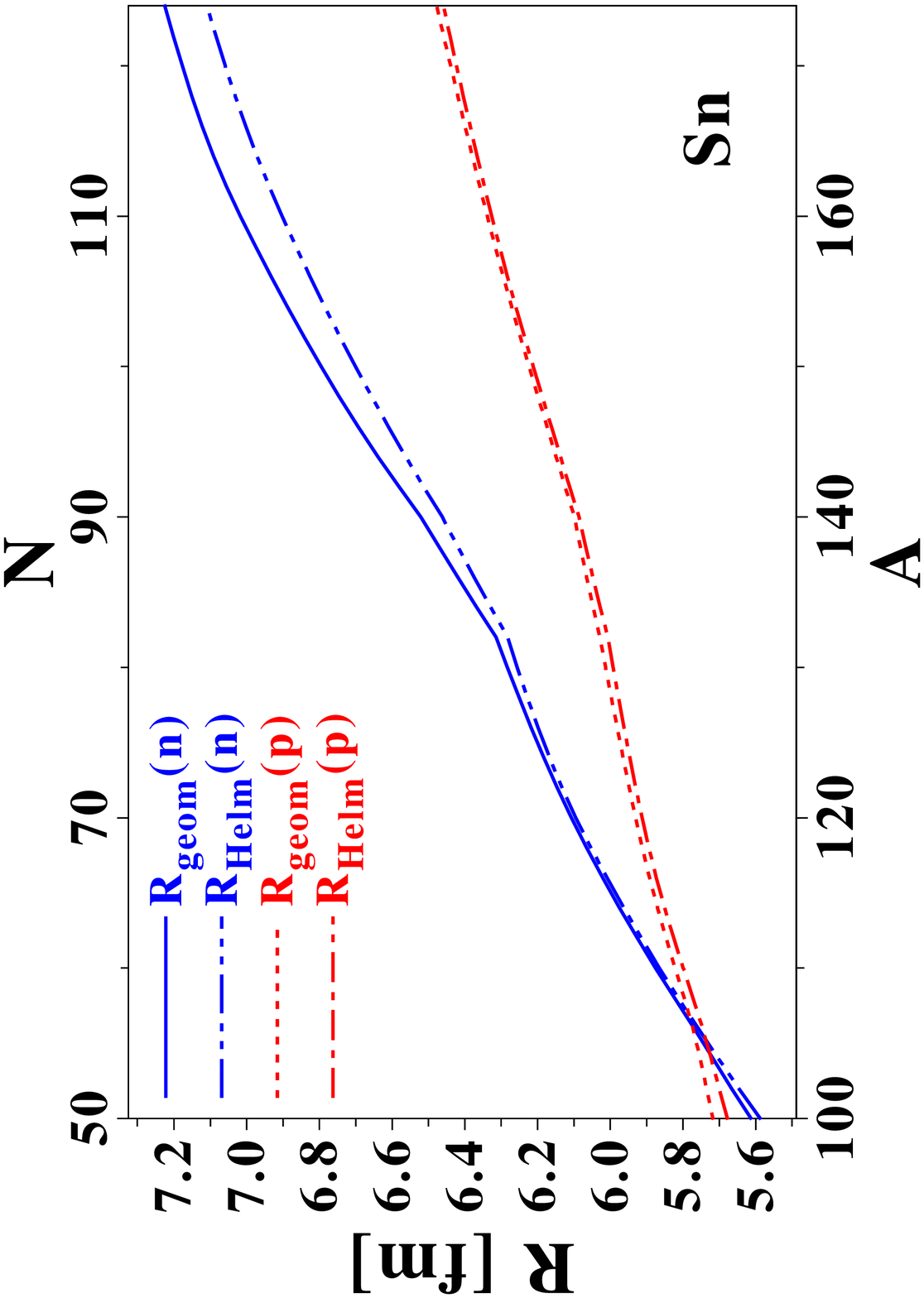}
\caption{ \label{fig:helm1} (Color Online) Geometric and Helm radii for Cr and Sn isotopes calculated in the spherical HFB approach with the
\{SLy4+REG-M\} functional.}
\end{minipage}
\end{figure}

Proton and neutron Helm radii are compared to geometric ones on \figu\ref{fig:helm1} for chromium and tin
isotopes. The behavior of $R_{\mathrm{geom}}(q)$ and $R_{\mathrm{Helm}}(q)$ for Sn isotopes is the same as in
\refer\cite{mizutori00}\footnoteb{Results differ slightly from \refer\protect\cite{mizutori00} because of the
different pairing functional and regularization scheme used, as well as the larger number of \mbox{$j$-shells}
taken into account in the present calculations.}.
For both isotopic chains, the sudden increase of the neutron
geometric radius beyond the last neutron shell closure might be interpreted as a signature of a halo formation.
However, $\Delta R_{\mathrm{halo}}$ is non-zero along the entire Cr isotopic chain, even on the proton-rich side. The
latter result is problematic as neutron halos can only be expected to exist at the neutron drip-line.

\begin{figure}[hptb]
{
\includegraphics[keepaspectratio, angle = -90, width = \columnwidth]{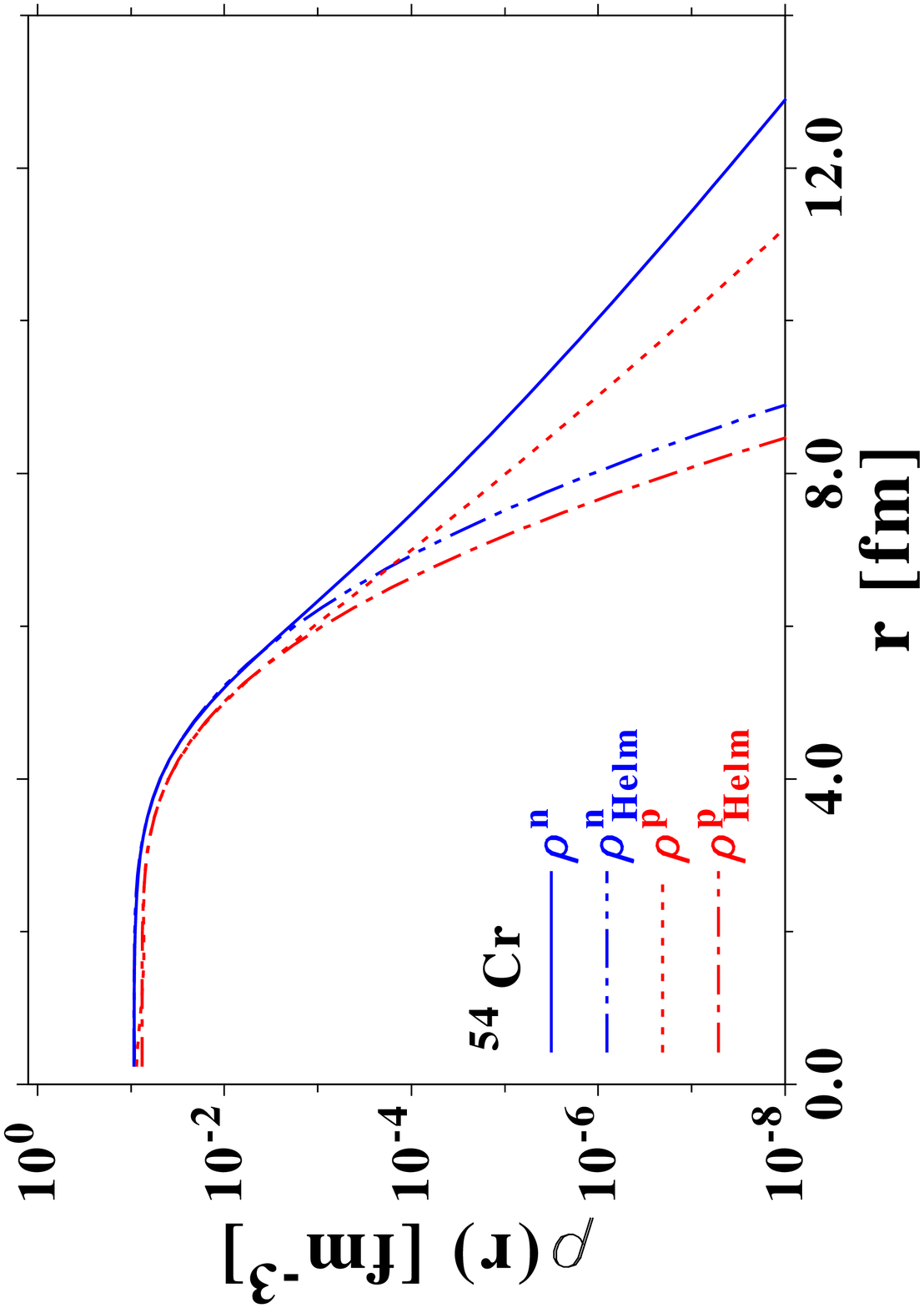}
\includegraphics[keepaspectratio, angle = -90, width = \columnwidth]{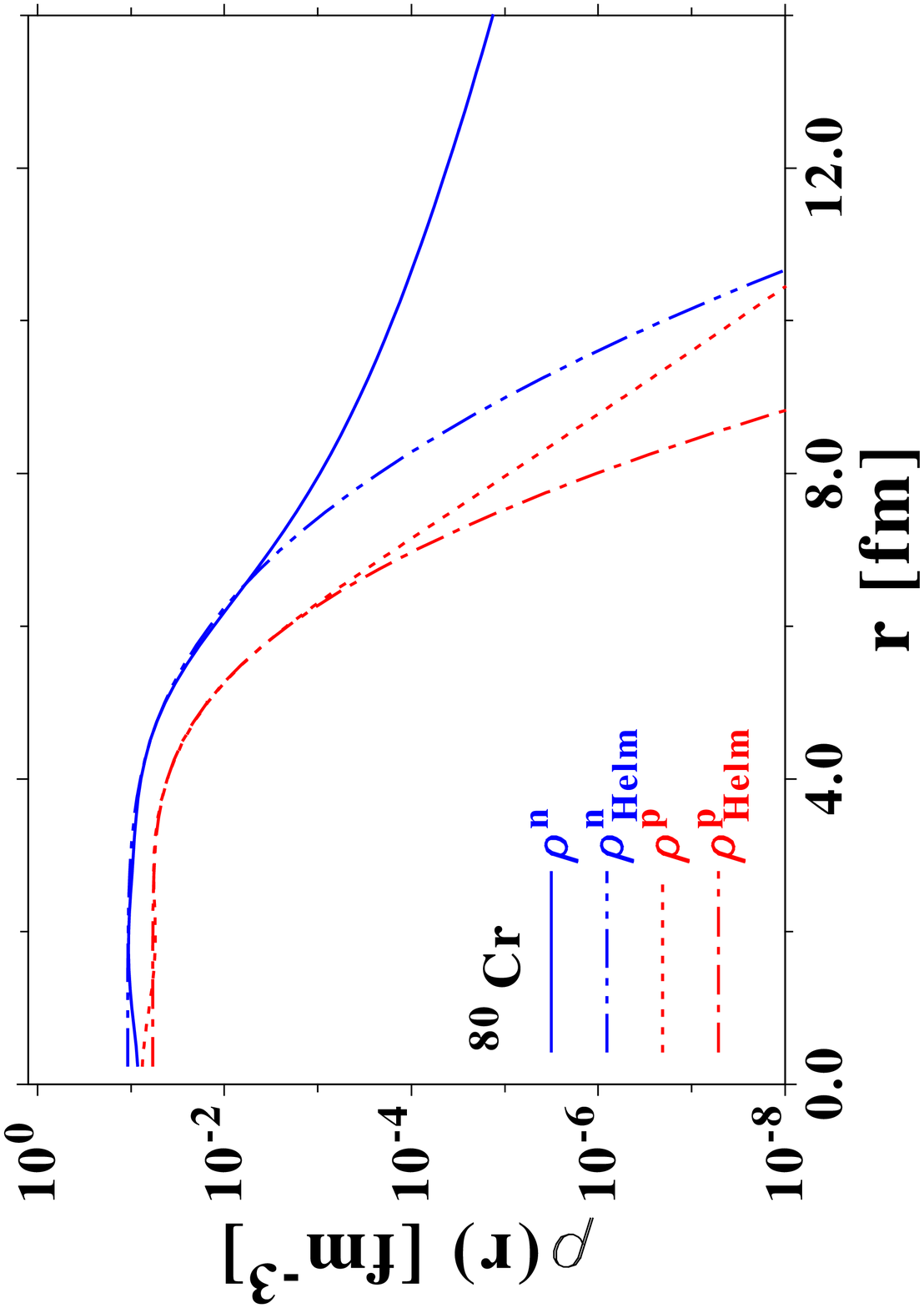}
}\caption{\label{fig:helm_Cr_dens} (Color Online) Realistic and Helm densities of \mbox{$^{54}$Cr} and \mbox{$^{80}$Cr}.}
\end{figure}

Such non-zero values for $\Delta R_{\mathrm{halo}}$ can be understood as a direct consequence of the gaussian folding in
the definition of the Helm density. The asymptotic decay of the Helm density is roughly
quadratic in logarithmic scale, instead of being linear~\cite{doba84a,doba96,tajima04}. To illustrate
this point, \figu\ref{fig:helm_Cr_dens} displays the realistic and Helm densities of \mbox{$^{54}$Cr} (in the
valley of stability) and \mbox{$^{80}$Cr} (drip-line nucleus). The difference in the asymptotic behaviors is
obvious. In particular, the Helm densities are unable to reproduce the correct long-range part of the non-halo
proton density, or the neutron density of nuclei in the valley of stability.

Such features lead to unsafe predictions for the halo parameter $\Delta R_{\mathrm{halo}}$ as the neutron skin and the
potential halo are not properly separated. Such problems, as well as a lack of flexibility to
account for finer details of the nuclear density had already been pointed out~\cite{friedrich82}.

One might thus question the fitting procedure introduced in \refer\cite{mizutori00}. The method naturally requires
$R^q_0$ and $\sigma^q$ to be adjusted on the form factor at sufficiently large $k$ so that the Helm density relates
to the core part of the density only. Of course, some flexibility remains, e.g. one could use the second zero $k_2^q$ of $F^q(k)$ to adjust
$R_0^q$. Following such arguments, four slightly different adjustment procedures $A_i$, $i=1,4$, all consistent with the general
requirement exposed above, have been tested to check the stability of the Helm model

\begin{description}
\item[$A_1$]: (i) \mbox{$F^q_H(k^q_1)=F(k^q_1)$} (ii) \mbox{$F^q_H(k^q_M)=F^q(k^q_M)$}\,,
\item[$A_2$]: (i)
\mbox{$F^q_H(k^q_1)=F^q(k^q_1)$} (ii) \mbox{${F^q_H}'(k^q_1)={F^q}'(k^q_1)$}\,,
\item[$A_3$]: (i)
\mbox{$F^q_H(k^q_2)=F^q(k^q_2)$} (ii) \mbox{${F^q_H}'(k^q_2)={F^q}'(k^q_2)$}\,,
\item[$A_4$]: (i)
\mbox{$F^q_H(k^q_1)=F^q(k^q_1)$} (ii) \mbox{${F^q_H}'(0.4\,k^q_1)={F^q}'(0.4\,k^q_1)$}\,.
\end{description}

\begin{figure}[hptb]
\includegraphics[keepaspectratio, angle = -90, width =\columnwidth]{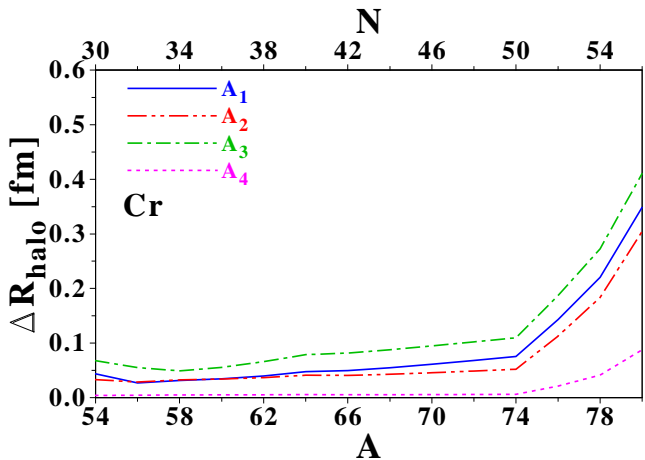}
\caption{\label{fig:helm3} (Color Online) Halo parameter $\Delta R_{\mathrm{halo}}$ for
chromium isotopes using different fitting procedures for the Helm
parameters $(R_0^q,\sigma^q)$ (see text).}
\end{figure}

\figu\ref{fig:helm3} shows the halo parameter $\Delta R_{\mathrm{halo}}$ obtained for Cr isotopes using protocols $A_1$ to
$A_4$. Note that protocol $A_1$ is the one proposed in \refer\protect\cite{mizutori00} and used earlier whereas the weight of the long distance part of the realistic density is more important in protocol $A_4$. Although the general
pattern remains unchanged, the halo parameter significantly depends on the fitting procedure used to determine
$(R^q_0,\sigma^q)$. Because of the wrong asymptotic behavior of the Helm density discussed above, one cannot make $\Delta
R_{\mathrm{halo}}$ to be zero for magic and proton-rich nuclei~(see protocol $A_4$), keeping unchanged its values for halo
candidates at the neutron drip-line\footnoteb{Helm densities obtained with the $A_4$ protocol still do not
match the realistic ones, even for protons.}. Such a fine tuning of the fitting procedure that would make use of
an a priori knowledge of non-halo nuclei is impractical and unsatisfactory.

As a next step, we tried to use other trial densities to improve on the standard Helm model. A key
feature is to obtain an analytical expression of the associated model form factor in order to adjust easily its free parameters. We could not find any form leading to both an analytical expression of $F^q_H$ and good asymptotic, with only two free
parameters\footnoteb{Using model densities depending on three parameters would make the Helm model even more
dependent on the fitting procedure.}.

Although the Helm model looked promising at first,
 we have shown the versatility of its predictions.
The inability of the model to describe the correct asymptotic of the
nuclear density in the valley of stability, as well as the too large
freedom in the fitting procedure, limit very much its predictive
power. Therefore a more robust analysis method is needed to
characterize medium-mass halo nuclei.

\section{New criterion for a quantitative analysis of halo systems}
\label{sec:newcrit}

Although deceiving, the previous attempts have underlined the following point: a useful method to study halos must
be able to characterize a \textit{spatially} decorrelated component in the nucleon density in a model-independent
fashion. We propose in the following a method which allows the identification of such a contribution to the internal
one-body density. Our starting point is a thorough analysis of medium-range and large-distance properties of the one-body internal density in \sect\ref{sec:newcrit_asympt}. Based on such an analysis, new quantitative criteria to identify and characterize halos are defined in \sect\ref{sec:newcrit_halo_char}. We already outline at this point that the analysis and the associated criteria are applicable to any finite many-fermion system, as long as the inter-fermion interaction is negligible beyond a certain relative distance. As done throughout the article, atomic nuclei are used as typical examples in the present section.

\subsection{Properties of the one-body density}
\label{sec:newcrit_asympt}

\subsubsection{Definitions and notations}
\label{sec:decomp_intro}

Let us start from the non-relativistic \mbox{$N$-body} Hamiltonian\footnoteb{The Coulomb interaction is
omitted here, as the focus is on neutron halos. The spin degrees of freedom are also not explicitly included as
their introduction would not change the final results. Finally, the Hamiltonian is restricted to a two-body
interaction. The conclusions would not change either with the introduction of the three-body force.}

\begin{equation}
{H}^{N}\equiv \sum_{i=1}^N\frac{p_i^2}{2m}+\sum_{\substack{i,j=1\\i<j}}^{N}
V(r_{ij}) \, \, ,
\label{eq:exact_hamil}
\end{equation}
where $p_{i}$ is the single-particle momentum, \mbox{$r_{ij}\equiv|\vec{r}_i-\vec{r}_j\,|$} and $V$ denotes the vacuum
nucleon-nucleon interaction. The nuclear Hamiltonian ${H}^{N}$ is invariant under translation and can be written as a sum of a center-of-mass
part ${H}^{N}_{\mathrm{c.m.}}$ and an internal part ${H}^{N}_{\mathrm{int}}$. Thus, eigenstates of ${H}^{N}$, denoted by
\mbox{${\Psi}^{N}_{i,\vec{K}}(\vec{r}_1\ldots \vec{r}_N)$}, can be factorized into the center-of-mass part (plane
wave) times the internal wave function
\begin{equation}
{\Psi}^{N}_{i,\vec{K}}(\vec{r}_1\ldots \vec{r}_N)=e^{i\,
\vec{K}\cdot\vec{R}_N}\,\Phi^{N}_i(\vec{\xi}_1\ldots\vec{\xi}_{N-1}) \, \, \, , \label{eq:Nbody_cm_decoupling}
\end{equation}
where $\vec{K}$ is the total momentum and $\vec{R}_N$ the
center-of-mass position

\begin{equation}
\vec{R}_N\equiv \frac{1}{N}\sum_{i=1}^N\vec{r}_i\,.
\end{equation}
The word {\it internal} relates to the fact that the
wave function \mbox{$\Phi^{N}_i$} can be expressed in terms of relative coordinates only, such as the {$(N-1)$} independent Jacobi variables

\begin{equation}
\vec{\xi}_i\equiv\vec{r}_{i+1}-\frac{1}{i}\sum_{j=1}^{i}\vec{r}_j\,,
\end{equation}
and is associated with the internal energy $E_i^N$. A consequence is that \mbox{$\Phi^{N}_i$} is invariant under translation of the system in the laboratory frame.

The ground-state internal wave function \mbox{$\Phi_0^N$} can be expanded in terms of the complete orthonormal
set of internal \mbox{$(N-1)$-body} wave functions \mbox{$\{\Phi^{N-1}_\nu\}$}, which are eigenstates of the
\mbox{$(N-1)$-body} internal Hamiltonian~\cite{vanneck96,vanneck98a,vanneck98b,escher01}

\begin{equation}
{H}^{N-1}_{\mathrm{int}}\,{\Phi}^{N-1}_{\nu}(\vec{r}_1\ldots\vec{r}_{N-1})={E}^{N-1}_\nu
\,{\Phi}^{N-1}_{\nu}(\vec{r}_1\ldots\vec{r}_{N-1})\,,
\end{equation}
such that

\begin{multline}
\Phi_0^{N}(\vec{r}_1\ldots\vec{r}_N)=\frac{1}{\sqrt{N}}\sum_{\nu}{\Phi^{N-1}_\nu}
(\vec{r}_1\ldots\vec{r}_{N-1})
\\\times\varphi_{\nu}(\vec{r}_N-\vec{R}_{N-1})\,. \label{eq:decomp_N_N-1_1}
\end{multline}

The states \mbox{$\Phi^{N-1}_\nu$} are ordered by increasing energies, \mbox{$\nu=0$} corresponding to
the ground state of the \mbox{$(N-1)$-body} system. The norm of the overlap functions
$\varphi_\nu(\vec{r}\,)$ provides the so-called spectroscopic factors~\cite{clement73a,clement73b}

\begin{equation}
S_\nu=\int d\vec{r}\,|\varphi_\nu(\vec{r}\,)|^2\,.
\end{equation}

Finally, the relevant object to be defined for self-bound systems is the internal one-body density
matrix~\cite{vanneck93,vanneck98b,shebeko06}

\begin{eqnarray}
\rho_{\mathrm{[1]}}(\vec{r},\vec{r}\,') &\equiv&\sum_{\nu}\varphi^*_\nu(\vec{r}\,')\varphi_\nu(\vec{r}\,)\,,
\label{eq:decomp_density}
\end{eqnarray}
which is completely determined by the overlap functions~\cite{vanneck93}. The actual internal one-body density \mbox{$\rho_{\mathrm{[1]}}(\vec{r}\,)=\rho_{\mathrm{[1]}}(r)$} is extracted as the local part of
the internal density matrix

\begin{equation}
\rho_{\mathrm{[1]}}(r)\equiv \sum_{\nu}|\varphi_\nu(\vec{r}\,)|^2
=\sum_{\nu}\frac{2\ell_\nu+1}{4\pi}|\bar{\varphi}_\nu(r)|^2\,, \label{eq:decomp_density4}
\end{equation}
where the energy degeneracy associated with the orbital momentum has been resolved through the summation over the
spherical harmonics.

\subsubsection{Long-distance behavior and ordering of the $\varphi_\nu(\vec{r}\,)$}
\label{sec:newcrit_dens_asympt}

For large distance, i.e. \mbox{$r>R$}, the nuclear interaction vanishes and the asymptotic radial part
$\bar{\varphi}_\nu$\footnoteb{In the following, the
radial part of a wave function \mbox{$f(\vec{r}\,)$} is noted \mbox{$\bar{f}(r)$}.} of the overlap function is solution of the free Schr\"{o}dinger equation with a reduced
mass \mbox{$m_{\mathrm{red}}=m (N-1)/N$}

\begin{equation}
\left[
\left(\frac{d^2}{dr^2}+\frac{2}{r}\frac{d}{dr}-\frac{\ell_{\nu}(\ell_{\nu}+1)}{r^2}\right)
-\kappa_\nu^2\right]\bar{\varphi}_{\nu}^{\infty}(r)=
0\label{eq:free_eqn_schrodinger}\,,
\end{equation}
with \mbox{$ \kappa_\nu=\sqrt{-2m_{\mathrm{red}}\epsilon_\nu/\hbar^2}$}, and
\mbox{$\epsilon_\nu=\left(E_0^{N}-E_\nu^{N-1}\right)$} is minus the one-nucleon separation energy to reach
\mbox{$\Phi_\nu^{N-1}$}. Solutions of the free Shr\"{o}dinger equation take the form

\begin{equation}
\varphi_\nu^{\infty}(\vec{r}\,)= B_{\nu} \, h_{\ell_{\nu}}(i\,\kappa_\nu\,r)
\,Y_{\ell_{\nu}}^{m_{\nu}}(\theta,\varphi)\,. \label{eq:overlap_asympt}
\end{equation}
As a result, the internal one-body density behaves at long distances as\footnoteb{Rigorously, this is true only if the convergence of the overlap functions to their asymptotic regime is uniform in the mathematical sense, i.e. if they reach the asymptotic regime at a common
distance $R$~\cite{vanneck93}. This is not actually proven in nuclear physics, but it has been shown to be true for the electron charge density in
atomic physics~\cite{levy84,dreizler90}.}

\begin{equation}
\label{eq:rho_asympt_decomp}
\rho_{\mathrm{[1]}}^{\infty}(r)=\sum_{\nu}\frac{B_{\nu}^2}{4\pi}(2\ell_\nu+1)|h_{\ell_\nu}(i\,\kappa_\nu
r)|^2\, \, .
\end{equation}

For very large arguments, the squared modulus of a Hankel function behaves as
\mbox{$e^{-2\kappa_i r}/(\kappa_i r)^2$}~\cite{abramowitz}. Thus the \mbox{$\nu=0$} component dominates and provides the usual
asymptotic behavior~\cite{doba84a,doba96,tajima04}\footnoteb{Note that the asymptotic of $\rho^p$ and
$\rho^n$ are different because of the charge factor (Hankel functions
for neutrons, Whittaker functions for protons).}

\begin{equation}
\label{eq:rho_asympt_usual}
\rho_{\mathrm{[1]}}^{\infty}(r)\underset{r\rightarrow+\infty}{\longrightarrow}\frac{B_{0}^2}{4\pi}(2\ell_0+1)\frac{e^{-2\kappa_0\,r}}{(\kappa_0\,r)^2}\,.
\end{equation}

The asymptotic form of the Hankel function is independent of the angular momentum which explains why
high-order moments \mbox{$\langle r^n\rangle$} of the density diverge when \mbox{high-$\ell$} states are loosely
bound, as discussed in \sect\ref{sec:lowlmanybody}. Thus, the contributions of the overlap
functions to $\rho_{\mathrm{[1]}}^{\infty}$ at very large distances are ordered according to their associated separation
energies \mbox{$|\epsilon_\nu|$}, independently of $\ell_{\nu}$. Corrections to this ordering at
smaller distances come from (i) the \mbox{$\ell$-dependence} of the Hankel functions due to the centrifugal
barrier, which favors low angular momentum states, and (ii) the \mbox{$(2\ell+1)$} degeneracy factor which favors
high angular momentum states. In any case, for extremely large distances the least bound component will always prevail,
although this may happen beyond simulation reach.
%\footnoteb{For instance, if a \mbox{$\ell=6$} component is less
%bound than a \mbox{$\ell=0$} one by only a very few keVs, the former will only become the leading component in the
%asymptotic density at very large distance.}.

\begin{figure}[hptb]
\includegraphics[keepaspectratio, angle = -90, width = \columnwidth]{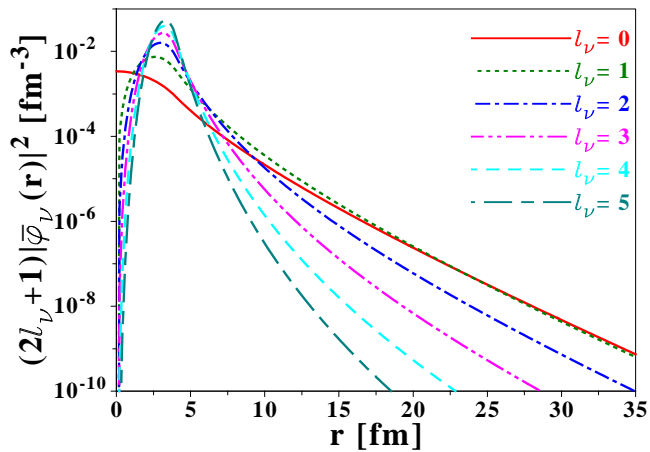}
\caption{\label{fig:sph_well} (Color Online) Squared components of the solutions of a finite spherical well of fixed radius \mbox{$a=4$}~fm,
multiplied by the degeneracy factor \mbox{$(2\ell_{\nu}+1)$}, for various angular momenta and fixed energy
\mbox{$\epsilon_{\nu}=-100$}~keV. The first state for each $\ell_\nu$ (nodeless component corresponding to a
primary quantum number equal to zero) is represented.}
\end{figure}

To characterize the net effect of corrections (i) and (ii) on the relative positioning of overlap functions at long distances, the contributions
\mbox{$(2\ell_\nu+1)|\bar{\varphi}_{\nu}(r)|^2$}, for a fixed energy but different angular momenta, are compared
in \figu\ref{fig:sph_well} for the solutions of a simple finite spherical well. Outside the well, Hankel functions
are exact solutions of the problem. The potential depth is adjusted to obtain identical eigenenergies for all
$\ell_\nu$. Although the \mbox{$(2\ell_\nu+1)$} factor reduces the gap between $s$ and $p$ components, the effect
of the centrifugal barrier is always the strongest at large $r$, where states are clearly ordered according to
$\ell_\nu$, favoring low angular momenta. In any case, the separation energy remains the leading factor as far as
the ordering of overlap functions at long distances is concerned.

\subsubsection{Crossing pattern in $\rho_{\mathrm{[1]}}^{\infty}(r)$}
\label{sec:newcrit_asympt_cons}

The (model-independent) ordering at long distances of individual components entering $\rho_{\mathrm{[1]}}^\infty$ has
interesting consequences on the properties of the density as a whole. As discussed below, this ordering induces a typical
crossing pattern between the individual components which will eventually be used to characterize halo nuclei.

Introducing normalized overlap functions \mbox{$\psi_\nu(\vec{r}\,)$}, \eq(\ref{eq:decomp_density4}) becomes

\begin{equation}
\rho_{\mathrm{[1]}}(r)=\sum_{\nu}
\frac{2\ell_\nu+1}{4\pi}\,S_\nu\,\left|\bar{\psi}_\nu(r)\right|^2\equiv\sum_{\nu}C_\nu(r)
\label{eq:decomp_density2}\,.
\end{equation}

Let us take all spectroscopic factors equal to one for now. The \mbox{$\nu=0$} component, corresponding to the
smallest separation energy, dominates at large distances. Because of continuity % \mbox{($r \,
%\bar{\psi}_{\nu}(r)\in\mathcal{L}^2(\mathbb{R}^+)$)}
and normalization conditions, this implies that \mbox{$\bar{\psi}_0(r)$} has to
cross all the other overlap functions as $r$ goes inward from $+\infty$ to zero. The position at which $\psi_0$
crosses each $\psi_\nu$ depends on the difference of their separation energies and on their angular momenta. In
particular, there will exist a crossing between \mbox{$|\bar{\psi}_0(r)|^{2}$} and the remaining density
\mbox{$\left[\rho^{[1]}(r)-C_0(r)\right]$}. The same is true about \mbox{$|\bar{\psi}_1(r)|^{2}$}: it must cross the
remaining density \mbox{$\left[\rho^{[1]}(r)-C_0(r)-C_1(r)\right]$}... As a result, any given individual component
must cross the sum of those that are more bound. Of course, the centrifugal barrier influences the position of such
crossings but not their occurrence because of the robustness of the (very) asymptotic ordering pattern discussed
in the previous section.

Let us now incorporate the role of spectroscopic factors. In practice, $S_\nu$ is known to increase with the
excitation energy of the corresponding eigenstate of the \mbox{$(N-1)$-body} system. Thus, the norm of $\varphi_0$
is smaller than those of the excited components $\varphi_\nu$, which mechanically ensures the existence of the
crossings discussed previously. A similar reasoning holds when going from $\varphi_0$ to $\varphi_1$ etc.

One should finally pay attention to the number of nodes of the overlap function \mbox{$\bar{\varphi}_\nu$}. This feature actually
favors low angular momentum states as far as the asymptotic positioning is concerned. If two components have the
same energy but different angular momenta, the one with the lowest $\ell$ will have a greater number of nodes. This will reduce the amplitude of the wave function in the nuclear interior. That
is, the weight of the asymptotic tail is increased, which favors its dominance at long distance. However, this
effect is expected to have a small impact in comparison with the other corrections discussed above. As a result, the crossing pattern between the components of the density is not jeopardized by the
existence of nodes in the overlap functions.

\subsection{Halo characterization} \label{sec:newcrit_halo_char}

\subsubsection{Definition}
\label{sec:newcrit_define}

The discussion of \sect\ref{sec:newcrit_asympt_cons} demonstrates how individual contributions to the one-body
density (i) are positioned with respect to each other (ii) display a typical crossing pattern. Such features
are now used to characterize halo systems.

As pointed out earlier, one general and model-independent definition of a halo relates to the existence of
nucleons which are spatially decorrelated from others, constituting the core. This can only be achieved if some
contributions to the internal density exhibit very long tails. Most importantly, the delocalization from the core requires
the latter to exist and to remain well localized. To achieve such a spatial decorrelation between a core and a tail
part, it is necessary to have a crossing between two well-identified groups of orbitals with significantly
different asymptotic slopes. This translates into a sharp crossing between those two groups of
orbitals and thus to a pronounced curvature in the density. Note that this explains the empirical observation that the first logarithmic derivative of the density invariably displays a minimum at some radius~\cite{stoitsov03a}. How much this feature is pronounced or not is key and will be used in
the following to design model-independent criteria to characterize halo systems.

\begin{figure}[hptb]
\includegraphics[keepaspectratio, angle = -90, width = \columnwidth]{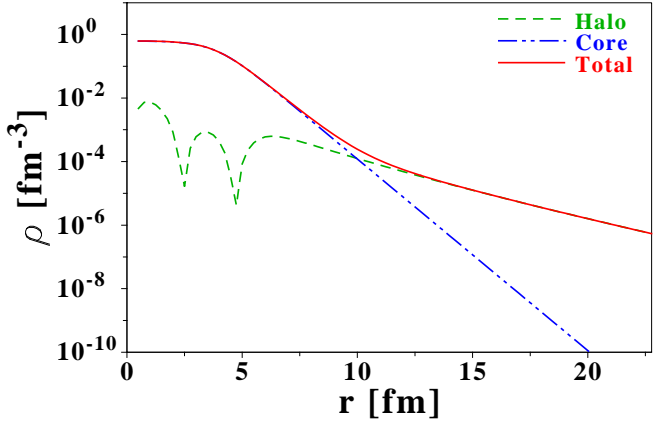}
\caption{\label{fig:model} (Color Online) "Core+tail" simplified model. The total density is the superposition of a well-bound component and a
loosely-bound one. A semi-phenomenological density (see Appendix~\ref{sec:def_r0_simu}) is used for the core
density, whereas the halo part is the realistic $3_{1/2}$ state of \mbox{$^{80}$Cr} obtained from spherical HFB
calculations with the \{SLy4+REG-M\} functional.}
\end{figure}

A pronounced crossing is illustrated in \figu\ref{fig:model} for a simple model where the halo is due to a
single orbital. Of course, more complex situations have to be considered where multiple states contribute to the
core and the halo. Indeed, the presence of collective motions in medium-mass systems implies that one hardly
expects a single orbital to be well separated from the others.

\subsubsection{Relevant energy scales}
\label{sec:newcrit_define2}

The need for the existence of two groups of orbitals characterized by significantly different asymptotic slopes provides critical conditions for the appearance of a halo: (i) the least bound component $\varphi_0$ must have a very small separation energy to extend far out, (ii)
several components \mbox{$\varphi_1, \varphi_2\ldots \varphi_m$} may contribute significantly to the density tail
if, and only if, they all have separation energies of the same order as that of $\varphi_0$, (iii) for this tail to
be spatially {decorrelated} from the rest of the density (the "core"), the components with \mbox{$\nu>\nu_m$} have
to be much more localized than those with \mbox{$\nu\le \nu_m$}. This third condition is fulfilled when the
crossing between the $m^{th}$ and \mbox{$(m+1)^{th}$} components in the density is sharp, which corresponds to
significantly different decay constants \mbox{$\kappa_m\ll\kappa_{m+1}$} at the crossing point.

The later situation translates eventually into specific patterns in the excitation energy spectrum of the \mbox{$(N-1)$-body}
system. It suggests that a halo appears when (i) the one-neutron separation energy
\mbox{$S_n=|\epsilon_0|$} is close to zero, (ii) a bunch of low energy states in the \mbox{$(N-1)$-body} system
have separation energies \mbox{$|\epsilon_\nu|$} close to zero, and (iii) a significant gap in the spectrum of the
\mbox{$(N-1)$-body} system exists, which separates the latter bunch of states $\varphi_\nu$ from higher
excitations.

\begin{figure}[hptb]
\subfigure[~Canonical neutron energy
spectrum $e_i$.]{
\includegraphics[keepaspectratio,width=\columnwidth]{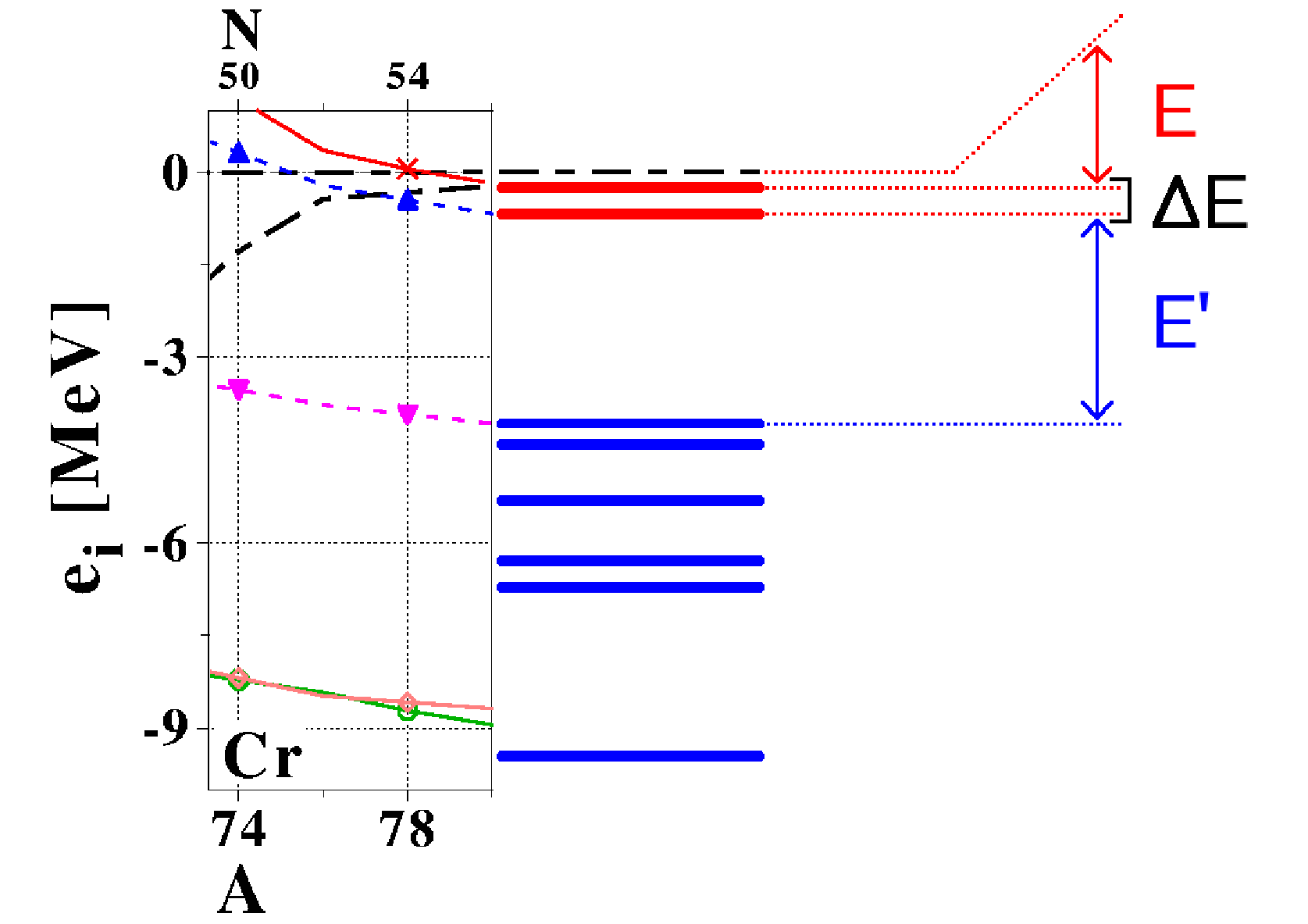}
\vspace{10pt} }
\\
\subfigure[~Separation energy spectrum \mbox{$|\epsilon_\nu|$}
 for the \mbox{$(N-1)$-body} system.]{
\includegraphics[keepaspectratio,width=\columnwidth]{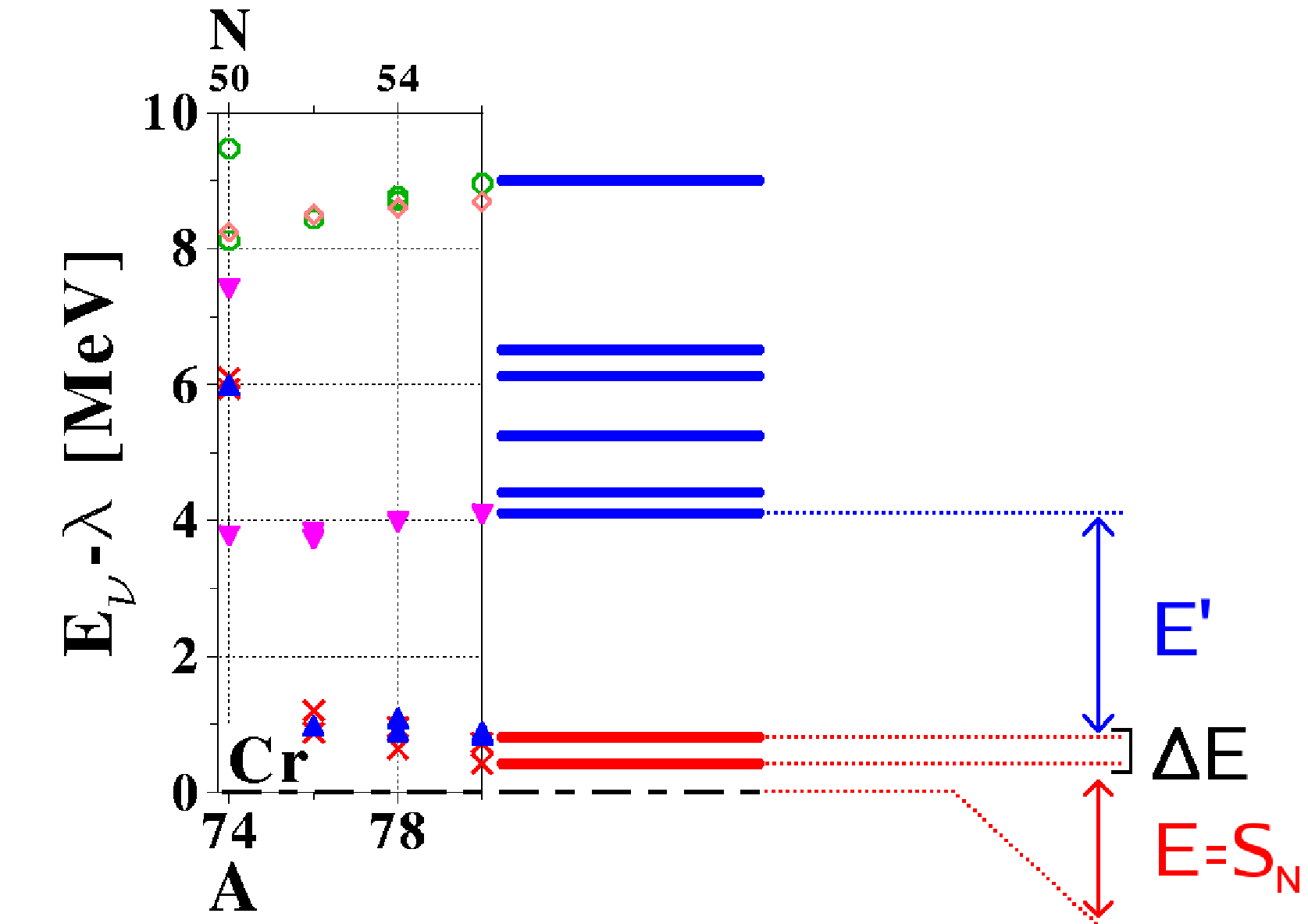}
} \caption{\label{fig:EFT_spectrum} (Color Online) Schematic display of the energy scales relevant for the appearance of halos (right-hand side). The
realistic spectra obtained through HFB calculations of the four last bound chromium isotopes are shown on the
left-hand sides.}
\end{figure}

A similar discussion was given in the context of designing an effective field theory (EFT) for weakly-bound
nuclei~\cite{bertulani02}, where two energy scales \mbox{$(E,E')$} were found to be relevant: (i) the nucleon separation
energy \mbox{$E=S_n$} which drives the asymptotic behavior of the one-body density, and (ii) the core excitation
energy \mbox{$E'=|\epsilon_{m+1}|$} which needs to be such as \mbox{$E'\gg E$}, in order for the tail orbitals to
be well decorrelated from the remaining core. The additional energy scale that we presently identify is the energy
spread \mbox{$\Delta E$} of the low-lying states in the $(N-1)$-body system, which becomes relevant when more than
one component is involved in the halo. The corresponding picture is displayed in the bottom panel of \figu\ref{fig:EFT_spectrum} and is also
translated in terms of canonical energies $e_i$ in the upper panel of the same figure.

More quantitatively, the ideal situation for the formation of a halo is obtained for (i) a very small separation
energy, in orders of a few hundred keVs, the empirical value of \mbox{$2$~MeV$/A^{2/3}$} from
\refers\cite{fedorov93,jensen00} giving a good approximation of expected values, (ii) a narrow bunch of low-lying
states, whose spread \mbox{$\Delta E$} should not exceed about one MeV, and (iii) a large gap $E'$ with the
remaining states, at least four or five times the separation energy $E$. Those are only indicative values, knowing
that there is no sharp limit between halo and non-halo domains.

\subsubsection{Halo region}
\label{sec:newcrit_def_region}

\begin{figure}[hptb]
\includegraphics[keepaspectratio, width = \columnwidth]{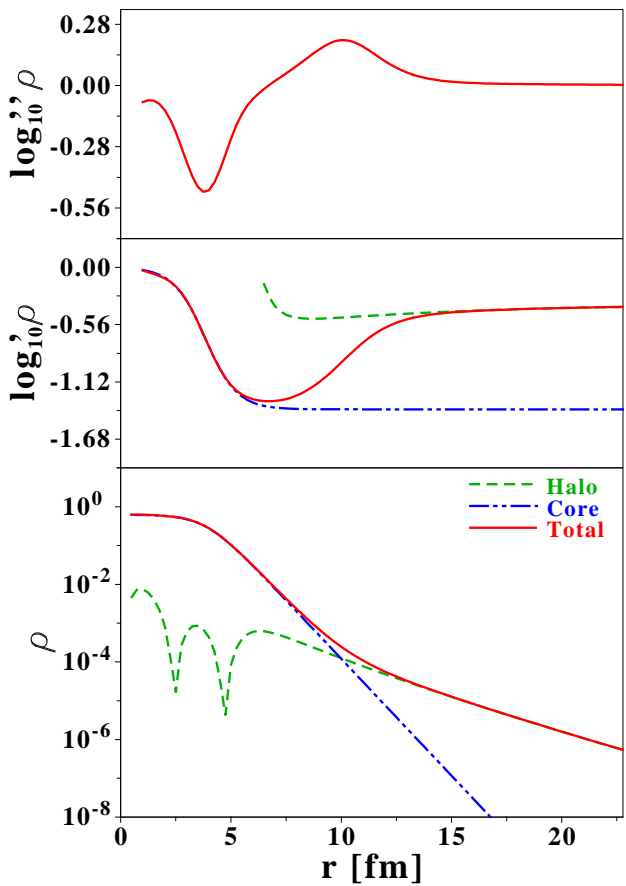}
\caption{\label{fig:crit_qualit_1} (Color Online) Ankle in the (base-$10$) log-density due to the presence of a low-lying state well separated from the remaining
ones: log-density (bottom panel), first (middle panel) and second (top panel) log-derivatives. The conventions
are the same as in \figu\ref{fig:model}.}
\end{figure}

As discussed in the previous section, a halo can be identified through a pronounced {ankle} in the density, due to
the sharp crossing between the aggregated low-lying components and the upper-lying ones. Such a large curvature
translates into a peak in the second derivative of the (base-$10$) logarithmic profile ($\log_{10}$) of the one-body density, as seen in \figu\ref{fig:crit_qualit_1} for a
schematic calculation.

At the radius \mbox{$r=r_{max}$} corresponding to the maximum of that peak, core and tail contributions cross;
i.e. they contribute equally to the total density. At larger radii, the halo, if it exists, dominates. Therefore,
\textit{we define the spatially decorrelated region as the region beyond the radius $r_0$ where the core density
is one order of magnitude smaller than the halo one}. In practice, the previous definition poses two problems.
First, in realistic calculations, one only accesses the total density. Second, the choice of one order of
magnitude is somewhat arbitrary.

\begin{figure}[hptb]
\includegraphics[keepaspectratio, width = \columnwidth]{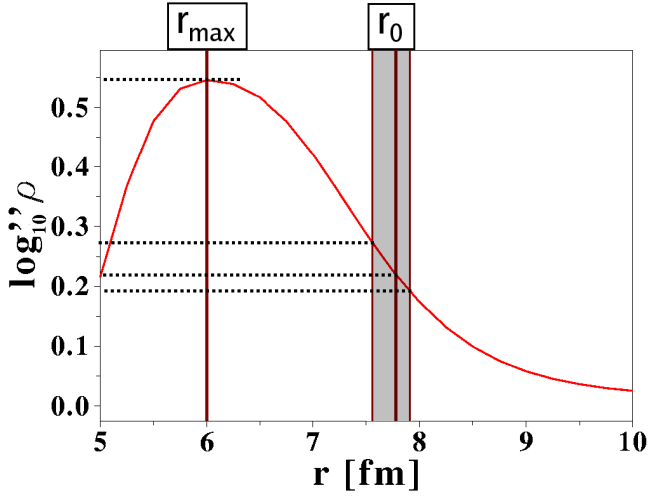}
\caption{(Color Online) Definition of $r_0$ through the
second derivative of the log-density, using the same model density as in
\figu\ref{fig:crit_qualit_1}. $r_0$ is represented by the central vertical line. The shaded area corresponds to the
tolerance margin on $r_0$ (see text).}
\label{fig:def_r_0}
\end{figure}

Extensive simulations have been performed to characterize $r_0$ unambiguously, using either one or several
contributions to the halo density, and covering large energy ranges for $E$, $E'$ and \mbox{$\Delta E$}. More
details on the method used to find the best approximation to $r_0$, as well as the corresponding theoretical
uncertainty, are given in Appendix~\ref{sec:def_r0_simu}. Given $r_{max}$, which can be extracted from the total
density, it has been found that $r_0$ can be reliably defined through

\begin{equation}
\left\{\begin{array}{l}
r_0>r_{max}\,,\\\\
\displaystyle\frac{\partial^2 \log_{10}{\rho(r)}}{\partial r^2}\biggl|_{\displaystyle
r=r_0}\equiv\frac{2}{5}\,\frac{\partial^2 \log_{10}{\rho(r)}}{\partial r^2}\biggl|_{\displaystyle r=r_{max}}\,,
\end{array}\right.\label{eq:def_r0}
\end{equation}
as exemplified in \figu\ref{fig:def_r_0}. Also, theoretical uncertainties
on the determination of $r_0$ are introduced, such that

\begin{equation}
0.35\le\frac{\log_{10}''(\rho(r_0))}{\log_{10}''(\rho(r_{max}))}\le0.50\,,\label{eq:def_r0_err}
\end{equation}
where $'$ denotes a compact notation for $\partial/\partial r$. \\\par Once validated by simulations, the method to
isolate the halo region only relies on the density as an input, and does not require an \textit{a priori}
separation of the one-body density into core and halo parts. Finally, one may note that our definition of the halo
region does not a priori exclude contributions from individual components with angular momenta greater than one.

\begin{figure}[hptb]
\includegraphics[keepaspectratio,angle = -90, width = \columnwidth]{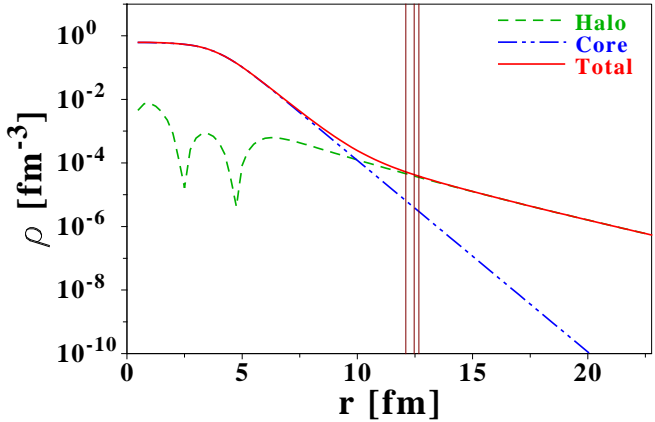}
\caption{(Color Online) Consequences of the definition of $r_0$ (vertical lines for the values of $r_0$ and the tolerance margin -
see text) in the same model as in \figu\ref{fig:crit_qualit_1}. The
halo density dominates the core part by around one order of magnitude.}
\label{fig:def_consec}
\end{figure}

\subsubsection{Halo criteria}
\label{sec:newcrit_crit}

We now introduce several criteria to characterize the halo in a quantitative way, by applying the previous analysis to the neutron one-body density\footnoteb{For neutron-rich medium-mass nuclei, protons are well confined in the nuclear interior, thus do not participate in the long-range part of the total density $\rho$. The two densities $\rho$ and $\rho^n$ can be used regardless to evaluate $N_{\mathrm{halo}}$ and $\delta R_{\mathrm{halo}}$.}.
First, the average number of
nucleons in the halo region can be extracted through

\begin{equation}
\label{eq:def_nhalo} N_{\mathrm{halo}}\equiv 4\pi\int_{r_0}^{+\infty}\!\!\rho^n(r)\,r^2\,dr \, \, \, .
\end{equation}

An important information is the effect of the halo region on the radial moments of the density. By definition, the
contribution of the core to any moment \mbox{$\langle r^n \rangle$} is negligible for \mbox{$r\ge
r_0$}. It has been checked in the case of the r.m.s. radius, and is all the more true as $n$ increases.
Thus, one can evaluate the effect of the decorrelated region on the nuclear extension through

\begin{eqnarray}
\delta R_{\mathrm{halo}}&\equiv&R^n_{\mathrm{r.m.s.},tot}-R^n_{\mathrm{r.m.s.},inner}\notag\\
&=&\!\!\!\!\sqrt{\frac{\int_{0}^{+\infty}\rho^n(r)r^4\,dr}
{\int_{0}^{+\infty}\rho^n(r)r^2\,dr}}-
\sqrt{\frac{\int_{0}^{r_0}\rho^n(r)r^4\,dr}
{\int_{0}^{r_0}\rho^n(r)r^2\,dr}}\,.
\label{eq:def_drhalo}
\end{eqnarray}

The quantity $\delta R_{\mathrm{halo}}$ is similar to $\Delta R_{\mathrm{halo}}$ defined within the Helm model
(\eq(\ref{eq:helm_halo_crit})). However, the former does not rely on any \textit{a priori} decomposition of the
density into core and halo components. That is of critical importance. Extensions to all radial moments of the
density can be envisioned\footnoteb{Numerical issues appear when going to high-order moments. Indeed,
\mbox{$\langle r^n \rangle$} is more and more sensitive to the upper limit of integration as $n$ increases. Thus,
the result may significantly depend on the box size used to discretize the continuum or on the size of the basis
used to expand quasiparticle wave-functions in HFB calculations.}.

The quantities $N_{\mathrm{halo}}$ and $\delta R_{\mathrm{halo}}$ are of course correlated, but they do not carry exactly the same
information.
Note that tolerance margins on $r_0$ from \eq(\ref{eq:def_r0_err}) propagate
into theoretical uncertainties on $N_{\mathrm{halo}}$ and $\delta R_{\mathrm{halo}}$.

\begin{figure}[hptb]
\includegraphics[keepaspectratio, angle = -90, width = \columnwidth]{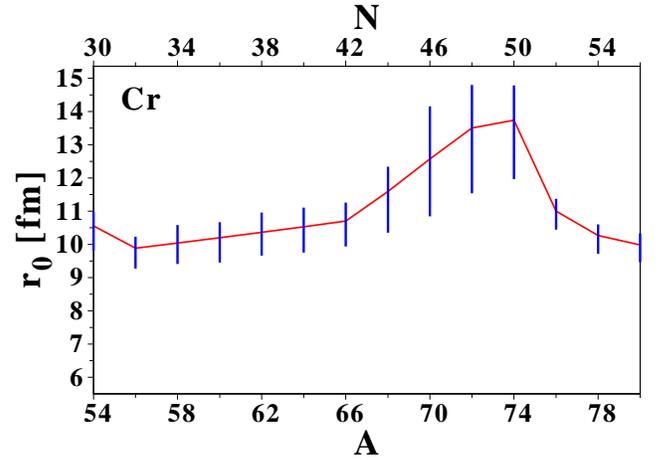}
\caption{ \label{fig:Cr_rint} (Color Online) Evolution of $r_0$ along the Cr isotopic chain, obtained
through spherical HFB calculations with the \{SLy4+REG-M\}
functional.}
\end{figure}

In the case of stable/non-halo nuclei, both quantities will be extremely small. There is still a slight curvature
in the density profile that provides a radius $r_0$ but the computed criteria will be consistent with zero. In the
particular case of magic neutron number, the curvature becomes particularly weak and translates into a broad
peak in the second log-derivative. As a result, the radius $r_0$ value is large and defines a region where the
density is particularly low. This is illustrated by \figu\ref{fig:Cr_rint}, where $r_0$ is plotted for chromium
isotopes as a function of $A$. The maximum of $r_0$ is attained for the magic shell \mbox{$N=50$}.

Finally, further characterization of the halo can be achieved by looking at the individual contributions of each
overlap function

\begin{equation}
N_{\mathrm{halo},\nu}\equiv4\pi \,
(2j^n_\nu+1)\,\int_{r_0}^{+\infty}|\bar{\varphi}^n_{\nu}(r)|^2\,r^2\,dr\,.\label{eq:def_nhalo_i}
\end{equation}
$N_{\mathrm{halo},\nu}$ provides a decomposition of the halo in terms of single-particle-like states. Note that the inner part of
an overlap function, i.e. for \mbox{$r<r_0$}, does not contribute to halo observables.\par By analogy
with the criterion used for light halo systems, the probability of each individual overlap function $\varphi_\nu$
to be in the \mbox{$r\ge r_0$} region can be defined through

\begin{equation}
P_\nu\equiv\frac{\displaystyle\int_{r_0}^{+\infty}|\bar{\varphi}_\nu(r)|^2\,r^2\,dr}
{\displaystyle\int_{0}^{+\infty}|\bar{\varphi}_\nu(r)|^2\,r^2\,dr}\,.\label{eq:def_drhalo_i}
\end{equation}

\section{Application to EDF calculations}
\label{sec:1stresults}

We apply the analysis method introduced in \sect\ref{sec:newcrit} to results obtained from self-consistent HFB
calculations of chromium and tin isotopes. In \sect\ref{sec:newcrit}, the energies $\epsilon^q_{\nu}$ that characterize internal overlap functions denote exact nucleon separation energies. No approximation to the nuclear many-body problem was involved in the analysis
conducted in Sec.~\ref{sec:newcrit}. The patterns of the internal one-body density thus extracted are fully general and
model-independent.

In practice of course, one uses an approximate treatment of the quantum many-body problem. This raises critical questions in the case of EDF calculations as discussed in Appendix~\ref{app:intrinsic}. Indeed, the one-body density at play in single-reference EDF calculations is an intrinsic density rather than the internal density, i.e. it is the laboratory density computed from a symmetry breaking state. As is customary in EDF methods though, one uses such an intrinsic density to approximate the internal density; e.g. when analyzing electron scattering data. Of course, such an identification is not rigourously justified and formulations of EDF methods directly in terms of the internal density are currently being considered~\cite{engel07}. Still, the asymptotic part of the lower component $V^q_{\nu}(\vec{r}\,)$ of the HFB quasiparticle wave-function satisfies the free Schr\"{o}dinger equation~\cite{bennaceur05a} (Eq.~\ref{eq:free_eqn_schrodinger}), just as the true internal overlap function $\varphi_\nu(\vec{r}\,)$ does. Considering in addition that the intrinsic HFB one-body density reads as

\begin{equation}
\rho^q(r)\equiv \sum_{\nu}\,\frac{2j^q_\nu+1}{4\pi}\,|\bar{V}^q_{\nu}(r)|^2\, \, , \label{HFBdensityQP}
\end{equation}
one realizes that the analysis performed in \sect\ref{sec:newcrit}, including the existence of the crossing pattern, applies directly to it\footnoteb{The method was developed in \sect\ref{sec:newcrit} for the exact internal density in order to demonstrate its generality and to eventually apply it to the results of other many-body methods dealing with a variety of finite many-fermion systems~\cite{rotival08b}.}.

\subsection{Implementation of the criteria}
\label{sec:res_implementation}

In the code \verb1HFBRAD1, the HFB problem is solved in a spherical box up to
a distance $R_{box}$ from the center of the nucleus on a radial mesh of step size \mbox{$\Delta
r=0.25$~fm}. For \mbox{$R_{box}=40$~fm}, the mesh has $160$ points in the radial direction, for both the
individual wave-functions and the densities. To obtain a satisfactory precision, the second order log-derivative
is computed using a five-points difference formula~\cite{abramowitz}. The precision of the formula is the
same as the intrinsic precision of the Numerov algorithm used for the integration of second-order differential
equations (which is \mbox{$\mathcal{O}\left(\Delta r^6\right)$})~\cite{dahlquist74,bennaceur05a}. Approximate
positions of the maximum of the second order log-derivative of $\rho^n(r)$ and of $r_0$ are first determined with a
simple comparison algorithm. To increase the precision, a \mbox{$11$-points} polynomial spline approximation to
the density and its second log-derivative around the two points of interest is performed. Because the functions
involved are regular enough, a spline approximation provides the radii $r_{max}$ and $r_0$ with a good precision,
as they are obtained using a dichotomy procedure up to a (arbitrary) precision of $10^{-5}$. Finally, the
integrations necessary to compute $N_{\mathrm{halo}}$ and $\delta R_{\mathrm{halo}}$ are performed with six-points Gaussian
integration.

In the definition of $\delta R_{\mathrm{halo}}$, the core contribution to the total r.m.s. radius is approximated as the
root-mean-square radius of the density distribution truncated to its \mbox{$r<r_0$} component. To check the
influence of this cut, the core density was extrapolated beyond the point where the second order log-derivative
crosses zero\footnoteb{This is the point where the halo contribution effect becomes significant.} using
\eq(\ref{eq:rho_asympt_usual}) and enforcing continuity of $\rho^n$ and ${\rho^n}'$. No difference was seen for $\delta
R_{\mathrm{halo}}$.

The individual contributions $N_{\mathrm{halo,i}}$, as well as the individual probabilities $P_{i}$, are evaluated in the
canonical basis. Equivalently, $N_{\mathrm{halo},\nu}$ and $P_\nu$ can be calculated in the quasiparticle basis.
Quasiparticle states are the best approximation to the overlap functions, but canonical and quasiparticle basis
really constitute two equivalent pictures. Indeed, each canonical state is, roughly speaking, split into
quasiparticle solutions of similar energies. A summation over quasiparticles having the same quantum
numbers in an appropriate energy window would recover the single-particle canonical approximation. The latter is
preferred here, as it is more intuitive to work in the natural basis.

\subsection{Cr isotopes}
\label{sec:res_cr}

\begin{figure}[hptb]
\includegraphics[keepaspectratio,angle = -90, width = \columnwidth]{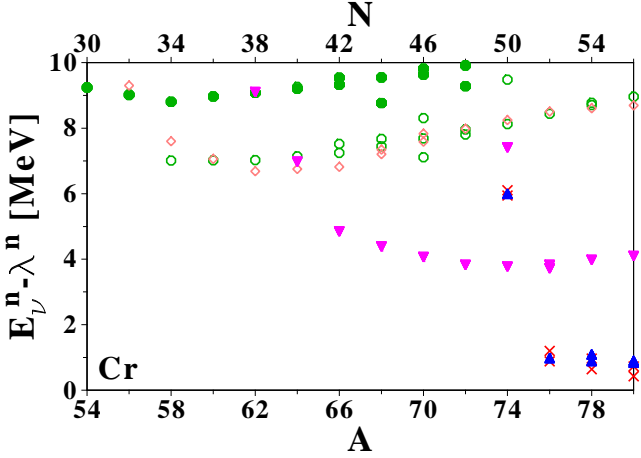}
\caption{ \label{fig:Cr_hfb_spectrum} (Color Online) Neutron separation energies \mbox{$|\epsilon^n_\nu|=E^n_\nu-\lambda^n$} along the
Cr isotopic chain, obtained through spherical HFB calculations with the \{SLy4+REG-M\} functional. Only relevant
quasiparticle energies \mbox{($N^n_\nu>0.01$)} are displayed. Conventions for labeling individual states are found
in \figu\ref{fig:ref}.}
\end{figure}

According to the analysis of \sect\ref{sec:newcrit_define2}, drip-line chromium isotopes appear to be
ideal halo candidates. The separation energy spectrum \mbox{$|\epsilon^n_\nu|=E^n_\nu-\lambda^n$} to the states in the
\mbox{$(N-1)$-body} system is shown in \figu\ref{fig:Cr_hfb_spectrum}.

\tab\ref{tab:Cr_spect} displays the canonical and quasiparticle spectra for the drip-line nucleus
\mbox{$^{80}$Cr}. In the canonical basis, $|e^n_0|$ is associated with a $3s_{1/2}$  state and is about $180$~keV.
The next low-lying state ($2d_{5/2}$) is within
an energy interval of \mbox{$\Delta E\approx500$}~keV. Those two states are separated from a core of orbitals by
\mbox{$E'\approx3.5$}~MeV. Equivalently, the separation energy in the quasiparticle basis is
\mbox{$|\epsilon^n_0|\approx430$}~keV, whereas four quasiparticle states ($s_{1/2}$ and $d_{5/2}$) are with an
energy spread of \mbox{$\Delta E\approx470$}~keV, and are further separated from higher-excited states by
\mbox{$E'\approx3.2$}~MeV. The separation energy $S_n$ for \mbox{$^{80}$Cr} is compatible with the
phenomenological binding energy necessary for the appearance of light halo nuclei, namely \mbox{$2$~MeV$/A^{2/3}\approx137$~keV}.
According to the discussion of \sect\ref{sec:newcrit_define2}, the energy scales at play in the three last bound
Cr isotopes correspond to ideal halo candidates.

\begin{table}
 \setlength{\extrarowheight}{4pt}
\begin{tabular}{|rc|c||rc|c|}
\hline \multicolumn{3}{|c}{Can. spectrum \mbox{$^{80}$Cr}} & \multicolumn{3}{c|}{Exc.
spectrum \mbox{$^{79}$Cr}}\\\hline
                \multicolumn{2}{|c|}{}             & $e^n_i$~[MeV] & \multicolumn{2}{|c|}{} & \mbox{$E^n_\nu-\lambda^n$}~[MeV]\\
\hline \hline
                \multicolumn{2}{|c|}{------------} & $>0$        &\multicolumn{2}{c|}{}   & $>10$ \\
                $E\updownarrow$                     &             &                       && $f_{5/2}$ & 8.694 \\
               \multirow{2}{*}{$\Delta E\left\{\vphantom{\begin{array}{l}a\\a\end{array}}\right.$}
                              & $3s_{1/2}$         &-0.178                            && $p_{1/2}$ & 8.960 \\

                            & $2d_{5/2}$ & -0.670                &
                                & $g_{9/2}$ & 4.103 \\
\multirow{3}{*}{$E'\left\updownarrow
\vphantom{\begin{array}{l}a\\a\\a\end{array}}\right.$} & & &
\multirow{2}{*}{$E'\left\updownarrow
\vphantom{\begin{array}{l}a\\a\end{array}}\right.$} & &\\
& & &&& \\
& & &\multirow{4}{*}{$\Delta E\left\{\vphantom{\begin{array}{l}a\\a\\a\\a\end{array}}\right.$}&$d_{5/2}$ & 0.893 \\
& $1g_{9/2}$ & -4.062 &&$d_{5/2}$ & 0.832  \\
& $1f_{5/2}$ & -8.676 &&$s_{1/2}$ & 0.728 \\
& $1f_{5/2}$ & -8.676 &&$s_{1/2}$ & 0.427 \\
& $2p_{1/2}$ & -8.942 & $E\updownarrow$& & \\
\multicolumn{2}{|c|}{} & $<-10$ &
\multicolumn{2}{c|}{------------} & $0$ \\
 \hline
\end{tabular}
\caption{\label{tab:Cr_spect} (Color Online) Neutron canonical energies $e^n_i$ in \mbox{$^{80}$Cr} and
separation energies \mbox{$|\epsilon^n_\nu|=E^n_\nu-\lambda^n$}, as predicted by the
\{SLy4+REG-M\} functional. Quasiparticle states with a spectroscopic factor smaller than $10^{-2}$ are not included.}
\end{table}

\begin{figure}[hptb]
\includegraphics[keepaspectratio,angle = -90, width = \columnwidth]{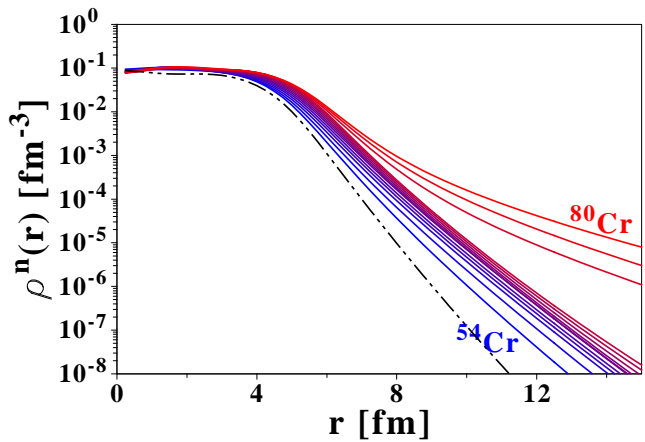}
\caption{(Color Online) Neutron densities for even-even Cr isotopes, from \mbox{$^{54}$Cr}
to \mbox{$^{80}$Cr}. The proton density of \mbox{$^{54}$Cr} is given (dashed-dotted line) as a reference for the neutron skin.} \label{fig:Cr_alldens}
\end{figure}

The criteria introduced in \sect\ref{sec:newcrit_crit} are now applied.
\begin{figure}[hptb]
\includegraphics[keepaspectratio,angle = -90, width = \columnwidth]{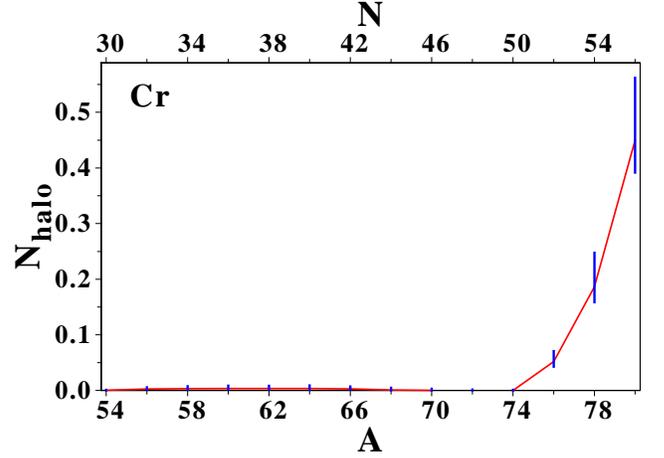}
\caption{(Color Online) Average number of nucleons participating in the halo along
the Cr isotopic chain, as a function of the nuclear mass, as
predicted by the \{SLy4+REG-M\} functional. Theoretical uncertainties are included (see text).} \label{fig:Cr_Nhalo}
\end{figure}
\figu\ref{fig:Cr_Nhalo} shows the average number of nucleons participating in the potential halo. Whereas
$N_{\mathrm{halo}}$ is consistent with zero for \mbox{$N\le50$}, a sudden increase is seen beyond the \mbox{$N=50$} shell
closure. The existence of a decorrelated region in the density of the last three Cr isotopes is consistent with the
evolution of the neutron densities along the isotopic chain in \figu\ref{fig:Cr_alldens}. For \mbox{$N>50$}, such
a behavior translates into a non-zero value of $N_{\mathrm{halo}}$. The value of $N_{\mathrm{halo}}$ remains small in comparison to the total neutron number, as the
decorrelated region is populated by \mbox{$\sim0.45$} nucleons on the average in \mbox{$^{80}$Cr}. In absolute value however, $N_{\mathrm{halo}}$ is comparable to what is
found in light \mbox{$s$-wave} halo nuclei like \mbox{$^{11}$Be}, where roughly $0.3$ nucleons constitute the
decorrelated part of the density~\cite{nunespriv}.

\begin{figure}[hptb]
\includegraphics[keepaspectratio,angle = -90, width = \columnwidth]{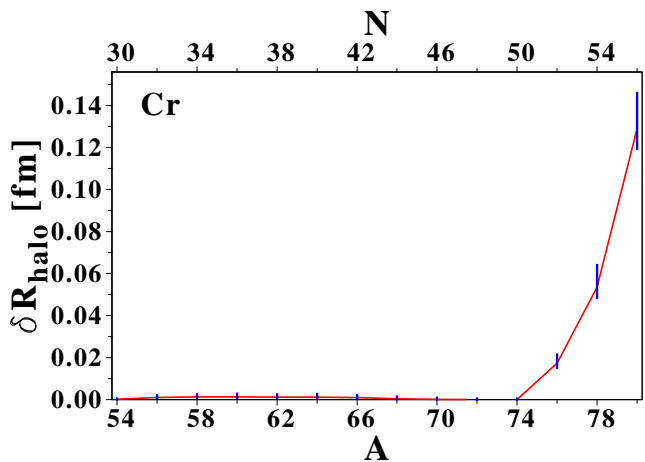}
\caption{(Color Online) Halo factor parameter $\delta{R}_{\mathrm{halo}}$ in the Cr isotopic
chain.} \label{fig:Cr_deltaR}
\end{figure}

\begin{figure}[hptb]
\includegraphics[keepaspectratio,angle = -90, width = \columnwidth]{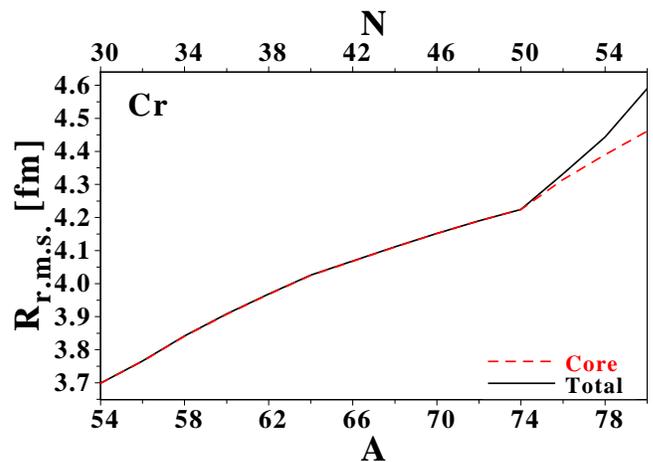}
\caption{(Color Online) Total neutron root-mean-square radius (solid line) and core contribution
(dashed line) for chromium isotopes, as predicted by the
\{SLy4+REG-M\} functional.} \label{fig:Cr_helmlike}
\end{figure}

The halo factor $\delta{R}_{\mathrm{halo}}$ is shown in \figu\ref{fig:Cr_deltaR} as a function of $A/N$. The halo contributes
significantly to the total neutron r.m.s. radius (up to $\sim0.13$~fm) beyond the \mbox{$N=50$} shell closure.
The latter result can be recast as a splitting of the total r.m.s. radius into a core and a halo contributions, as
displayed in \figu\ref{fig:Cr_helmlike}. In contrast to the Helm model, shell effects are here properly
separated from halo ones, e.g. the core r.m.s. radius includes a kink at \mbox{$N=50$} which is due to the
filling of least bound states and not to the halo per se. Only the physics related to the existence of truly
decorrelated neutrons is extracted by $N_{\mathrm{halo}}$ and $\delta R_{\mathrm{halo}}$. The kink of the neutron r.m.s radius (i)
was not assumed as a halo signature a priori~\cite{meng98,meng04} but recovered a posteriori
(ii) must be corroborated using finer tools such as $N_{\mathrm{halo}}$ and $\delta R_{\mathrm{halo}}$ to extract quantitatively
the contribution of the halo to that kink.

To characterize further this halo region, individual contributions $N_{\mathrm{halo,i}}$ are evaluated. The results are
summarized in \tab\ref{tab:Cr_indiv}. As expected, the main contributions to the halo come from the most
weakly-bound states, while for non-halo nuclei, like \mbox{$^{74}$Cr}, all contributions are consistent with zero.
At the neutron drip-line, important contributions are found from {both} $3s_{1/2}$ and $2d_{5/2}$ states. The
latter \mbox{$\ell=2$} states contribute for almost $50\%$ of the total number of nucleons in the decorrelated
region, although this state is more localized than the $3s_{1/2}$ because of its binding energy and of the effect
of the centrifugal barrier. Such hindrance effects are compensated by the larger canonical occupation of the
$d_{5/2}$ states and the larger intrinsic degeneracy of the shell. The significant contribution of the $\ell=2$
states could not be expected from the standard qualitative analysis presented in \sect\ref{sec:lowlmanybody} or,
with a few exceptions~\cite{kanungo00}, from the experience acquired in light nuclei.

\begin{table}[htbp]
\setlength{\extrarowheight}{2pt}
\begin{tabular}{|l||c|c|c|c|}
\hline & \multicolumn{4}{c|}{$^{74}$Cr}\\
\hline $N_{\mathrm{halo}}$ & \multicolumn{4}{c|}{$1.7.10^{-4}$}\\
\hline & $e^n_i$~[MeV] & $v_i^{n\,2}$ & $N_{\mathrm{halo,i}}$ & $P_i$ \\
\hline
$3s_{1/2}$ & $+0.036$ & $0.000$ & $0.000$            & $0.0\%$ \\
$2d_{5/2}$ & $-0.024$ & $0.000$ & $0.000$            & $0.0\%$\\
$1g_{9/2}$ & $-3.618$ & $1.000$ & $0.001$            & $0.1\%$\\
$2p_{1/2}$ & $-8.100$ & $1.000$ & $0.000$            & $0.0\%$\\
$1f_{5/2}$ & $-8.400$ & $1.000$ & $0.000$            & $0.0\%$\\
$Other$ & $<-10.0$    & ---    & $\sim 1.7.10^{-4}$ & ---  \\
\hline \hline
& \multicolumn{4}{c|}{$^{76}$Cr}\\
\hline $N_{\mathrm{halo}}$ & \multicolumn{4}{c|}{$5.2.10^{-2}$}\\
\hline & $e^n_i$~[MeV]  & $v_i^{n\,2}$ & $N_{\mathrm{halo,i}}$ & $P_i$ \\
\hline
$3s_{1/2}$ & $+0.356$ & $0.050$ & $0.007$            & $14.8\%$ \\
$2d_{5/2}$ & $-0.209$ & $0.311$ & $0.039$            & $12.6\%$\\
$1g_{9/2}$ & $-3.764$ & $0.991$ & $0.002$            & $0.2\%$\\
$2p_{1/2}$ & $-8.416$ & $0.998$ & $0.000$            & $0.0\%$\\
$1f_{5/2}$ & $-8.477$ & $0.998$ & $0.000$            & $0.0\%$\\
$Other$ & $<-10.0$    & ---    & $\sim 2.2.10^{-3}$ & ---  \\
\hline \hline
& \multicolumn{4}{c|}{$^{78}$Cr}\\
\hline $N_{\mathrm{halo}}$ & \multicolumn{4}{c|}{$0.186$}\\
\hline & $e^n_i$~[MeV]  & $v_i^{n\,2}$ & $N_{\mathrm{halo,i}}$& $P_i$ \\
\hline
$3s_{1/2}$ & $+0.052$ & $0.147$          & $0.045$            & $30.4\%$ \\
$\mathbf{2d_{5/2}}$   & $\mathbf{-0.450}$  & $\mathbf{0.604}$ &  $\mathbf{0.128}$ &$\mathbf{21.2\%}$\\
$1g_{9/2}$ & $-3.919$ & $0.991$          & $0.005$            & $0.5\%$\\
$1f_{5/2}$ & $-8.576$ & $0.998$          & $0.001$            & $0.1\%$\\
$2p_{1/2}$ & $-8.714$ & $0.998$          & $0.001$            & $0.1\%$\\
$Other$ & $<-10.0$    & ---             & $\sim 6.2.10^{-3}$ & ---  \\
\hline \hline
& \multicolumn{4}{c|}{$^{80}$Cr}\\
\hline $N_{\mathrm{halo}}$ & \multicolumn{4}{c|}{$0.450$}\\
\hline & $e^n_i$~[MeV] & $v_i^{n\,2}$ & $N_{\mathrm{halo,i}}$ & $P_i$ \\
\hline
$\mathbf{3s_{1/2}}$ & $\mathbf{-0.178}$  & $\mathbf{0.421}$   & $\mathbf{0.207}$ & $\mathbf{49.3\%}$ \\
$\mathbf{2d_{5/2}}$ & $\mathbf{-0.670}$  & $\mathbf{0.843}$   & $\mathbf{0.223}$ & $\mathbf{26.4\%}$\\
$1g_{9/2}$ & $-4.062$ & $0.995$          & $0.008$            & $0.8\%$\\
$1f_{5/2}$ & $-8.676$ & $0.999$          & $0.001$            & $0.1\%$\\
$2p_{1/2}$ & $-8.942$ & $0.999$          & $0.002$            & $0.2\%$\\
$Other$ & $<-10.0$    & ---             & $\sim 9.4.10^{-2}$ & ---  \\
\hline %
\end{tabular} \caption{\label{tab:Cr_indiv}Contributions of the least
bound canonical orbitals to the number of nucleons in the decorrelated region, and probabilities for those states
to be in the outer region \mbox{$r\ge r_0$}. The data are provided for the four last (predicted) bound Cr
isotopes.}
\end{table}

Finally, the probability $P_i$ for nucleons occupying the canonical state $\phi^n_{i}$ to be in the outer region
\mbox{$r\ge r_0$} in \mbox{$^{80}$Cr} is typical of $s-$wave  halo systems; i.e. $49\%$ for the $3s_{1/2}$ state
and a little bit lower for the $2d_{5/2}$ state, around $26\%$.

The analysis method applied to neutron-rich Cr isotopes demonstrates unambiguously that a halo is predicted for
the last three bound isotopes. We have indeed been able to characterize the existence of a decorrelated region in
the density profile for isotopes beyond the \mbox{$N=50$} shell closure. Such a region contains a small
fraction of neutrons which impact significantly the extension of the nucleus. It is generated by an admixture of
$s_{1/2}$ and $d_{5/2}$ states, whose probabilities to be in the halo region \mbox{$r\ge r_0$} are compatible
with what is seen in light halo nuclei. This provides the picture of a rather \textit{collective} halo building up
at the neutron drip-line for Cr isotopes.

\subsection{Sn isotopes}
\label{sec:res_sn}

\begin{figure}[hptb]
\includegraphics[keepaspectratio,angle = -90, width = \columnwidth]{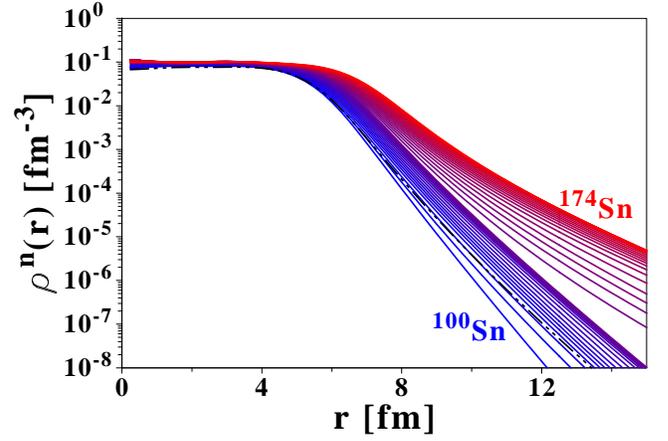}
\caption{(Color Online) Same as \figu\ref{fig:Cr_alldens} for the Sn isotopes. The "separation" between the two groups of neutron densities occurs for \mbox{$N=82$}.
Proton density of \mbox{$^{100}$Sn} is given as a reference in dashed-dotted line.}
 \label{fig:Sn_alldens}
\end{figure}

So far, the prediction of halos in tin isotopes beyond the \mbox{$N=82$} shell closure~\cite{mizutori00} have been
based on the Helm model, whose limitations have been pointed out in \sect\ref{sec:helm_results}. The robust
analysis tools introduced in the present work are expected to give more reliable results. Neutron densities of Sn
isotopes do exhibit a qualitative change for \mbox{$N>82$}, as seen in \figu\ref{fig:Sn_alldens}. However, the
transition is smoother than in the case of chromium isotopes (\figu\ref{fig:Cr_alldens}). This is partly due the
increase of collectivity associated with the higher mass. There are also specific nuclear-structure features that
explain the absence of halo in drip-line Sn isotopes.

\begin{figure}[hptb]
\includegraphics[keepaspectratio,angle = -90, width = \columnwidth]{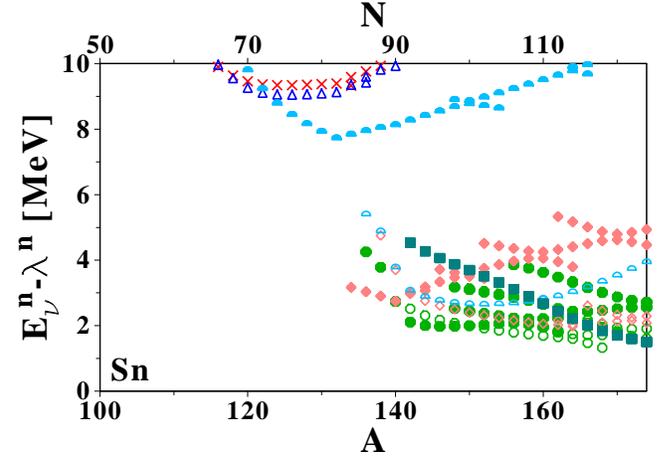}
\caption{ \label{fig:Sn_hfb_spectrum} (Color Online) Same as \figu\ref{fig:Cr_hfb_spectrum} for neutron separation energies of
Sn isotopes.}
\end{figure}

\tab\ref{tab:Sn_spect} displays the canonical and quasiparticle spectra for the drip-line nucleus
\mbox{$^{174}$Sn}. The energy scales at play are not compliant with the definition of a halo, as can also be seen
from \figu\ref{fig:Sn_hfb_spectrum}. In the canonical basis, the separation energy $E$ is roughly $1.2$~MeV,
whereas six states with an energy spread \mbox{$\Delta E\approx3.8$}~MeV are separated from a core of orbitals by
a gap \mbox{$E'\approx5.5$}~MeV. Equivalently in the quasiparticle basis one has (i)
\mbox{$S_{n}=E\approx1.5$}~MeV, (ii) four low-lying quasiparticles
 with a spread \mbox{$\Delta E\approx3.4$}~MeV (iii) separated from higher excitations
by \mbox{$E'\approx5.6$}~MeV. The energy spread of the low-lying states \mbox{$\Delta E$} is too
large to favor the formation of a halo. Also, according to the phenomenological criterion extracted for light halo nuclei,
the separation energy of \mbox{$^{174}$Sn} should have been of the order of \mbox{$2$~MeV$/A^{2/3}\approx64$~keV}
for a halo to emerge.

\begin{table}
 \setlength{\extrarowheight}{4pt}
\begin{tabular}{|rc|c||rc|c|}
\hline \multicolumn{3}{|c}{Can. spectrum \mbox{$^{174}$}} & \multicolumn{3}{c|}{Exc.
spectrum \mbox{$^{173}$Sn}}\\\hline
\multicolumn{2}{|c|}{} & $e^n_i$~[MeV] & \multicolumn{2}{|c|}{} & $E^n_\nu-\lambda^n$~[MeV]\\
\hline \hline \multicolumn{2}{|c|}{------------} & $>0$ &
\multicolumn{2}{c|}{} & $>15$ \\
\multirow{3}{*}{$E\left\updownarrow\vphantom{\begin{array}{l}a\\a\\a\end{array}}\right.\!\!$}
&&&& $d_{5/2}$ & 14.169 \\
& &&& $d_{3/2}$ & 12.026 \\
& &&& $s_{1/2}$ & 11.967 \\
\multirow{6}{*}{$\Delta E\left\{
\vphantom{\begin{array}{l}a\\a\\a\\a\\a\\a\end{array}}\right.$}
& $1i_{13/2}$& -1.208  && $1h_{11/2}$ & 10.603 \\
& $3p_{1/2}$ & -1.855  &
\multirow{4}{*}{$E'\left\updownarrow\vphantom{\begin{array}{l}a\\a\\a\\a\end{array}}\right.\!\!$}
& & \\
& $2d_{5/2}$ & -2.227  &&& \\
& $3p_{3/2}$ & -2.665  &&& \\
& $1h_{9/2}$ & -3.823  &&& \\
& $2f_{7/2}$ & -5.014  &\multirow{11}{*}{$\Delta E\left\{
\vphantom{\begin{array}{l}a\\a\\a\\a\\a\\a\\a\\a\\a\\a\\a\end{array}}\right.$}
& $f_{7/2}$  & 4.937 \\
\multirow{8}{*}{$E'\left\updownarrow\vphantom{\begin{array}{l}a\\a\\a\\a\\a\\a\\a\\a\end{array}}\right.\!\!$}
& &                    && $f_{7/2}$  & 4.463 \\
& &                    && $h_{9/2}$  & 3.890 \\
& &                    && $p_{3/2}$  & 2.722 \\
&&                     && $p_{1/2}$  & 2.648 \\
&&                     && $p_{3/2}$  & 2.559 \\
&&                     && $f_{5/2}$  & 2.290 \\
&&                     && $f_{5/2}$  & 2.082 \\
&&                     && $p_{1/2}$  & 1.905 \\
& $1h_{11/2}$& -10.575 && $p_{1/2}$  & 1.610 \\
& $2d_{3/2}$ & -12.581 && $i_{13/2}$ & 1.502 \\
& $3s_{1/2}$ & -12.747 & \multirow{2}{*}{$E\left\updownarrow\vphantom{\begin{array}{l}a\\a\end{array}}\right.\!\!$}
& & \\
& $2d_{5/2}$ & -14.944 & & & \\
\multicolumn{2}{|c|}{} & $<-15$ &
\multicolumn{2}{c|}{------------} & $0$ \\
 \hline
\end{tabular}
\caption{\label{tab:Sn_spect} Same as \tab\ref{tab:Cr_spect} for the neutron canonical
energies of \mbox{$^{174}$Sn}, and associated separation energies \mbox{$|\epsilon^n_\nu|$ of $^{173}$Sn}.}
\end{table}
\begin{table}[htbp]
 \setlength{\extrarowheight}{2pt}
\begin{tabular}{|l||c|c|c|c|}
\hline & \multicolumn{4}{c|}{$^{132}$Sn}\\
\hline $N_{\mathrm{halo}}$ & \multicolumn{4}{c|}{$0.13.10^{-2}$}\\
\hline & $e^n_i$~[MeV] & $v_i^{n\,2}$ &  $N_{\mathrm{halo,i}}$ & $P_i$ \\
\hline
$1i_{13/2}$ & $+2.648$ & $0.000$ & $0.000$ & $0.0\%$ \\
$3p_{1/2}$ & $+2.489$ & $0.000$ & $0.000$ & $0.0\%$ \\
$2f_{5/2}$ & $+1.661$ & $0.000$ & $0.000$ & $0.0\%$\\
$3p_{3/2}$ & $+1.240$ & $0.000$ & $0.000$ & $0.0\%$\\
$1h_{9/2}$ & $+1.141$ & $0.000$ & $0.000$ & $0.0\%$\\
$2f_{7/2}$ & $-1.785$ & $0.000$ & $0.000$ & $0.0\%$\\
$Other$ & $<-7.0$ & ---  & $\sim 0.13.10^{-2}$ & ---  \\
\hline \hline
& \multicolumn{4}{c|}{$^{146}$Sn}\\
\hline $N_{\mathrm{halo}}$ & \multicolumn{4}{c|}{$0.71.10^{-1}$}\\
\hline & $e^n_i$~[MeV] & $v_i^{n\,2}$ & $N_{\mathrm{halo,i}}$ & $P_i$ \\
\hline
$1i_{13/2}$& $+1.435$  & $0.064$ & $0.000$& $0.2\%$ \\
$2f_{5/2}$ & $-0.056$  & $0.155$ & $0.004$& $2.4\%$\\
$3p_{1/2}$ & $-0.202$  & $0.143$ & $0.005$& $3.8\%$ \\
$1h_{9/2}$ & $-0.401$  & $0.262$ & $0.001$& $0.3\%$\\
$3p_{3/2}$ & $-1.050$  & $0.442$ & $0.040$& $9.0\%$\\
$2f_{7/2}$ & $-3.037$  & $0.923$ & $0.017$& $1.9\%$\\
$Other$ & $<-7.0$  & --- & $\sim 3.1.10^{-3}$& ---  \\
\hline \hline
& \multicolumn{4}{c|}{$^{164}$Sn}\\
\hline $N_{\mathrm{halo}}$ & \multicolumn{4}{c|}{$0.179$}\\
\hline & $e^n_i$~[MeV] & $v_i^{n\,2}$ & $N_{\mathrm{halo,i}}$ & $P_i$ \\
\hline
$1i_{13/2}$ & $-0.216$& $0.349$ & $0.002$            & $0.5\%$ \\
$3p_{1/2}$ & $-1.347$ & $0.804$ & $0.052$            & $6.6\%$ \\
$2f_{5/2}$ & $-1.481$ & $0.155$ & $0.032$            & $4.0\%$\\
$3p_{3/2}$ & $-2.143$ & $0.923$ & $0.072$            & $7.8\%$\\
$1h_{9/2}$ & $-2.503$ & $0.894$ & $0.003$            & $0.4\%$\\
$2f_{7/2}$ & $-4.301$ & $0.975$ & $0.014$            & $1.4\%$\\
$Other$ & $<-7.0$     & ---    & $\sim 4.7.10^{-3}$ & ---  \\
\hline \hline
& \multicolumn{4}{c|}{$^{174}$Sn}\\
\hline $N_{\mathrm{halo}}$ & \multicolumn{4}{c|}{$0.149$}\\
\hline & $e^n_i$~[MeV] & $v_i^{n\,2}$  & $N_{\mathrm{halo,i}}$& $P_i$ \\
\hline
$1i_{13/2}$ & $-1.208$& $0.872$ & $0.005$            & $0.5\%$ \\
$3p_{1/2}$ & $-1.854$ & $0.979$ & $0.049$            & $5.0\%$ \\
$2f_{5/2}$ & $-2.227$ & $0.977$ & $0.028$            & $2.9\%$\\
$3p_{3/2}$ & $-2.665$ & $0.989$ & $0.054$            & $5.5\%$\\
$1h_{9/2}$ & $-3.823$ & $0.989$ & $0.002$            & $0.2\%$\\
$2f_{7/2}$ & $-5.014$ & $0.996$ & $0.009$            & $0.9\%$\\
$Other$ & $<-7.0$     & ---    & $\sim 2.3.10^{-3}$ & ---  \\
\hline
\end{tabular} \caption{Same as \tab\ref{tab:Cr_indiv}
for Sn isotopes.} \label{tab:Sn_indiv}
\end{table}

\begin{figure}[hptb]
\includegraphics[keepaspectratio,angle = -90, width = \columnwidth]{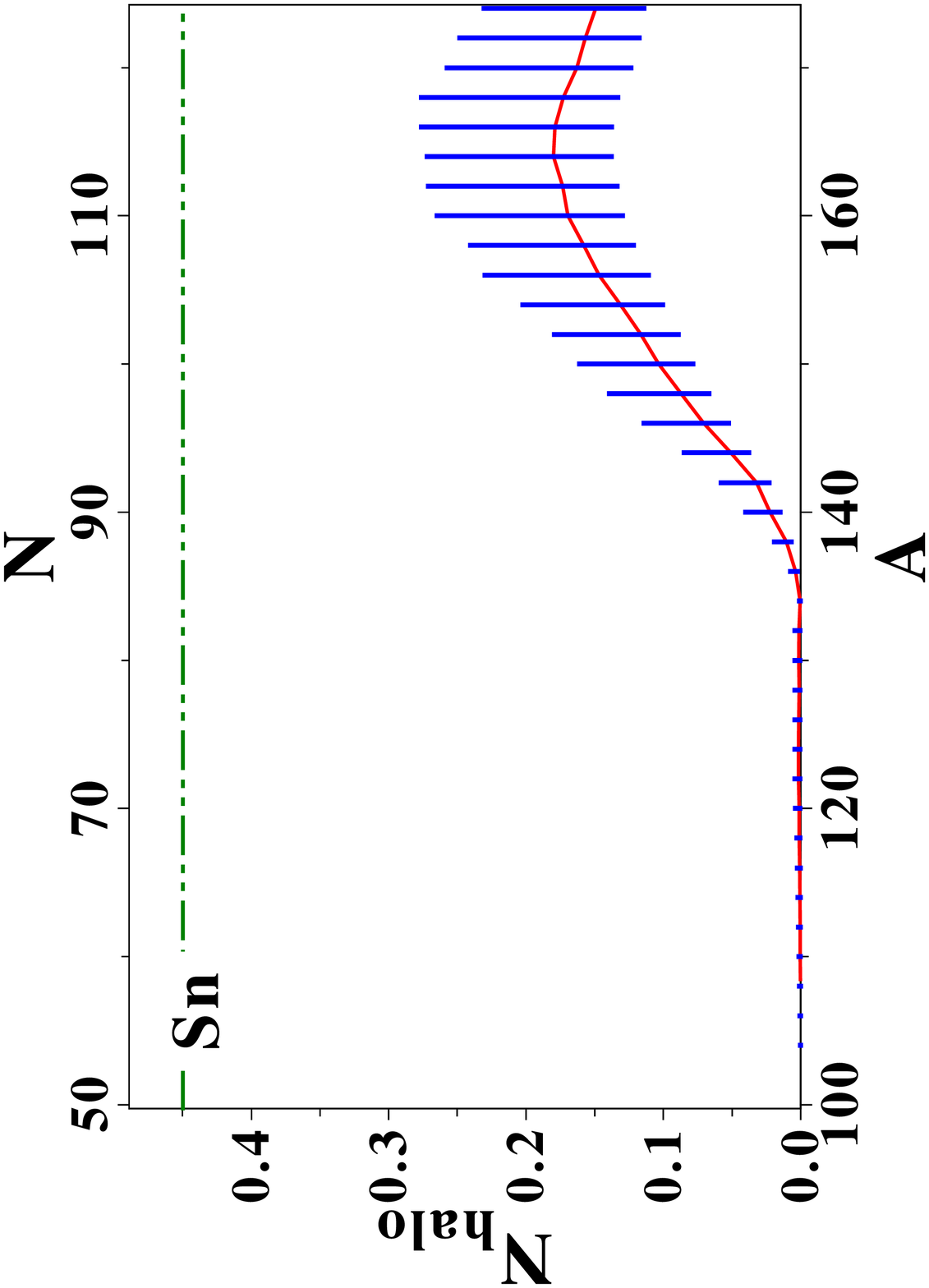}
\caption{(Color Online) Average number of nucleons in the spatially decorrelated region for Sn isotopes. For comparison,
\mbox{$N_{\mathrm{halo}}(^{80}$Cr$)$} is shown as a horizontal dashed-dotted line.} \label{fig:Sn_Nhalo}
\end{figure}

The $N_{\mathrm{halo}}$ parameter is displayed in \figu\ref{fig:Sn_Nhalo}. The maximum value of $N_{\mathrm{halo}}$, around $0.18$,
is very small compared to the total number of nucleons. The absolute numbers are also smaller than the ones
obtained in (lighter) Cr halos. We may add that the value of $N_{\mathrm{halo}}$ found here is of the same order of magnitude as those encountered for a non-halo $p$-wave nucleus such as $^{13}$N, where around $0.12$ neutron out of
six reside in average in the classically forbidden region~\cite{nunespriv}. An interesting feature is the decrease
of $N_{\mathrm{halo}}$ for \mbox{$N>166$}. This is a consequence of the filling of the highly degenerate $1i_{13/2}$ state
right at the drip-line (see \figu\ref{fig:Sn_cano_spectrum}). As the number of neutrons occupying the $1i_{13/2}$
shell increases, the depth of the one-body potential also increases and the shells become more bound, thus more
localized. As this happens over a significant number of neutrons, the effect on $N_{\mathrm{halo}}$ is visible. This
constitutes an additional hindrance to the formation of halos from low-lying high angular-momentum states.

\begin{figure}[hptb]
\includegraphics[keepaspectratio,angle = -90, width = \columnwidth]{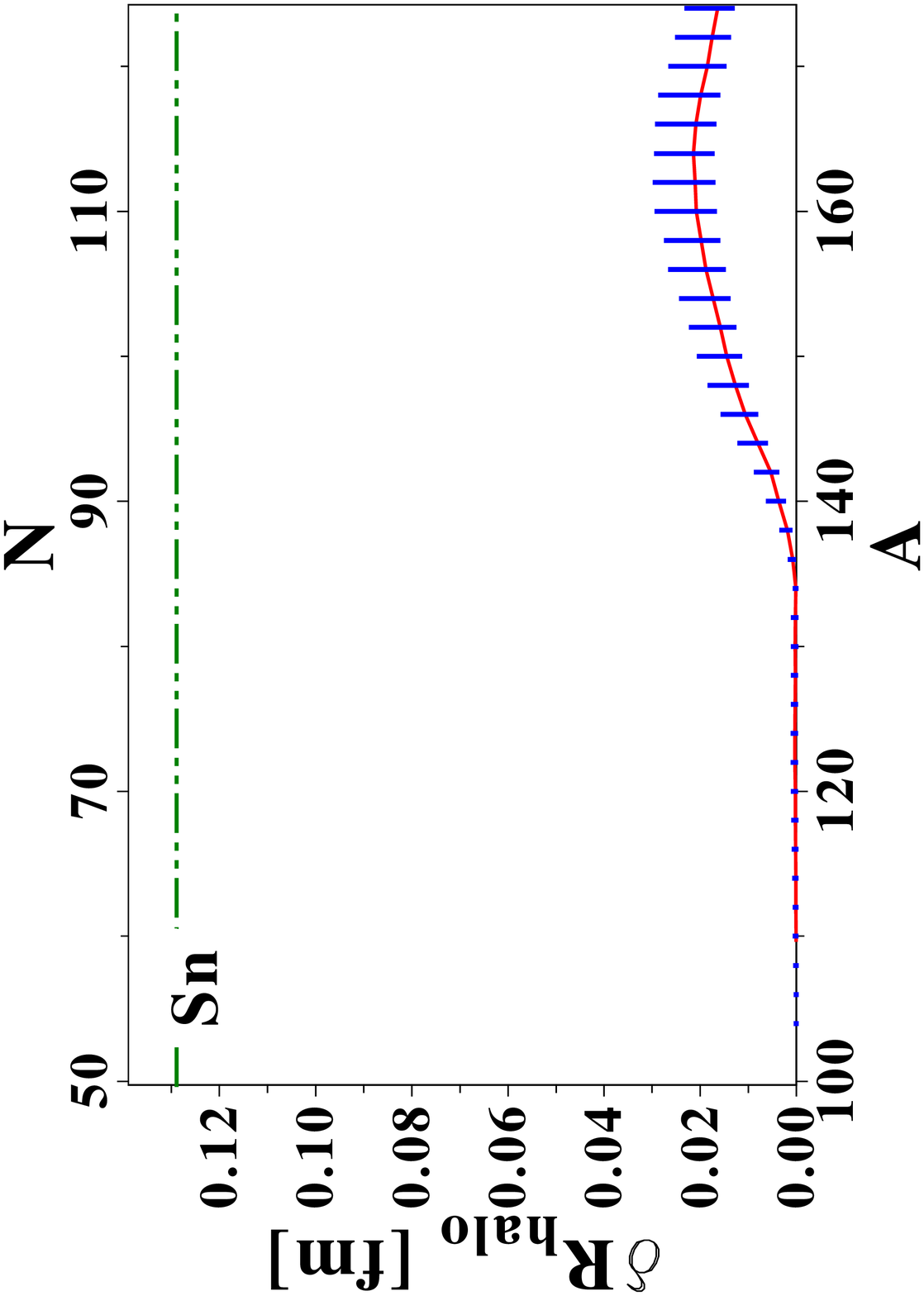}
\caption{(Color Online) Halo factor parameter $\delta{R}_{\mathrm{halo}}$ in the Sn isotopic chain. For comparison purposes, the maximum value of
$\delta R_{\mathrm{halo}}$ obtained for Cr isotopes is represented as a horizontal dashed-dotted line.}
\label{fig:Sn_deltaR}
\end{figure}

The second halo parameter $\delta{R}_{\mathrm{halo}}$ displayed in \figu\ref{fig:Sn_deltaR} shows that the decorrelated
region has little influence on the nuclear extension, of the order of $0.02$~fm. Its contribution is found to be
much less than predicted by the Helm model. The heavy mass of tin isotopes hinders the possibility of a sharp
separation of core and tail contributions in the total density and thus, of the formation of a halo.

\begin{figure}[hptb]
\includegraphics[keepaspectratio,angle = -90, width = \columnwidth]{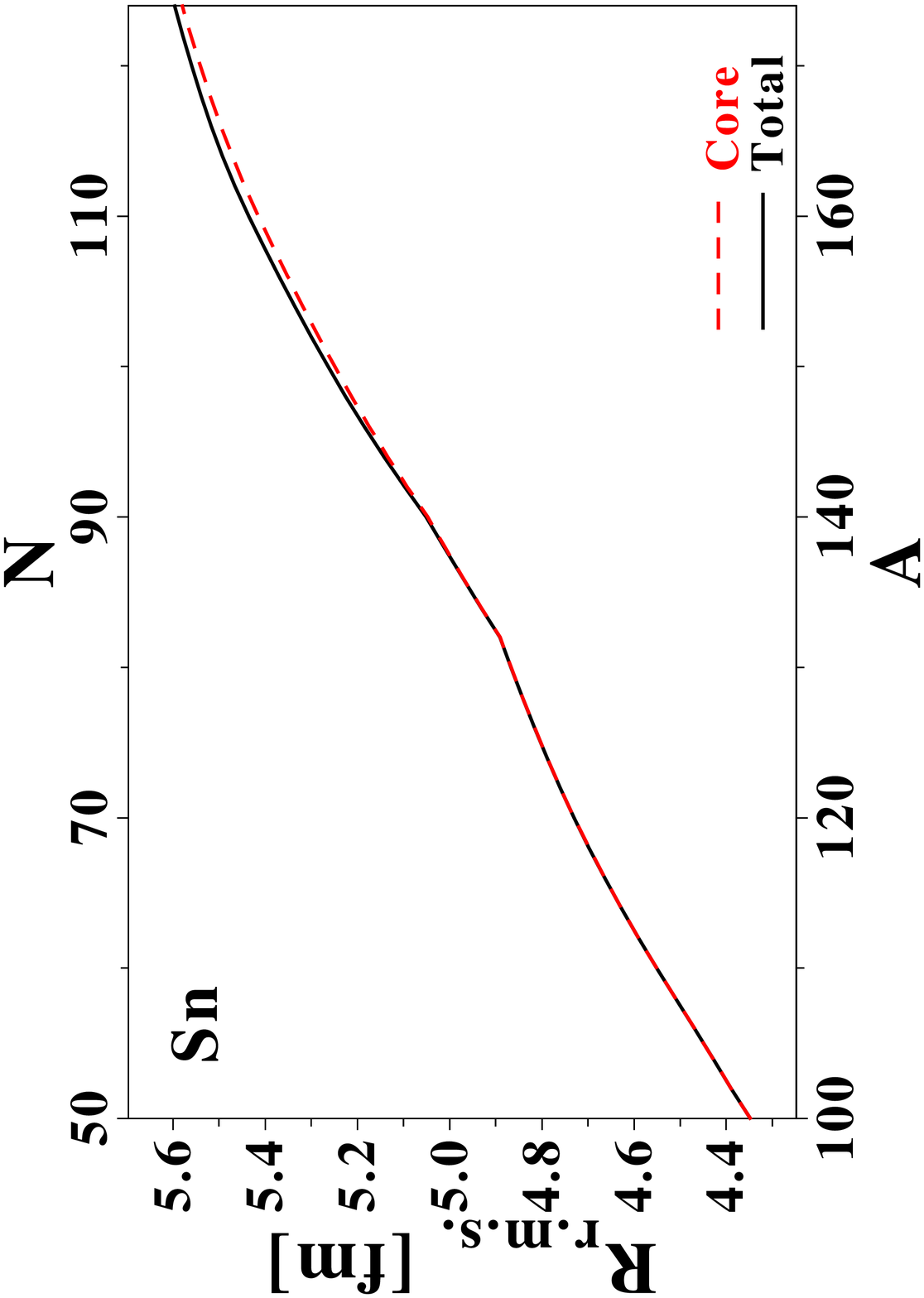}
\caption{(Color Online) Same as \figu\ref{fig:Cr_helmlike} for Sn isotopes.} \label{fig:Sn_helmlike}
\end{figure}

The analysis of single-particle contributions, summarized in \tab\ref{tab:Sn_indiv}, confirms the latter analysis.
First, $3p_{1/2}$, $3p_{3/2}$ and $2f_{7/2}$ ($\ell=3$) states contribute roughly the same to $N_{\mathrm{halo}}$. For
higher angular-momentum orbitals, the effect of the centrifugal barrier is seen: the $1h_{9/2}$ and $1i_{13/2}$
orbitals, the latter being the least bound orbital, do not contribute significantly to the decorrelated region.
Finally, individual probabilities $P_i$ remain very small, and do not exceed a few percent.

For all the reasons exposed above, only a neutron skin effect is seen in tin isotopes, and no
significant halo formation is envisioned. Of course, all results presented
here have been obtained with a particular EDF and it is of interest
to probe the sensitivity of the predictions to the different ingredients
of the method~\cite{rotival08b}.

In any case, the two previous examples already provide a coherent picture regarding the properties of halo or
non-halo medium-mass nuclei. In particular, it is rather obvious that the notion of {\it giant
halo}~\cite{meng98,sandulescu03,geng04,kaushik05,grasso06,terasaki06} constituted of six to eight neutrons is
misleading. Indeed, such a picture was obtained by summing up the {\it total} occupations of loosely bound
orbitals. Although loosely bound orbitals are indeed responsible for the formation of the halo, nucleons
occupying them still reside mostly inside the nuclear volume. It is thus unappropriate to simply sum up their occupations to characterize the halo. The identification of the halo region in the presently proposed method led us to define the more meaningful quantity $N_{\mathrm{halo}}$.

\section{Conclusions}
\label{sec:concl}

The formation of halo in finite many-fermion systems is a quantum phenomenon
caused by the possibility for non-classical systems to
expand in the classically forbidden region. One difficulty to further understand this phenomenon resides in the absence of tools to characterize halo properties in a quantitative way. Light nuclei constitute an exception considering that the quantification of halo properties in terms of
the dominance of a cluster configuration and of the probability of the weakly-bound clusters to extend beyond
the classical turning point is well acknowledged~\cite{fedorov93,jensen00,riisager00,jensen01}. Several attempts
to characterize halos in systems constituted of tens of fermions have been made but were based on loose definitions and quantitative criteria. Such a situation is unsatisfactory because important questions, such as the very existence of halos at the neutron drip-line of medium-mass nuclei, are still open.

After demonstrating the inability of the Helm model to provide reliable predictions, a new quantitative analysis method has been developed to identify and characterize halos in finite many-fermion systems in a
model-independent fashion. It is based on the decomposition of the internal one-body density in terms of overlap
functions. The definition of the halo, as a region where nucleons are spatially decorrelated from the others, has been shown to be
connected to specific patterns of the internal one-body density and of the energy spectrum of the
\mbox{$(N-1)$-body} system. In particular, halos can be characterized by the existence of a
small nucleon separation energy $E$, a small energy spread \mbox{$\Delta E$} of low-lying excitations, and a
large excitation energy $E'$ of the upper-lying states with respect to low-lying bunched ones, in the excitation spectrum of the
\mbox{$(N-1)$-body} system.

Based on the new analysis method, it is possible to extract the radius $r_0$ beyond which the halo, if it exists, dominates over the core. Such an identification of $r_0$ has been validated by extensive simulations. It is important to stress that the method does not rely on an \textit{a priori}
separation of the density into core and halo components. The latter are extracted from the analysis itself, using the
total matter density as the only input. Several quantitative observables are then introduced, namely (i) the average
number of fermions participating in the halo, (ii) the influence of the halo region on the total
extension of the system, and (iii) the contributions of individual overlap functions to the halo.

The new analysis method has been applied to the results obtained from energy density functional calculations of
chromium and tin isotopes using the code \verb1HFBRAD1~\cite{bennaceur05a}. Drip-line Cr isotopes appear as ideal
halo candidates whereas tin isotopes do not.

For drip-line Cr isotopes, the average fraction of nucleon participating in the halo is of the order of $\sim0.5$.
Such a value is compliant with those found for light halo systems~\cite{nunespriv}. The halo region was also found to influence
significantly the nuclear extension. Contributions from several individual components, including \mbox{$\ell=2$} ones,
were identified, contradicting the standard picture arising from few-body models. The notion of
{collective halos} in medium-mass nuclei has been introduced.

In the case of Sn isotopes, the average number of nucleons participating in the halo is very small and has
no influence on the nuclear extension. Thus, the drip-line phenomenon discussed previously for tin
isotopes~\cite{mizutori00} is rather a pronounced neutron skin effect. Such skin effects are of course of
interest as they emphasize the isovector dependence of the energy density functionals. However, they should not be
confused with halo systems which display an additional long tail of low density matter.

This preliminary study on two isotopic series gives promising results and validates the theoretical grounds of the
analysis. With upcoming new radioactive beam facilities, interaction cross-sections are expected to be measurable
in the drip-line region of \mbox{$Z\approx26$} elements~\cite{whitepapernscl}. This would constitute a giant leap
towards an extensive comparison between theoretical and experimental works on drip-line physics.

\begin{acknowledgments}
We wish to thank K. Bennaceur for his help with using \verb1HFBRAD1, F.M. Nunes for useful discussions on light
halo nuclei and D. Van Neck for his valuable input regarding the internal one-body density. The proofreading of
the manuscript by K. Bennaceur, J.-F. Berger, D. Lacroix and H. Goutte is greatly acknowledged. Finally, V. R.
wishes to thank the NSCL for its hospitality and support. This work was
supported by the U.S. National Science Foundation under Grant No. PHY-0456903.
\end{acknowledgments}
\begin{appendix}

\section{Internal one-body density}
\label{sec:intr_dens_def}

\subsection{Definition}

In the laboratory frame, the one-body density is the expectation value of the operator

\begin{equation}
\hat{\rho}(\vec{r}\,)=\sum_{i=1}^{N}\,\delta(\vec{r}-\hat{\vec{r}}_i)
\label{eq:def_oper_dens_lab}\,,
\end{equation}
which leads for the \mbox{$N$-body} ground state to

\begin{eqnarray}
\notag\!\!\!\!\rho(\vec{r}\,) \!&=& \!N\!\!\int
d\vec{r}_1\ldots d\vec{r}_{N-1}\,|{\Psi}^{N}_{0}(\vec{r}_1\ldots
\vec{r}_{N-1},\vec{r}\,)|^2\\&=&\!N\!\!\int d\vec{\xi}_1\ldots
d\vec{\xi}_{N-2}\,d\vec{R}_{N-1}\notag\\&&~~~~~~~~\times|{\tilde{\Phi}}^{N}_{0}(\vec{\xi}_1\ldots
\vec{\xi}_{N-2},\vec{r}-\vec{R}_{N-1})|^2\label{eq:lab_dens}\, \, \,
\end{eqnarray}
where \mbox{$\Phi^{N}_i(\vec{r}_1\ldots\vec{r}_{N})\equiv\tilde{\Phi}^{N}_i(\vec{\xi}_1\ldots\vec{\xi}_{N-1})$}. Using that $\tilde{\Phi}^{N}_i$ is invariant under translation of the system, one easily proves that the one-body density in the laboratory frame is also
translationally invariant, \mbox{$\rho(\vec{r}+\vec{a}\,)=\rho(\vec{r}\,)$}, and thus is uniform. This is a
general property of translationally invariant systems which underlines that the density in the laboratory frame is
not the proper tool to study self-bound systems.

The relevant object for self-bound systems is the internal one-body density matrix, defined as the expectation
value of the operator

\begin{multline}
\hat{\rho}_{\mathrm{[1]}}(\vec{r},\vec{r}\,')=
\delta(\vec{R}_N)\sum_{i=1}^N\delta(\vec{r}-\hat{\vec{r}}_i+\hat{\vec{R}}_{N-1}^{\,i})\\
\times\delta(\vec{R}\,'_N)\sum_{j=1}^N\delta(\vec{r}\,'-\hat{\vec{r}}\,'_j+\hat{\vec{R}}_{N-1}^{\,j'})\\
\times \prod_{\substack{k,l=1..N\\k,l\ne
i,j}}\delta(\hat{\vec{r}}_k-\hat{\vec{r}}_l)\,,
\end{multline}
where
\begin{equation}
\hat{\vec{R}}_{N-1}^{\,i}=\frac{1}{N-1}\sum_{\substack{j=1\\j\ne
i}}^N\hat{\vec{r}}_j\,.
\end{equation}

The internal density defined with respect to the center-of-mass of the remaining $(N-1)$-body\footnoteb{One could define another internal one-body density taking the
center-of-mass of the \mbox{$N$-body} system as a pivot point. This is a more relevant choice to analyze electron scattering data.} is of direct relevance to knockout reactions~\cite{clement73a,clement73b,dieperink74}. Using the orthogonality relationship~\cite{vanneck96}

\begin{equation}
\int\,d\vec{r}_1\ldots d\vec{r}_N\, {{\Phi}^{N}_i}^*(\vec{r}_1\ldots
\vec{r}_N)\,\delta(\vec{R}_N)\,{\Phi}^{N}_j(\vec{r}_1\ldots
\vec{r}_N)=\delta_{ij}\, \, , \label{eq:orthogonality_N}
\end{equation}
and (\ref{eq:decomp_N_N-1_1}), one obtains~\cite{vanneck93,vanneck98b,shebeko06}

\begin{eqnarray}
\rho_{\mathrm{[1]}}(\vec{r},\vec{r}\,') &=& N\int d\vec{r}_1\ldots
d\vec{r}_{N-1}\,{{\Phi}^{N}_{0}}^*(\vec{r}_1\ldots
\vec{r}_{N-1},\vec{r}\,')\notag\\&&~~~~~~\times\delta(\vec{R}_{N-1})
{\Phi}^{N}_{0}(\vec{r}_1\ldots \vec{r}_{N-1},\vec{r}\,)\notag\\
&=&\sum_{\nu}\varphi^*_\nu(\vec{r}\,')\varphi_\nu(\vec{r}\,)\,,
\label{eq:decomp_density1b}
\end{eqnarray}
which shows that the internal one-body density matrix is completely determined by internal overlap
functions~\cite{vanneck93}.

The internal one-body density \mbox{$\rho_{\mathrm{[1]}}(\vec{r}\,)$} is the local part of the internal density matrix,
and is the expectation value of the operator
\begin{equation}
\hat{\rho}_{\mathrm{[1]}}(\vec{r}\,)=
\delta(\vec{R}_N)\sum_{i=1}^N\delta(\vec{r}-\hat{\vec{r}}_i+\hat{\vec{R}}_{N-1}^{\,i})\,.
\label{eq:localdens}
\end{equation}

According to \eq(\ref{eq:decomp_density1b}), one has
\begin{equation}
\rho_{\mathrm{[1]}}(\vec{r}\,)=\sum_{\nu}|\varphi_\nu(\vec{r}\,)|^2
=\sum_{\nu}\frac{2\ell_\nu+1}{4\pi}|\bar{\varphi}_\nu(r)|^2\,. \label{encoreune}
\end{equation}

\subsection{Nuclear EDF calculations}
\label{app:intrinsic}

The behavior of the internal one-body density highlighted in \sect\ref{sec:newcrit} is general and
model-independent. It is valid for any finite many-fermion system, as long as the inter-fermion interaction is negligible beyond a certain relative distance. Of course, when an approximate treatment of the $N$-body system is used, a certain deterioration of the properties of the density can be observed. In the case of EDF calculations however, some more profound issues are raised.

First, an important clarification regarding the physical interpretation of the quantities at play in the
calculations must be carried out. In single-reference
implementations of the nuclear EDF method, one manipulates the so-called "intrinsic" one body density,
in the sense that it is built from an \textit{auxiliary} state that
breaks symmetries of the nuclear Hamiltonian, e.g. translational, rotational and gauge invariance.
The intrinsic density is associated with a wave packet from which true eigenstates, and their laboratory and internal densities, can be recovered by restoring broken symmetries through multi-reference EDF calculations~\cite{bender03b}. In practice, the intrinsic density is used as a good approximation to the internal density, e.g. when analyzing electron scattering data. Still, the intrinsic density of a symmetry breaking state and
the internal density associated with the true eigenstate of interest are different\footnoteb{In shell model, the internal wave-function is explicitly computed when the center-of-mass part of the $N$ body wave function can be mapped onto a 0s state.}~\cite{giraud77}. As a result, EDF methods\footnoteb{The SR-EDF method, as
it is currently applied to self-bound nuclei, is not related to an existence theorem \textit{\`a la} Hohenberg-Kohn.}~\cite{engel07}.
 expressed directly in terms of the internal density are currently being considered~\cite{engel07}.

As just said, the EDF intrinsic density has been shown in many cases to be a good approximation of the internal density extracted through electron scattering. In practice, one identifies the lower component of the intrinsic HFB wave-function $V^q_{\nu}(\vec{r}\,)$ with the internal overlap function $\varphi^{q}_\nu(\vec{r}\,)$ leading from the ground
state of the \mbox{$N$-body} system to the corresponding excited state of the \mbox{$(N-1)$-body}
system\footnoteb{It can be shown that the perturbative one-quasiparticle state
\mbox{${{\eta}_{i}}^\dagger|\Phi\rangle$} contains \mbox{$N+{u_i}^2-{v_i}^2$} particles on the average if
\mbox{$|\Phi\rangle$} is constrained to $N$ particles on the average. It is only for deep-hole quasiparticle
excitations \mbox{($v_i^2\approx 1$)} that the final state will be a good approximation of the \mbox{$(N-1)$-body}
system. The correct procedure, that also contains some of the rearrangement terms alluded to above, consists of
constructing each one-quasiparticle state self-consistently by breaking time-reversal invariance and requiring
\mbox{$(N-1)$} particles in average, or of creating the quasiparticle excitation on top of a fully paired vacuum
designed such that the final state has the right average particle number~\cite{duguet02a,duguet02b}. The overlap
functions and spectroscopic factors can be computed explicitly in such a context.
}. In particular, and this is key to the present discussion, the asymptotic part of $V^n_{\nu}(\vec{r}\,)$ satisfies the free Schr\"{o}dinger equation~\cite{bennaceur05a}, just as the asymptotic part of $\varphi^{n}_\nu(\vec{r}\,)$ does. The smallest energy \mbox{$|\epsilon^n_0|$} thus extracted relates to the exact separation energy, i.e.  an analogue to Koopmans' theorem derived originally in the case of Hartree-Fock approximation applies. Given that the intrinsic density (Eq.~\ref{HFBdensityQP}) expressed in terms of the lower component of HFB quasiparticle wave-functions reads the same as the internal density expressed in terms of overlap functions (Eq.~\ref{encoreune}), the analysis method developed in \sect\ref{sec:newcrit}, including the occurrence of crossing patterns, applies directly to the former.

\subsubsection{Slater determinant as an auxiliary state}
\label{sec:multimodel_hf}

In the implementation of the EDF method based on a Slater determinant, {\it explicit} spectroscopic factors
are either zero or one, and behave according to a step function \mbox{$S^q_\nu=\Theta(\epsilon^q_F-e^q_\nu)$}. The
single-particle orbitals $\varphi^q_\nu$ are identified with overlap functions and the density takes the form
given by \eq(\ref{encoreune}).

\subsubsection{Quasiparticle vacuum as an auxiliary state}
\label{sec:multimodel_hfb}

In the implementation of the EDF method based on a quasiparticle vacuum, the one-body density can be evaluated
using either the canonical states $\phi^q_i$ or the lower components $V^q_\nu$ of the quasiparticle states

\begin{equation}
\rho^{q}(r)=
\sum_{i} \frac{2j^{q}_\nu+1}{4\pi} v^{q \, 2}_{i}|\bar{\phi}^{q}_{i}(r)|^2
=\sum_{\nu}\frac{2j^{q}_\nu+1}{4\pi}|\bar{V}^{q}_\nu(r)|^2\, \, ,  \label{encoreune2}
\end{equation}
where $j^{q}$ relates to the total angular momentum. In the present case, the
spectroscopic factor $S^q_\nu$ identifies with the quasiparticle occupation $N^q_\nu$ defined by
\eq(\ref{eq:def_qp_occ}). This underlines that implementation of the EDF approach based on a quasiparticle vacuum
incorporates {\it explicitly} parts of the spreading of the single-particle
strength~\protect\cite{vanneck06}.

\begin{figure}[htbp]
\includegraphics[keepaspectratio, angle = -90, width =\columnwidth]{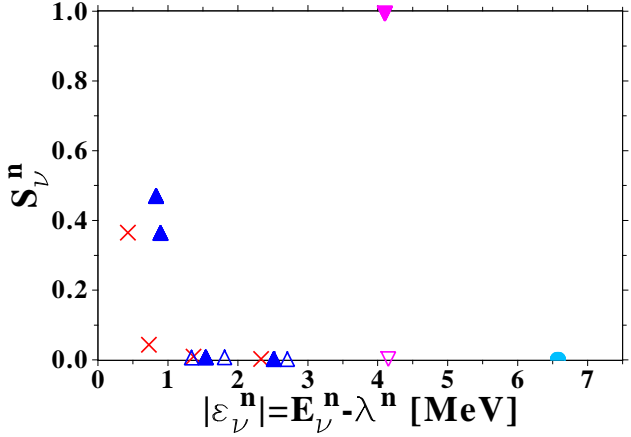}
\caption{\label{fig:Spectro_epsilon} (Color Online) Neutron quasiparticle
occupation $N^n_\nu$ as a function of the separation energy in \mbox{$^{80}$Cr}, calculated with the \{SLy4+REG-M\}
functional. Conventions from \figu\ref{fig:ref} are used to label individual quasiparticle states.
Only quasiparticles with occupations greater than $10^{-3}$ are displayed.}
\end{figure}

The function \mbox{$S^n_{\nu}=f(|\epsilon^n_\nu|)$}, whose typical behavior is presented in
\figu\ref{fig:Spectro_epsilon} for \mbox{$^{80}$Cr}, takes values between zero and one. The difference between
hole-like quasiparticle excitations and particle-like ones is visible. Indeed, $S^n_\nu$ increases with excitation
energy \mbox{$|\epsilon^n_\nu|$} for hole-like excitations. This constitutes the main branch which tends towards a
step function when correlations are not explicitly included into the auxiliary state; i.e. for the EDF approach
based on an auxiliary Slater determinant. Spectroscopic factors of particle-like quasiparticle
excitations remain small and go to zero for high-lying excitations.

\section{Determination of the halo region}
\label{sec:def_r0_simu}

Let us start with a very crude toy model, where everything is analytical. The total density $\rho$ is assumed to
be a superposition of a core $\rho_c$ and a tail $\rho_h$, both taking the form
\begin{equation}
\rho_i(r)=A_i\,\kappa_i\,e^{-\kappa_i\,r}\,.
\end{equation}

This amounts to considering that the asymptotic regime is reached in the region of the crossing between $\rho_c$
and $\rho_h$, and we neglect for now the $r^{-2}$ factor. In this model the second-order (base-$10$) log-derivative of the
total density is analytical, as well as the exact positions of (i) its maximum $r_{\mathrm{max}}$ (ii) the point $r_{0}$
where the halo density is exactly equal to ten times the core one. Then, the ratio
\mbox{$\mathcal{R}(r_{0})=\log_{10}''\rho(r_{0})/\log_{10}''\rho(r_{\mathrm{max}})$} can be evaluated and becomes in the weak binding limit of
interest \mbox{$\kappa_h/\kappa_c\rightarrow0$}

\begin{equation}
\mathcal{R}(r_{0})\underset{\kappa_h/\kappa_c\rightarrow0}{\longrightarrow}
\frac{40}{121}+\mathcal{O}\left[\left(\frac{\kappa_h}{\kappa_c}\right)^2\right]\,.
\end{equation}

This shows that the position where there is a factor of ten between $\rho_c$ and $\rho_h$ is equivalently obtained
by finding the position where there is a given ratio between the value of the second-order log-derivative of the
density and its maximal value. The critical value \mbox{$40/121\approx0.33$}
 found in the toy model is not believed to be accurate for complex nuclei, as (i) the
asymptotic regime is not reached at the crossing point and is more complicated because of the $r^{-2}$ factor (ii)
the total density is a superposition of more than two components. However, we expect the one-to-one correspondence
between ratios on the densities and ratios on \mbox{$\log_{10}''\rho$} to hold in realistic cases. Thus,
\textit{the position where the halo dominates the core by one order of magnitude can be found using
\mbox{$\log_{10}''\rho$} as the only input}.

More realistic model calculations have been used to characterize the position of $r_0$. The total density is taken
as a linear combination of core and halo contributions. Their relative normalization are free parameters in this
simulation, allowing to artificially change the fraction of halo in the total density

\begin{equation}
\rho_{\mathrm{tot}}(r)=\displaystyle N_c \, \rho_c(r)+\sum_{\nu=1}^{m}N_\nu \, \rho_{\nu}(r)\,,
\end{equation}
where $N_c$ and \mbox{$\displaystyle N_h=\sum_{\nu=1}^{m} N_\nu$} are the number of nucleons in the core part and
in the halo part, respectively. The densities $\rho_c$ and $\rho_\nu$ are normalized to one. We considered (i)
simple models, where the core and each halo components are defined as
\begin{equation}
\left\{
\begin{array}{lcl}
\displaystyle \rho_i(r)=\frac{1}{\mathcal{N}_i}&&r<R_0\,,\\\\
\displaystyle
\rho_i(r)=\frac{1}{\mathcal{N}_i}e^{\frac{R_0-r}{a_i}}&&r>R_0\,,
\end{array}
\right.
\end{equation}
$\mathcal{N}_i$ standing for a normalization constant. This model only accounts for the basic features of the
nuclear density: a uniform core of radius $R_0$ and a spatial extension becoming larger as \mbox{$a_i\rightarrow
0$} (ii) {double Fermi} models, where the un-physical sharp edge in the logarithmic representation of the previous
density is smoothened out
\begin{equation}
\rho_i(r)=\frac{\rho_0}{1+e^{\frac{r-R_0}{a_i}}}
\end{equation}
(iii) {semi-phenomenological} models, which fulfill the
asymptotic behavior of \eq(\ref{eq:rho_asympt_usual}). Core and tail densities vanish at
\mbox{$r=0$}, as well as their derivatives with respect to $x$, $y$ and
$z$, in order to avoid singularities at the nucleus
center~\cite{berdich82}. In
\refers\cite{gambhir85,gambhir86,bhagwat00}, such densities were
adjusted on experimental data. The core part was defined as
\begin{equation}
\rho_c(r)=\frac{\rho_{0,c}}{1+\left[\frac{1+\left(\frac{r}{R_{0,c}}\right)^2
}{2}\right]^{\alpha}\left[e^{\frac{r-R_{0,c}}{a_c}}+e^{\frac{-r-R_{0,c}}{a_c}}\right]} \, \, ,
\label{eq:def_semipheno}
\end{equation}
where \mbox{$\alpha=1$} for neutrons, and the halo density as
\begin{eqnarray}
\rho_h(r)=\rho_{0,t}\left[\frac{r^2}{\left(r^2+R_{0,t}^2\right)}\right] e^{-\frac{r}{a_t}} \, \, , \\\notag
\end{eqnarray}
(iv) more realistic models, where the core density is still defined as in \eq(\ref{eq:def_semipheno}),
 but the halo contributions are
realistic wave functions taken from self-consistent EDF
calculations of Cr and Sn isotopes.\\
\begin{figure}[htbp]
\includegraphics[keepaspectratio, angle=-90,width = \columnwidth]{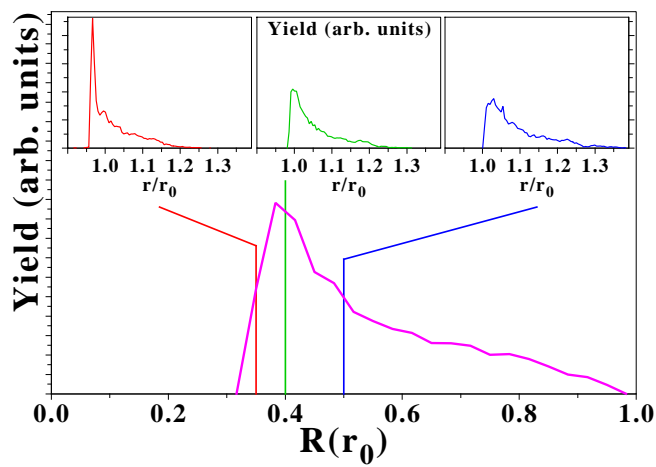}
\caption{ \label{fig:stat_r0} (Color Online) [Main panel] Ratio between the second-order log-density at $r_{0}$ and its peak
value \mbox{$\log_{10}''\rho(r_{\mathrm{max}})$} [Top panels] Distribution of $r/r_{0}$ for which \mbox{$\mathcal{R}(r)$} is
equal to a given value (left panel: \mbox{$\mathcal{R}(r)=0.35$}, middle panel: \mbox{$\mathcal{R}(r)=0.4$}, right
panel: \mbox{$\mathcal{R}(r)=0.5$}), and $r_{0}$.}
\end{figure}

The results from a wide panel of test cases are presented in \figu\ref{fig:stat_r0}. We recall that $r_0$ is
defined as the radius for which $\rho_{h}(r_{0}) = 10 \rho_{c}(r_{0})$. The goal is to determine such a radius
through the value of another ratio: $\mathcal{R}(r_{0})$. For each simulation, the position $r_{0}$ and the ratio
\mbox{$\mathcal{R}(r_{0})$} are computed. The main panel of \figu\ref{fig:stat_r0} shows the distribution of
\mbox{$\mathcal{R}(r_{0})$} which is peaked around $0.4$. This value is greater than in the toy model case for the
reasons detailed above.  The distribution of \mbox{$\mathcal{R}(r_{0})$} is asymmetric, but the tail towards high
values corresponds precisely to non-halo systems. In any case, we are going to reflect such an asymmetry into the
choice of theoretical uncertainties in the determination of $r_0$.

The inserts of \figu\ref{fig:stat_r0} display the distribution of ratio $r/r_{0}$ corresponding to a given value
of \mbox{$\mathcal{R}(r)$}. The ratio \mbox{$\mathcal{R}(r)=2/5$} (top-center panel) indeed picks out quite
consistently the radius $r_{0}$. For \mbox{$\mathcal{R}(r)=40/121$} (top-left panel), the position $r$ is in most
cases below $r_{0}$. As a consequence, the average ratio between tail and core components in the density will be
consistently below ten in this case. On the contrary for \mbox{$\mathcal{R}(r)=1/2$} (top-right panel), $r$ is
systematically larger than $r_{0}$, meaning that the \mbox{$\rho_h/\rho_c$} larger than ten on the average. In the
end, it appears that $r_0$ is indeed well picked out through the condition

\begin{equation}
\mathcal{R}(r_{0})=\frac{2}{5}\,. \label{defr0}
\end{equation}

For those reasons, we use those values of \mbox{$\mathcal{R}(r)$} to set the error bars on the determination of
$r_0$.

Of course, we need to account for the fact that a difference by one order of magnitude between core and halo
densities to define the halo region is somewhat arbitrary and that the corresponding radius $r_0$ cannot be
perfectly picked out in all cases through \eq(\ref{defr0}). As a result, we add a tolerance margin to the
definition of $r_0$ by allowing \mbox{$\mathcal{R}(r_{0})$} to vary between $40/121\approx0.35$ and $1/2$. The
upper margin is greater than the lower one to account for the asymmetry of the peak in \figu\ref{fig:stat_r0}.
Note that the procedure chosen to determine $r_0$ combined with that asymmetry put us on the safe side, i.e. the
radius found through that procedure, if not perfect, is likely to be too large, leading to a slight
underestimation of the halo factors $N_{\mathrm{halo}}$ and $\delta R_{\mathrm{halo}}$.

\end{appendix}
\bibliography{bib-data}

\begin{thebibliography}{126}
\expandafter\ifx\csname natexlab\endcsname\relax\def\natexlab#1{#1}\fi
\expandafter\ifx\csname bibnamefont\endcsname\relax
  \def\bibnamefont#1{#1}\fi
\expandafter\ifx\csname bibfnamefont\endcsname\relax
  \def\bibfnamefont#1{#1}\fi
\expandafter\ifx\csname citenamefont\endcsname\relax
  \def\citenamefont#1{#1}\fi
\expandafter\ifx\csname url\endcsname\relax
  \def\url#1{\texttt{#1}}\fi
\expandafter\ifx\csname urlprefix\endcsname\relax\def\urlprefix{URL }\fi
\providecommand{\bibinfo}[2]{#2}
\providecommand{\eprint}[2][]{\url{#2}}

\bibitem[{\citenamefont{Hansen and Jonson}(1987)}]{hansen87}
\bibinfo{author}{\bibfnamefont{P.~G.} \bibnamefont{Hansen}} \bibnamefont{and}
  \bibinfo{author}{\bibfnamefont{B.}~\bibnamefont{Jonson}},
  \bibinfo{journal}{Europhys. Lett.} \textbf{\bibinfo{volume}{4}},
  \bibinfo{pages}{409} (\bibinfo{year}{1987}).

\bibitem[{\citenamefont{Tanihata
  et~al.}(1985{\natexlab{a}})\citenamefont{Tanihata, Hamagaki, Hashimoto,
  Shida, Yoshikawa, Sugimoto, Yamakawa, Kobayashi, and
  Takahashi}}]{tanihata85a}
\bibinfo{author}{\bibfnamefont{I.}~\bibnamefont{Tanihata}},
  \bibinfo{author}{\bibfnamefont{H.}~\bibnamefont{Hamagaki}},
  \bibinfo{author}{\bibfnamefont{O.}~\bibnamefont{Hashimoto}},
  \bibinfo{author}{\bibfnamefont{Y.}~\bibnamefont{Shida}},
  \bibinfo{author}{\bibfnamefont{N.}~\bibnamefont{Yoshikawa}},
  \bibinfo{author}{\bibfnamefont{K.}~\bibnamefont{Sugimoto}},
  \bibinfo{author}{\bibfnamefont{O.}~\bibnamefont{Yamakawa}},
  \bibinfo{author}{\bibfnamefont{T.}~\bibnamefont{Kobayashi}},
  \bibnamefont{and}
  \bibinfo{author}{\bibfnamefont{N.}~\bibnamefont{Takahashi}},
  \bibinfo{journal}{Phys. Rev. Lett.} \textbf{\bibinfo{volume}{55}},
  \bibinfo{pages}{2676} (\bibinfo{year}{1985}{\natexlab{a}}).

\bibitem[{\citenamefont{Tanihata
  et~al.}(1985{\natexlab{b}})\citenamefont{Tanihata, Hamagaki, Hashimoto,
  Nagamiya, Shida, Yoshikawa, Yamakawa, Sugimoto, Kobayashi, Greiner, Takahashi, Nojiri}}]{tanihata85b}
\bibinfo{author}{\bibfnamefont{I.}~\bibnamefont{Tanihata}},
  \bibinfo{author}{\bibfnamefont{H.}~\bibnamefont{Hamagaki}},
  \bibinfo{author}{\bibfnamefont{O.}~\bibnamefont{Hashimoto}},
  \bibinfo{author}{\bibfnamefont{S.}~\bibnamefont{Nagamiya}},
  \bibinfo{author}{\bibfnamefont{Y.}~\bibnamefont{Shida}},
  \bibinfo{author}{\bibfnamefont{N.}~\bibnamefont{Yoshikawa}},
  \bibinfo{author}{\bibfnamefont{O.}~\bibnamefont{Yamakawa}},
  \bibinfo{author}{\bibfnamefont{K.}~\bibnamefont{Sugimoto}},
  \bibinfo{author}{\bibfnamefont{T.}~\bibnamefont{Kobayashi}},
  \bibinfo{author}{\bibfnamefont{D.~E.} \bibnamefont{Greiner}}, 
  \bibinfo{author}{\bibfnamefont{N.} \bibnamefont{Takahashi}},  
  \bibinfo{author}{\bibfnamefont{Y.} \bibnamefont{Nojiri}}, 
  \bibinfo{journal}{Phys. Lett. B}
  \textbf{\bibinfo{volume}{160}}, \bibinfo{pages}{380}
  (\bibinfo{year}{1985}{\natexlab{b}}).

\bibitem[{\citenamefont{Zhukov et~al.}(1993)\citenamefont{Zhukov, Danilin,
  Fedorov, Bang, Thompson, and Vaagen}}]{zhukov93}
\bibinfo{author}{\bibfnamefont{M.~V.} \bibnamefont{Zhukov}},
  \bibinfo{author}{\bibfnamefont{B.~V.} \bibnamefont{Danilin}},
  \bibinfo{author}{\bibfnamefont{D.~V.} \bibnamefont{Fedorov}},
  \bibinfo{author}{\bibfnamefont{J.~M.} \bibnamefont{Bang}},
  \bibinfo{author}{\bibfnamefont{I.~J.} \bibnamefont{Thompson}},
  \bibnamefont{and} \bibinfo{author}{\bibfnamefont{J.~S.}
  \bibnamefont{Vaagen}}, \bibinfo{journal}{Phys. Rep.}
  \textbf{\bibinfo{volume}{231}}, \bibinfo{pages}{151} (\bibinfo{year}{1993}).

\bibitem[{\citenamefont{Tanihata
  et~al.}(1985{\natexlab{c}})\citenamefont{Tanihata, Kobayashi, Yamakawa,
  Shimoura, Ekuni, Sugimoto, Takahashi, Shimoda, and Sato}}]{tanihata88}
\bibinfo{author}{\bibfnamefont{I.}~\bibnamefont{Tanihata}},
  \bibinfo{author}{\bibfnamefont{T.}~\bibnamefont{Kobayashi}},
  \bibinfo{author}{\bibfnamefont{O.}~\bibnamefont{Yamakawa}},
  \bibinfo{author}{\bibfnamefont{S.}~\bibnamefont{Shimoura}},
  \bibinfo{author}{\bibfnamefont{K.}~\bibnamefont{Ekuni}},
  \bibinfo{author}{\bibfnamefont{K.}~\bibnamefont{Sugimoto}},
  \bibinfo{author}{\bibfnamefont{N.}~\bibnamefont{Takahashi}},
  \bibinfo{author}{\bibfnamefont{T.}~\bibnamefont{Shimoda}}, \bibnamefont{and}
  \bibinfo{author}{\bibfnamefont{H.}~\bibnamefont{Sato}},
  \bibinfo{journal}{Phys. Lett. B} \textbf{\bibinfo{volume}{206}},
  \bibinfo{pages}{592} (\bibinfo{year}{1985}{\natexlab{c}}).

\bibitem[{\citenamefont{Fukuda et~al.}(1991)\citenamefont{Fukuda, Ichihara,
  Inabe, Kubo, Kumagai, Nakagawa, Yano, Tanihata, Adachi, Asahi, Kougichi, Ishihara, Sagawa, Shimoura}}]{fukuda91}
\bibinfo{author}{\bibfnamefont{M.}~\bibnamefont{Fukuda}},
  \bibinfo{author}{\bibfnamefont{T.}~\bibnamefont{Ichihara}},
  \bibinfo{author}{\bibfnamefont{N.}~\bibnamefont{Inabe}},
  \bibinfo{author}{\bibfnamefont{T.}~\bibnamefont{Kubo}},
  \bibinfo{author}{\bibfnamefont{H.}~\bibnamefont{Kumagai}},
  \bibinfo{author}{\bibfnamefont{T.}~\bibnamefont{Nakagawa}},
  \bibinfo{author}{\bibfnamefont{Y.}~\bibnamefont{Yano}},
  \bibinfo{author}{\bibfnamefont{I.}~\bibnamefont{Tanihata}},
  \bibinfo{author}{\bibfnamefont{M.}~\bibnamefont{Adachi}},
  \bibinfo{author}{\bibfnamefont{K.}~\bibnamefont{Asahi}},
  \bibinfo{author}{\bibfnamefont{M.}~\bibnamefont{Kougichi}},
  \bibinfo{author}{\bibfnamefont{M.}~\bibnamefont{Ishihara}},
  \bibinfo{author}{\bibfnamefont{H.}~\bibnamefont{Sagawa}},
  \bibinfo{author}{\bibfnamefont{S.}~\bibnamefont{Shimoura}},
  \bibinfo{journal}{Phys. Lett. B}
  \textbf{\bibinfo{volume}{268}}, \bibinfo{pages}{339} (\bibinfo{year}{1991}).
 
\bibitem[{\citenamefont{Zahar et~al.}(1993)\citenamefont{Zahar, Belbot, Kolata,
  Lamkin, Thompson, Orr, Kelley, Kryger, Morrissey, Sherrill, Winger, Winfield, Wuosama}}]{zahar93}
\bibinfo{author}{\bibfnamefont{M.}~\bibnamefont{Zahar}},
  \bibinfo{author}{\bibfnamefont{M.}~\bibnamefont{Belbot}},
  \bibinfo{author}{\bibfnamefont{J.~J.} \bibnamefont{Kolata}},
  \bibinfo{author}{\bibfnamefont{K.}~\bibnamefont{Lamkin}},
  \bibinfo{author}{\bibfnamefont{R.}~\bibnamefont{Thompson}},
  \bibinfo{author}{\bibfnamefont{N.~A.} \bibnamefont{Orr}},
  \bibinfo{author}{\bibfnamefont{J.~H.} \bibnamefont{Kelley}},
  \bibinfo{author}{\bibfnamefont{R.~A.} \bibnamefont{Kryger}},
  \bibinfo{author}{\bibfnamefont{D.~J.} \bibnamefont{Morrissey}},
  \bibinfo{author}{\bibfnamefont{B.~M.} \bibnamefont{Sherrill}},
  \bibinfo{author}{\bibfnamefont{J.~A.} \bibnamefont{Winger}},
  \bibinfo{author}{\bibfnamefont{J.~S.} \bibnamefont{Winfield}},
  \bibinfo{author}{\bibfnamefont{A.~H.} \bibnamefont{Wuosama}},
  \bibinfo{journal}{Phys. Rev. C}
  \textbf{\bibinfo{volume}{48}}, \bibinfo{pages}{R1484} (\bibinfo{year}{1993}).
 
\bibitem[{\citenamefont{Thompson and Zhukov}(1996)}]{thompson96}
\bibinfo{author}{\bibfnamefont{I.~J.} \bibnamefont{Thompson}} \bibnamefont{and}
  \bibinfo{author}{\bibfnamefont{M.~V.} \bibnamefont{Zhukov}},
  \bibinfo{journal}{Phys. Rev. C} \textbf{\bibinfo{volume}{53}},
  \bibinfo{pages}{708} (\bibinfo{year}{1996}).

\bibitem[{\citenamefont{Bazin et~al.}(1995)\citenamefont{Bazin, Brown, Brown,
  Fauerbach, Hellstr{\"o}m, Hirzebruch, Helley, Kryger, Morrissey, Pfaff, Powell, Sherill, Thoenessen}}]{bazin95}
\bibinfo{author}{\bibfnamefont{D.}~\bibnamefont{Bazin}},
  \bibinfo{author}{\bibfnamefont{B.~A.} \bibnamefont{Brown}},
  \bibinfo{author}{\bibfnamefont{J.}~\bibnamefont{Brown}},
  \bibinfo{author}{\bibfnamefont{M.}~\bibnamefont{Fauerbach}},
  \bibinfo{author}{\bibfnamefont{M.}~\bibnamefont{Hellstr{\"o}m}},
  \bibinfo{author}{\bibfnamefont{S.~E.} \bibnamefont{Hirzebruch}},
  \bibinfo{author}{\bibfnamefont{J.~H.} \bibnamefont{Helley}},
  \bibinfo{author}{\bibfnamefont{R.~A.} \bibnamefont{Kryger}},
  \bibinfo{author}{\bibfnamefont{D.}~\bibnamefont{Morrissey}},
  \bibinfo{author}{\bibfnamefont{R.}~\bibnamefont{Pfaff}},
  \bibinfo{author}{\bibfnamefont{C.~F.}~\bibnamefont{Powell}},
  \bibinfo{author}{\bibfnamefont{B.~M.}~\bibnamefont{Sherill}},
  \bibinfo{author}{\bibfnamefont{M.}~\bibnamefont{Thoenessen}},
  \bibinfo{journal}{Phys. Rev. Lett.}
  \textbf{\bibinfo{volume}{74}}, \bibinfo{pages}{3569} (\bibinfo{year}{1995}).
 
\bibitem[{\citenamefont{Kanungo et~al.}(2002)\citenamefont{Kanungo, Tanihata,
  Ogawa, Toki, and Ozawa}}]{kanungo00}
\bibinfo{author}{\bibfnamefont{R.}~\bibnamefont{Kanungo}},
  \bibinfo{author}{\bibfnamefont{I.}~\bibnamefont{Tanihata}},
  \bibinfo{author}{\bibfnamefont{Y.}~\bibnamefont{Ogawa}},
  \bibinfo{author}{\bibfnamefont{H.}~\bibnamefont{Toki}}, \bibnamefont{and}
  \bibinfo{author}{\bibfnamefont{A.}~\bibnamefont{Ozawa}},
  \bibinfo{journal}{Nucl. Phys.} \textbf{\bibinfo{volume}{A701}},
  \bibinfo{pages}{378} (\bibinfo{year}{2002}).

\bibitem[{\citenamefont{Zhukov and Thompson}(1995)}]{zhukov95}
\bibinfo{author}{\bibfnamefont{M.~V.} \bibnamefont{Zhukov}} \bibnamefont{and}
  \bibinfo{author}{\bibfnamefont{I.~J.} \bibnamefont{Thompson}},
  \bibinfo{journal}{Phys. Rev. C} \textbf{\bibinfo{volume}{52}},
  \bibinfo{pages}{3505} (\bibinfo{year}{1995}).

\bibitem[{\citenamefont{Minamisono et~al.}(1992)\citenamefont{Minamisono,
  Ohtsubo, Minami, Fukuda, Kitagawa, Fukuda, Matsua, Nojiri, Takeda, Sagawa, Kitagawa}}]{minamisono92}
\bibinfo{author}{\bibfnamefont{T.}~\bibnamefont{Minamisono}},
  \bibinfo{author}{\bibfnamefont{T.}~\bibnamefont{Ohtsubo}},
  \bibinfo{author}{\bibfnamefont{I.}~\bibnamefont{Minami}},
  \bibinfo{author}{\bibfnamefont{S.}~\bibnamefont{Fukuda}},
  \bibinfo{author}{\bibfnamefont{A.}~\bibnamefont{Kitagawa}},
  \bibinfo{author}{\bibfnamefont{M.}~\bibnamefont{Fukuda}},
  \bibinfo{author}{\bibfnamefont{K.}~\bibnamefont{Matsua}},
  \bibinfo{author}{\bibfnamefont{Y.}~\bibnamefont{Nojiri}},
  \bibinfo{author}{\bibfnamefont{S.}~\bibnamefont{Takeda}},
  \bibinfo{author}{\bibfnamefont{H.}~\bibnamefont{Sagawa}},
  \bibinfo{author}{\bibfnamefont{H.}~\bibnamefont{Kitagawa}},
  \bibinfo{journal}{Phys. Rev. Lett.}
  \textbf{\bibinfo{volume}{69}}, \bibinfo{pages}{2058} (\bibinfo{year}{1992}).

\bibitem[{\citenamefont{Schwab et~al.}(1995)\citenamefont{Schwab, Geissel,
  Lenske, Behr, Br{\"u}nle, Burkard, Irnich, Kobayashi, Kraus, Magel, M{\"u}nzenberg, Nickel, Riisager, Scheidenberger, Sherrill, Suzuki, Voss}}]{schwab95}
\bibinfo{author}{\bibfnamefont{W.}~\bibnamefont{Schwab}},
  \bibinfo{author}{\bibfnamefont{H.}~\bibnamefont{Geissel}},
  \bibinfo{author}{\bibfnamefont{H.}~\bibnamefont{Lenske}},
  \bibinfo{author}{\bibfnamefont{K.-H.} \bibnamefont{Behr}},
  \bibinfo{author}{\bibfnamefont{A.}~\bibnamefont{Br{\"u}nle}},
  \bibinfo{author}{\bibfnamefont{K.}~\bibnamefont{Burkard}},
  \bibinfo{author}{\bibfnamefont{H.}~\bibnamefont{Irnich}},
  \bibinfo{author}{\bibfnamefont{T.}~\bibnamefont{Kobayashi}},
  \bibinfo{author}{\bibfnamefont{G.}~\bibnamefont{Kraus}},
  \bibinfo{author}{\bibfnamefont{A.}~\bibnamefont{Magel}},
  \bibinfo{author}{\bibfnamefont{G.}~\bibnamefont{M{\"u}nzenberg}},
  \bibinfo{author}{\bibfnamefont{F.}~\bibnamefont{Nickel}},
  \bibinfo{author}{\bibfnamefont{K.}~\bibnamefont{Riisager}},
  \bibinfo{author}{\bibfnamefont{C.}~\bibnamefont{Scheidenberger}},
  \bibinfo{author}{\bibfnamefont{B.~M.}~\bibnamefont{Sherrill}},
  \bibinfo{author}{\bibfnamefont{T.}~\bibnamefont{Suzuki}},
  \bibinfo{author}{\bibfnamefont{B.}~\bibnamefont{Voss}},
  \bibinfo{journal}{Z. Phys.}
  \textbf{\bibinfo{volume}{A350}}, \bibinfo{pages}{283} (\bibinfo{year}{1995}).

\bibitem[{\citenamefont{Warner et~al.}(1995)\citenamefont{Warner, Kelley,
  Zecher, Becchetti, Brown, Carpenter, Aglonsky, Kruse, Muthukrishnan, Nadasen, Ronnigen, Schwandt, Sherill, Wang, Winfield}}]{warner95}
\bibinfo{author}{\bibfnamefont{R.~E.} \bibnamefont{Warner}},
  \bibinfo{author}{\bibfnamefont{J.~H.} \bibnamefont{Kelley}},
  \bibinfo{author}{\bibfnamefont{P.}~\bibnamefont{Zecher}},
  \bibinfo{author}{\bibfnamefont{F.~D.} \bibnamefont{Becchetti}},
  \bibinfo{author}{\bibfnamefont{J.~A.} \bibnamefont{Brown}},
  \bibinfo{author}{\bibfnamefont{C.~L.} \bibnamefont{Carpenter}},
  \bibinfo{author}{\bibfnamefont{A.}~\bibnamefont{Aglonsky}},
  \bibinfo{author}{\bibfnamefont{J.}~\bibnamefont{Kruse}},
  \bibinfo{author}{\bibfnamefont{A.}~\bibnamefont{Muthukrishnan}},
  \bibinfo{author}{\bibfnamefont{A.}~\bibnamefont{Nadasen}},
  \bibinfo{author}{\bibfnamefont{R.~M.}~\bibnamefont{Ronnigen}},
  \bibinfo{author}{\bibfnamefont{P.}~\bibnamefont{Schwandt}},
  \bibinfo{author}{\bibfnamefont{B.~M.}~\bibnamefont{Sherill}},
  \bibinfo{author}{\bibfnamefont{J.}~\bibnamefont{Wang}},
  \bibinfo{author}{\bibfnamefont{J.~S.}~\bibnamefont{Winfield}},
  \bibinfo{journal}{Phys. Rev. C}
  \textbf{\bibinfo{volume}{52}}, \bibinfo{pages}{R1166} (\bibinfo{year}{1995}).

\bibitem[{\citenamefont{Negoita et~al.}(1996)\citenamefont{Negoita, Borcea,
  Carstoiu, Lewitowicz, Saint-Laurent, Anne, Bazin, Corre, Roussel-Chomaz,
  Borrel, Guillemaud-Mueller, Keller, Mueller, Pougheon, Sorlin, Lukyanov, Penionzhkevich, Fomichev, 
  Skobelev, Tarasov, Dlouhy, Kordyasz}}]{negoita96}
\bibinfo{author}{\bibfnamefont{F.}~\bibnamefont{Negoita}},
  \bibinfo{author}{\bibfnamefont{C.}~\bibnamefont{Borcea}},
  \bibinfo{author}{\bibfnamefont{F.}~\bibnamefont{Carstoiu}},
  \bibinfo{author}{\bibfnamefont{M.}~\bibnamefont{Lewitowicz}},
  \bibinfo{author}{\bibfnamefont{M.~G.} \bibnamefont{Saint-Laurent}},
  \bibinfo{author}{\bibfnamefont{R.}~\bibnamefont{Anne}},
  \bibinfo{author}{\bibfnamefont{D.}~\bibnamefont{Bazin}},
  \bibinfo{author}{\bibfnamefont{J.~M.} \bibnamefont{Corre}},
  \bibinfo{author}{\bibfnamefont{P.}~\bibnamefont{Roussel-Chomaz}},
  \bibinfo{author}{\bibfnamefont{V.}~\bibnamefont{Borrel}},
  \bibinfo{author}{\bibfnamefont{D.}~\bibnamefont{Guillemaud-Mueller}},
  \bibinfo{author}{\bibfnamefont{H.}~\bibnamefont{Keller}},
  \bibinfo{author}{\bibfnamefont{A.~C.}~\bibnamefont{Mueller}},
  \bibinfo{author}{\bibfnamefont{F.}~\bibnamefont{Pougheon}},
  \bibinfo{author}{\bibfnamefont{O.}~\bibnamefont{Sorlin}},
  \bibinfo{author}{\bibfnamefont{S.}~\bibnamefont{Lukyanov}},
  \bibinfo{author}{\bibfnamefont{Y.}~\bibnamefont{Penionzhkevich}},
  \bibinfo{author}{\bibfnamefont{A.}~\bibnamefont{Fomichev}},
  \bibinfo{author}{\bibfnamefont{N.}~\bibnamefont{Skobelev}},
  \bibinfo{author}{\bibfnamefont{O.}~\bibnamefont{Tarasov}},
  \bibinfo{author}{\bibfnamefont{Z.}~\bibnamefont{Dlouhy}},
  \bibinfo{author}{\bibfnamefont{A.}~\bibnamefont{Kordyasz}},
  \bibinfo{journal}{Phys. Rev. C}
  \textbf{\bibinfo{volume}{54}}, \bibinfo{pages}{1787} (\bibinfo{year}{1996}).

\bibitem[{\citenamefont{Kanungo et~al.}(2003)\citenamefont{Kanungo, Chiba,
  Adhikari, Fang, Isawa, Kimura, Maeda, Nishimura, Ogawa, Ohnishi, Ozawa, Samanta, Suda, Suzuki, Wang, Wu, Yamaguchi, 
  Yamada, Yoshida, Zheng, Tanihata}}]{kanungo03}
\bibinfo{author}{\bibfnamefont{R.}~\bibnamefont{Kanungo}},
  \bibinfo{author}{\bibfnamefont{M.}~\bibnamefont{Chiba}},
  \bibinfo{author}{\bibfnamefont{S.}~\bibnamefont{Adhikari}},
  \bibinfo{author}{\bibfnamefont{D.}~\bibnamefont{Fang}},
  \bibinfo{author}{\bibfnamefont{N.}~\bibnamefont{Isawa}},
  \bibinfo{author}{\bibfnamefont{K.}~\bibnamefont{Kimura}},
  \bibinfo{author}{\bibfnamefont{K.}~\bibnamefont{Maeda}},
  \bibinfo{author}{\bibfnamefont{S.}~\bibnamefont{Nishimura}},
  \bibinfo{author}{\bibfnamefont{Y.}~\bibnamefont{Ogawa}},
  \bibinfo{author}{\bibfnamefont{T.}~\bibnamefont{Ohnishi}},
  \bibinfo{author}{\bibfnamefont{A.}~\bibnamefont{Ozawa}},
  \bibinfo{author}{\bibfnamefont{C.}~\bibnamefont{Samanta}},
  \bibinfo{author}{\bibfnamefont{T.}~\bibnamefont{Suda}},
  \bibinfo{author}{\bibfnamefont{T.}~\bibnamefont{Suzuki}},
  \bibinfo{author}{\bibfnamefont{Q.}~\bibnamefont{Wang}},
  \bibinfo{author}{\bibfnamefont{C.}~\bibnamefont{Wu}},
  \bibinfo{author}{\bibfnamefont{Y.}~\bibnamefont{Yamaguchi}},
  \bibinfo{author}{\bibfnamefont{K.}~\bibnamefont{Yamada}},
  \bibinfo{author}{\bibfnamefont{A.}~\bibnamefont{Yoshida}},
  \bibinfo{author}{\bibfnamefont{T.}~\bibnamefont{Zheng}},
  \bibinfo{author}{\bibfnamefont{I.}~\bibnamefont{Tanihata}},
  \bibinfo{journal}{Phys. Lett. B}
  \textbf{\bibinfo{volume}{571}}, \bibinfo{pages}{21} (\bibinfo{year}{2003}).

\bibitem[{\citenamefont{Jeppesen et~al.}(2004)\citenamefont{Jeppesen, Kanungo,
  Abu-Ibrahim, Adhikiri, Chiba, Fang, Isawa, Kimura, Maeda, Nishimura, Ohnishi, Ozawa, Samanta, Suda, Suzuki, Tanihata, Wang, Wu, Yamaguchi, 
  Yamada, Yoshida, Zheng}}]{jeppesen04}
\bibinfo{author}{\bibfnamefont{H.}~\bibnamefont{Jeppesen}},
  \bibinfo{author}{\bibfnamefont{R.}~\bibnamefont{Kanungo}},
  \bibinfo{author}{\bibfnamefont{B.}~\bibnamefont{Abu-Ibrahim}},
  \bibinfo{author}{\bibfnamefont{S.}~\bibnamefont{Adhikiri}},
  \bibinfo{author}{\bibfnamefont{M.}~\bibnamefont{Chiba}},
  \bibinfo{author}{\bibfnamefont{D.}~\bibnamefont{Fang}},
  \bibinfo{author}{\bibfnamefont{N.}~\bibnamefont{Isawa}},
  \bibinfo{author}{\bibfnamefont{K.}~\bibnamefont{Kimura}},
  \bibinfo{author}{\bibfnamefont{K.}~\bibnamefont{Maeda}},
  \bibinfo{author}{\bibfnamefont{S.}~\bibnamefont{Nishimura}},
  \bibinfo{author}{\bibfnamefont{T.}~\bibnamefont{Ohnishi}},
  \bibinfo{author}{\bibfnamefont{A.}~\bibnamefont{Ozawa}},
  \bibinfo{author}{\bibfnamefont{C.}~\bibnamefont{Samanta}},
  \bibinfo{author}{\bibfnamefont{T.}~\bibnamefont{Suda}},
  \bibinfo{author}{\bibfnamefont{T.}~\bibnamefont{Suzuki}},
  \bibinfo{author}{\bibfnamefont{I.}~\bibnamefont{Tanihata}},
  \bibinfo{author}{\bibfnamefont{Q.}~\bibnamefont{Wang}},
  \bibinfo{author}{\bibfnamefont{C.}~\bibnamefont{Wu}},
  \bibinfo{author}{\bibfnamefont{Y.}~\bibnamefont{Yamaguchi}},
  \bibinfo{author}{\bibfnamefont{K.}~\bibnamefont{Yamada}},
  \bibinfo{author}{\bibfnamefont{A.}~\bibnamefont{Yoshida}},
  \bibinfo{author}{\bibfnamefont{T.}~\bibnamefont{Zheng}},
  \bibinfo{journal}{Nucl. Phys.}
  \textbf{\bibinfo{volume}{A739}}, \bibinfo{pages}{57} (\bibinfo{year}{2004}).

\bibitem[{\citenamefont{Morlock et~al.}(1997)\citenamefont{Morlock, Kunz,
  Mayer, Jaeger, M\"uller, Hammer, Mohr, Oberhummer, Staudt, and
  K\"olle}}]{morlock97}
\bibinfo{author}{\bibfnamefont{R.}~\bibnamefont{Morlock}},
  \bibinfo{author}{\bibfnamefont{R.}~\bibnamefont{Kunz}},
  \bibinfo{author}{\bibfnamefont{A.}~\bibnamefont{Mayer}},
  \bibinfo{author}{\bibfnamefont{M.}~\bibnamefont{Jaeger}},
  \bibinfo{author}{\bibfnamefont{A.}~\bibnamefont{M\"uller}},
  \bibinfo{author}{\bibfnamefont{J.~W.} \bibnamefont{Hammer}},
  \bibinfo{author}{\bibfnamefont{P.}~\bibnamefont{Mohr}},
  \bibinfo{author}{\bibfnamefont{H.}~\bibnamefont{Oberhummer}},
  \bibinfo{author}{\bibfnamefont{G.}~\bibnamefont{Staudt}}, \bibnamefont{and}
  \bibinfo{author}{\bibfnamefont{V.}~\bibnamefont{K\"olle}},
  \bibinfo{journal}{Phys. Rev. Lett.} \textbf{\bibinfo{volume}{79}},
  \bibinfo{pages}{3837} (\bibinfo{year}{1997}).

\bibitem[{\citenamefont{Ren et~al.}(1998)\citenamefont{Ren, Faessler, and
  Bobyk}}]{ren98}
\bibinfo{author}{\bibfnamefont{Z.}~\bibnamefont{Ren}},
  \bibinfo{author}{\bibfnamefont{A.}~\bibnamefont{Faessler}}, \bibnamefont{and}
  \bibinfo{author}{\bibfnamefont{A.}~\bibnamefont{Bobyk}},
  \bibinfo{journal}{Phys. Rev C} \textbf{\bibinfo{volume}{57}},
  \bibinfo{pages}{2752} (\bibinfo{year}{1998}).

\bibitem[{\citenamefont{Lin et~al.}(2001)\citenamefont{Lin, Liu, Zhang, Wu,
  Yang, and Ruan}}]{lin01}
\bibinfo{author}{\bibfnamefont{C.~J.} \bibnamefont{Lin}},
  \bibinfo{author}{\bibfnamefont{Z.~H.} \bibnamefont{Liu}},
  \bibinfo{author}{\bibfnamefont{H.~Q.} \bibnamefont{Zhang}},
  \bibinfo{author}{\bibfnamefont{Y.~W.} \bibnamefont{Wu}},
  \bibinfo{author}{\bibfnamefont{F.}~\bibnamefont{Yang}}, \bibnamefont{and}
  \bibinfo{author}{\bibfnamefont{M.}~\bibnamefont{Ruan}},
  \bibinfo{journal}{Chin. Phys. Lett.} \textbf{\bibinfo{volume}{18}},
  \bibinfo{pages}{1183} (\bibinfo{year}{2001}).

\bibitem[{\citenamefont{Liu et~al.}(2001)\citenamefont{Liu, Lin, Zhang, Li,
  Zhang, Wu, Yang, Ruan, Liu, Li, Peng}}]{liu01}
\bibinfo{author}{\bibfnamefont{Z.~H.} \bibnamefont{Liu}},
  \bibinfo{author}{\bibfnamefont{C.~J.} \bibnamefont{Lin}},
  \bibinfo{author}{\bibfnamefont{H.~Q.} \bibnamefont{Zhang}},
  \bibinfo{author}{\bibfnamefont{Z.~C.} \bibnamefont{Li}},
  \bibinfo{author}{\bibfnamefont{J.~S.} \bibnamefont{Zhang}},
  \bibinfo{author}{\bibfnamefont{Y.~W.} \bibnamefont{Wu}},
  \bibinfo{author}{\bibfnamefont{F.}~\bibnamefont{Yang}},
  \bibinfo{author}{\bibfnamefont{M.}~\bibnamefont{Ruan}},
  \bibinfo{author}{\bibfnamefont{J.~C.} \bibnamefont{Liu}},
  \bibinfo{author}{\bibfnamefont{S.~Y.} \bibnamefont{Li}},
  \bibinfo{author}{\bibfnamefont{Z.~H.} \bibnamefont{Peng}},
  \bibinfo{journal}{Phys. Rev. C}
  \textbf{\bibinfo{volume}{64}}, \bibinfo{pages}{034312}
  (\bibinfo{year}{2001}).

\bibitem[{\citenamefont{Chen et~al.}(2005)\citenamefont{Chen, Cai, Shen, Ma,
  Ren, Zhang, Jiang, Zhong, Wei, Guo, Zhou, Wang, Ma}}]{chen05}
\bibinfo{author}{\bibfnamefont{J.~G.} \bibnamefont{Chen}},
  \bibinfo{author}{\bibfnamefont{X.~Z.} \bibnamefont{Cai}},
  \bibinfo{author}{\bibfnamefont{W.~Q.} \bibnamefont{Shen}},
  \bibinfo{author}{\bibfnamefont{Y.~G.} \bibnamefont{Ma}},
  \bibinfo{author}{\bibfnamefont{Z.~Z.} \bibnamefont{Ren}},
  \bibinfo{author}{\bibfnamefont{H.~Y.} \bibnamefont{Zhang}},
  \bibinfo{author}{\bibfnamefont{W.~J.} \bibnamefont{Jiang}},
  \bibinfo{author}{\bibfnamefont{C.}~\bibnamefont{Zhong}},
  \bibinfo{author}{\bibfnamefont{Y.~B.} \bibnamefont{Wei}},
  \bibinfo{author}{\bibfnamefont{W.}~\bibnamefont{Guo}},
  \bibinfo{author}{\bibfnamefont{X.~F.}~\bibnamefont{Zhou}},
  \bibinfo{author}{\bibfnamefont{K.}~\bibnamefont{Wang}},
  \bibinfo{author}{\bibfnamefont{G.~L.}~\bibnamefont{Ma}},
  \bibinfo{journal}{Eur. Phys. J. A} \textbf{\bibinfo{volume}{23}},
  \bibinfo{pages}{11} (\bibinfo{year}{2005}).

\bibitem[{\citenamefont{Li et~al.}(2006)\citenamefont{Li, Gou, and Shi}}]{li06}
\bibinfo{author}{\bibfnamefont{Y.}~\bibnamefont{Li}},
  \bibinfo{author}{\bibfnamefont{Q.}~\bibnamefont{Gou}}, \bibnamefont{and}
  \bibinfo{author}{\bibfnamefont{T.}~\bibnamefont{Shi}},
  \bibinfo{journal}{Phys. Rev. A} \textbf{\bibinfo{volume}{74}},
  \bibinfo{pages}{032502} (\bibinfo{year}{2006}).

\bibitem[{\citenamefont{Sch{\"o}llkopf and Toensies}(1996)}]{schollkopf96}
\bibinfo{author}{\bibfnamefont{W.}~\bibnamefont{Sch{\"o}llkopf}}
  \bibnamefont{and} \bibinfo{author}{\bibfnamefont{J.~P.}
  \bibnamefont{Toensies}}, \bibinfo{journal}{J. Chem. Phys.}
  \textbf{\bibinfo{volume}{104}}, \bibinfo{pages}{3} (\bibinfo{year}{1996}).

\bibitem[{\citenamefont{Nielsen et~al.}(1998)\citenamefont{Nielsen, Fedorov,
  and Jensen}}]{nielsen98}
\bibinfo{author}{\bibfnamefont{E.~D.} \bibnamefont{Nielsen}},
  \bibinfo{author}{\bibfnamefont{D.~V.} \bibnamefont{Fedorov}},
  \bibnamefont{and} \bibinfo{author}{\bibfnamefont{A.~S.}
  \bibnamefont{Jensen}}, \bibinfo{journal}{J. Phys. B: At. Mol. Opt. Phys.}
  \textbf{\bibinfo{volume}{31}}, \bibinfo{pages}{4035} (\bibinfo{year}{1998}).

\bibitem[{\citenamefont{Grisenti et~al.}(2000)\citenamefont{Grisenti,
  Sch\"ollkopf, Toennies, Hegerfeldt, K\"ohler, and Stoll}}]{grisenti00}
\bibinfo{author}{\bibfnamefont{R.~E.} \bibnamefont{Grisenti}},
  \bibinfo{author}{\bibfnamefont{W.}~\bibnamefont{Sch\"ollkopf}},
  \bibinfo{author}{\bibfnamefont{J.~P.} \bibnamefont{Toennies}},
  \bibinfo{author}{\bibfnamefont{G.~C.} \bibnamefont{Hegerfeldt}},
  \bibinfo{author}{\bibfnamefont{T.}~\bibnamefont{K\"ohler}}, \bibnamefont{and}
  \bibinfo{author}{\bibfnamefont{M.}~\bibnamefont{Stoll}},
  \bibinfo{journal}{Phys. Rev. Lett.} \textbf{\bibinfo{volume}{85}},
  \bibinfo{pages}{2284} (\bibinfo{year}{2000}).

\bibitem[{\citenamefont{Bressanini et~al.}(2002)\citenamefont{Bressanini,
  Morosi, Bertini, and Mella}}]{bressanini02}
\bibinfo{author}{\bibfnamefont{D.}~\bibnamefont{Bressanini}},
  \bibinfo{author}{\bibfnamefont{G.}~\bibnamefont{Morosi}},
  \bibinfo{author}{\bibfnamefont{L.}~\bibnamefont{Bertini}}, \bibnamefont{and}
  \bibinfo{author}{\bibfnamefont{M.}~\bibnamefont{Mella}},
  \bibinfo{journal}{Few-Body Syst.} \textbf{\bibinfo{volume}{31}},
  \bibinfo{pages}{199} (\bibinfo{year}{2002}).

\bibitem[{\citenamefont{Mitroy}(2005)}]{mitroy05}
\bibinfo{author}{\bibfnamefont{J.}~\bibnamefont{Mitroy}},
  \bibinfo{journal}{Phys. Rev. Lett.} \textbf{\bibinfo{volume}{94}},
  \bibinfo{pages}{033402} (\bibinfo{year}{2005}).

\bibitem[{\citenamefont{Cobis et~al.}(1997)\citenamefont{Cobis, Jensen, and
  Fedorov}}]{cobis97}
\bibinfo{author}{\bibfnamefont{A.}~\bibnamefont{Cobis}},
  \bibinfo{author}{\bibfnamefont{A.~S.} \bibnamefont{Jensen}},
  \bibnamefont{and} \bibinfo{author}{\bibfnamefont{D.~V.}
  \bibnamefont{Fedorov}}, \bibinfo{journal}{J. Phys. G: Nucl. and Part. Phys.}
  \textbf{\bibinfo{volume}{23}}, \bibinfo{pages}{401} (\bibinfo{year}{1997}).

\bibitem[{\citenamefont{Fedorov
  et~al.}(1994{\natexlab{a}})\citenamefont{Fedorov, Jensen, and
  Riisager}}]{fedorov94}
\bibinfo{author}{\bibfnamefont{D.~V.} \bibnamefont{Fedorov}},
  \bibinfo{author}{\bibfnamefont{A.~S.} \bibnamefont{Jensen}},
  \bibnamefont{and} \bibinfo{author}{\bibfnamefont{K.}~\bibnamefont{Riisager}},
  \bibinfo{journal}{Phys. Rev. C} \textbf{\bibinfo{volume}{50}},
  \bibinfo{pages}{2372} (\bibinfo{year}{1994}{\natexlab{a}}).

\bibitem[{\citenamefont{Nunes et~al.}(1996{\natexlab{a}})\citenamefont{Nunes,
  Thompson, and Johnson}}]{nunes96a}
\bibinfo{author}{\bibfnamefont{F.}~\bibnamefont{Nunes}},
  \bibinfo{author}{\bibfnamefont{I.~J.} \bibnamefont{Thompson}},
  \bibnamefont{and} \bibinfo{author}{\bibfnamefont{R.~C.}
  \bibnamefont{Johnson}}, \bibinfo{journal}{Nucl. Phys.}
  \textbf{\bibinfo{volume}{A609}}, \bibinfo{pages}{43}
  (\bibinfo{year}{1996}{\natexlab{a}}).

\bibitem[{\citenamefont{Nunes et~al.}(1996{\natexlab{b}})\citenamefont{Nunes,
  Christley, Thompson, Johnson, and Efros}}]{nunes96b}
\bibinfo{author}{\bibfnamefont{F.}~\bibnamefont{Nunes}},
  \bibinfo{author}{\bibfnamefont{J.~A.} \bibnamefont{Christley}},
  \bibinfo{author}{\bibfnamefont{I.~J.} \bibnamefont{Thompson}},
  \bibinfo{author}{\bibfnamefont{R.~C.} \bibnamefont{Johnson}},
  \bibnamefont{and} \bibinfo{author}{\bibfnamefont{V.~D.} \bibnamefont{Efros}},
  \bibinfo{journal}{Nucl. Phys.} \textbf{\bibinfo{volume}{A596}},
  \bibinfo{pages}{171} (\bibinfo{year}{1996}{\natexlab{b}}).

\bibitem[{\citenamefont{Bang}(1996)}]{bang96}
\bibinfo{author}{\bibfnamefont{J.~M.} \bibnamefont{Bang}},
  \bibinfo{journal}{Phys. Rep.} \textbf{\bibinfo{volume}{264}},
  \bibinfo{pages}{27} (\bibinfo{year}{1996}).

\bibitem[{\citenamefont{Riisager et~al.}(2000)\citenamefont{Riisager, Fedorov,
  and Jensen}}]{riisager00}
\bibinfo{author}{\bibfnamefont{K.}~\bibnamefont{Riisager}},
  \bibinfo{author}{\bibfnamefont{D.~V.} \bibnamefont{Fedorov}},
  \bibnamefont{and} \bibinfo{author}{\bibfnamefont{A.~S.}
  \bibnamefont{Jensen}}, \bibinfo{journal}{Europhys. Lett.}
  \textbf{\bibinfo{volume}{49}}, \bibinfo{pages}{547} (\bibinfo{year}{2000}).

\bibitem[{\citenamefont{Jensen and Zhukov}(2001)}]{jensen01}
\bibinfo{author}{\bibfnamefont{A.~S.} \bibnamefont{Jensen}} \bibnamefont{and}
  \bibinfo{author}{\bibfnamefont{M.~V.} \bibnamefont{Zhukov}},
  \bibinfo{journal}{Nucl. Phys.} \textbf{\bibinfo{volume}{A693}},
  \bibinfo{pages}{411} (\bibinfo{year}{2001}).

\bibitem[{\citenamefont{Fedorov et~al.}(1993)\citenamefont{Fedorov, Jensen, and
  Riisager}}]{fedorov93}
\bibinfo{author}{\bibfnamefont{D.~V.} \bibnamefont{Fedorov}},
  \bibinfo{author}{\bibfnamefont{A.~S.} \bibnamefont{Jensen}},
  \bibnamefont{and} \bibinfo{author}{\bibfnamefont{K.}~\bibnamefont{Riisager}},
  \bibinfo{journal}{Phys. Lett. B} \textbf{\bibinfo{volume}{312}},
  \bibinfo{pages}{1} (\bibinfo{year}{1993}).

\bibitem[{\citenamefont{Jensen and Riisager}(2000)}]{jensen00}
\bibinfo{author}{\bibfnamefont{A.~S.} \bibnamefont{Jensen}} \bibnamefont{and}
  \bibinfo{author}{\bibfnamefont{K.}~\bibnamefont{Riisager}},
  \bibinfo{journal}{Phys. Lett. B} \textbf{\bibinfo{volume}{480}},
  \bibinfo{pages}{39} (\bibinfo{year}{2000}).

\bibitem[{\citenamefont{Fedorov
  et~al.}(1994{\natexlab{b}})\citenamefont{Fedorov, Jensen, and
  Riisager}}]{fedorov94b}
\bibinfo{author}{\bibfnamefont{D.~V.} \bibnamefont{Fedorov}},
  \bibinfo{author}{\bibfnamefont{A.~S.} \bibnamefont{Jensen}},
  \bibnamefont{and} \bibinfo{author}{\bibfnamefont{K.}~\bibnamefont{Riisager}},
  \bibinfo{journal}{Phys. Rev. C} \textbf{\bibinfo{volume}{49}},
  \bibinfo{pages}{201} (\bibinfo{year}{1994}{\natexlab{b}}).

\bibitem[{\citenamefont{Liang et~al.}()\citenamefont{Liang, Li, Deng, Li, Bian,
  F.-S, Liu, and Zhou}}]{liang07}
\bibinfo{author}{\bibfnamefont{Y.-J.} \bibnamefont{Liang}},
  \bibinfo{author}{\bibfnamefont{Y.-S.} \bibnamefont{Li}},
  \bibinfo{author}{\bibfnamefont{F.-G.} \bibnamefont{Deng}},
  \bibinfo{author}{\bibfnamefont{X.-H.} \bibnamefont{Li}},
  \bibinfo{author}{\bibfnamefont{B.-A.} \bibnamefont{Bian}},
  \bibinfo{author}{\bibfnamefont{Z.}~\bibnamefont{F.-S}},
  \bibinfo{author}{\bibfnamefont{Z.-H.} \bibnamefont{Liu}}, \bibnamefont{and}
  \bibinfo{author}{\bibfnamefont{H.-Y.} \bibnamefont{Zhou}},
  \eprint{arXiv:0708.0071}.

\bibitem[{\citenamefont{Ring and Schuck}(1980)}]{ring80a}
\bibinfo{author}{\bibfnamefont{P.}~\bibnamefont{Ring}} \bibnamefont{and}
  \bibinfo{author}{\bibfnamefont{P.}~\bibnamefont{Schuck}},
  \emph{\bibinfo{title}{{The Nuclear Many-Body Problem}}}
  (\bibinfo{publisher}{Springer-Verlag}, \bibinfo{address}{New-York},
  \bibinfo{year}{1980}).

\bibitem[{\citenamefont{Bender et~al.}(2003)\citenamefont{Bender, Heenen, and
  Reinhard}}]{bender03b}
\bibinfo{author}{\bibfnamefont{M.}~\bibnamefont{Bender}},
  \bibinfo{author}{\bibfnamefont{P.-H.} \bibnamefont{Heenen}},
  \bibnamefont{and} \bibinfo{author}{\bibfnamefont{P.-G.}
  \bibnamefont{Reinhard}}, \bibinfo{journal}{Rev. Mod. Phys.}
  \textbf{\bibinfo{volume}{75}}, \bibinfo{pages}{121} (\bibinfo{year}{2003}).

\bibitem[{\citenamefont{Skyrme}(1956)}]{skyrme56}
\bibinfo{author}{\bibfnamefont{T.~H.~R.} \bibnamefont{Skyrme}},
  \bibinfo{journal}{Phil. Mag.} \textbf{\bibinfo{volume}{1}},
  \bibinfo{pages}{1043} (\bibinfo{year}{1956}).

\bibitem[{\citenamefont{Vautherin and Brink}(1972)}]{vautherin72a}
\bibinfo{author}{\bibfnamefont{D.}~\bibnamefont{Vautherin}} \bibnamefont{and}
  \bibinfo{author}{\bibfnamefont{D.~M.} \bibnamefont{Brink}},
  \bibinfo{journal}{Phys. Rev. C} \textbf{\bibinfo{volume}{5}},
  \bibinfo{pages}{626} (\bibinfo{year}{1972}).

\bibitem[{\citenamefont{Decharg\'e and Gogny}(1980)}]{decharge80a}
\bibinfo{author}{\bibfnamefont{J.}~\bibnamefont{Decharg\'e}} \bibnamefont{and}
  \bibinfo{author}{\bibfnamefont{D.}~\bibnamefont{Gogny}},
  \bibinfo{journal}{Phys. Rev. C} \textbf{\bibinfo{volume}{21}},
  \bibinfo{pages}{1568} (\bibinfo{year}{1980}).

\bibitem[{\citenamefont{Bouyssy et~al.}(1987)\citenamefont{Bouyssy, Mathiot,
  {Van Giai}, and Marcos}}]{bouyssy87}
\bibinfo{author}{\bibfnamefont{A.}~\bibnamefont{Bouyssy}},
  \bibinfo{author}{\bibfnamefont{J.-F.} \bibnamefont{Mathiot}},
  \bibinfo{author}{\bibfnamefont{N.}~\bibnamefont{{Van Giai}}},
  \bibnamefont{and} \bibinfo{author}{\bibfnamefont{S.}~\bibnamefont{Marcos}},
  \bibinfo{journal}{Phys. Rev. C} \textbf{\bibinfo{volume}{36}},
  \bibinfo{pages}{380} (\bibinfo{year}{1987}).

\bibitem[{\citenamefont{Reinhard}(1989)}]{reinhard89}
\bibinfo{author}{\bibfnamefont{P.-G.} \bibnamefont{Reinhard}},
  \bibinfo{journal}{Rep. Prog. Phys.} \textbf{\bibinfo{volume}{52}},
  \bibinfo{pages}{439} (\bibinfo{year}{1989}).

\bibitem[{\citenamefont{Gambhir et~al.}(1990)\citenamefont{Gambhir, Ring, and
  Thimet}}]{gambhir90}
\bibinfo{author}{\bibfnamefont{Y.~K.} \bibnamefont{Gambhir}},
  \bibinfo{author}{\bibfnamefont{P.}~\bibnamefont{Ring}}, \bibnamefont{and}
  \bibinfo{author}{\bibfnamefont{A.}~\bibnamefont{Thimet}},
  \bibinfo{journal}{Ann. Phys.} \textbf{\bibinfo{volume}{198}},
  \bibinfo{pages}{132} (\bibinfo{year}{1990}).

\bibitem[{\citenamefont{Ring}(1993)}]{ring93}
\bibinfo{author}{\bibfnamefont{P.}~\bibnamefont{Ring}}, \bibinfo{journal}{Prog.
  Part. Nucl. Phys.} \textbf{\bibinfo{volume}{37}}, \bibinfo{pages}{193}
  (\bibinfo{year}{1993}).

\bibitem[{\citenamefont{Todd and Piekarewicz}(2003)}]{todd03}
\bibinfo{author}{\bibfnamefont{B.~G.} \bibnamefont{Todd}} \bibnamefont{and}
  \bibinfo{author}{\bibfnamefont{J.}~\bibnamefont{Piekarewicz}},
  \bibinfo{journal}{Phys. Rev. C} \textbf{\bibinfo{volume}{67}},
  \bibinfo{pages}{044317} (\bibinfo{year}{2003}).

\bibitem[{\citenamefont{Samyn et~al.}(2002)\citenamefont{Samyn, Goriely,
  Heenen, Pearson, and Tondeur}}]{samyn02}
\bibinfo{author}{\bibfnamefont{M.}~\bibnamefont{Samyn}},
  \bibinfo{author}{\bibfnamefont{S.}~\bibnamefont{Goriely}},
  \bibinfo{author}{\bibfnamefont{P.-H.} \bibnamefont{Heenen}},
  \bibinfo{author}{\bibfnamefont{J.~M.} \bibnamefont{Pearson}},
  \bibnamefont{and} \bibinfo{author}{\bibfnamefont{J.~F.}
  \bibnamefont{Tondeur}}, \bibinfo{journal}{Nucl. Phys.}
  \textbf{\bibinfo{volume}{A700}}, \bibinfo{pages}{142} (\bibinfo{year}{2002}).

\bibitem[{\citenamefont{Goriely et~al.}(2002)\citenamefont{Goriely, Samyn,
  Heenen, Pearson, and Tondeur}}]{goriely02}
\bibinfo{author}{\bibfnamefont{S.}~\bibnamefont{Goriely}},
  \bibinfo{author}{\bibfnamefont{M.}~\bibnamefont{Samyn}},
  \bibinfo{author}{\bibfnamefont{P.-H.} \bibnamefont{Heenen}},
  \bibinfo{author}{\bibfnamefont{J.~M.} \bibnamefont{Pearson}},
  \bibnamefont{and} \bibinfo{author}{\bibfnamefont{F.}~\bibnamefont{Tondeur}},
  \bibinfo{journal}{Phys. Rev. C} \textbf{\bibinfo{volume}{66}},
  \bibinfo{pages}{024326} (\bibinfo{year}{2002}).

\bibitem[{\citenamefont{Goriely et~al.}(2003)\citenamefont{Goriely, Samyn,
  Bender, and Pearson}}]{goriely03}
\bibinfo{author}{\bibfnamefont{S.}~\bibnamefont{Goriely}},
  \bibinfo{author}{\bibfnamefont{M.}~\bibnamefont{Samyn}},
  \bibinfo{author}{\bibfnamefont{M.}~\bibnamefont{Bender}}, \bibnamefont{and}
  \bibinfo{author}{\bibfnamefont{J.~M.} \bibnamefont{Pearson}},
  \bibinfo{journal}{Phys. Rev. C} \textbf{\bibinfo{volume}{68}},
  \bibinfo{pages}{054325} (\bibinfo{year}{2003}).

\bibitem[{\citenamefont{Samyn et~al.}(2004)\citenamefont{Samyn, Goriely,
  Bender, and Pearson}}]{samyn04}
\bibinfo{author}{\bibfnamefont{M.}~\bibnamefont{Samyn}},
  \bibinfo{author}{\bibfnamefont{S.}~\bibnamefont{Goriely}},
  \bibinfo{author}{\bibfnamefont{M.}~\bibnamefont{Bender}}, \bibnamefont{and}
  \bibinfo{author}{\bibfnamefont{J.~M.} \bibnamefont{Pearson}},
  \bibinfo{journal}{Phys. Rev. C} \textbf{\bibinfo{volume}{70}},
  \bibinfo{pages}{044309} (\bibinfo{year}{2004}).

\bibitem[{\citenamefont{Dobaczewski and Nazarewicz}(2002)}]{doba02b}
\bibinfo{author}{\bibfnamefont{J.}~\bibnamefont{Dobaczewski}} \bibnamefont{and}
  \bibinfo{author}{\bibfnamefont{W.}~\bibnamefont{Nazarewicz}},
  \bibinfo{journal}{Prog. Theor. Phys. Suppl.} \textbf{\bibinfo{volume}{146}},
  \bibinfo{pages}{70} (\bibinfo{year}{2002}).

\bibitem[{\citenamefont{Bender and Heenen}(2003)}]{bender03c}
\bibinfo{author}{\bibfnamefont{M.}~\bibnamefont{Bender}} \bibnamefont{and}
  \bibinfo{author}{\bibfnamefont{P.-H.} \bibnamefont{Heenen}},
  \bibinfo{journal}{Nucl. Phys.} \textbf{\bibinfo{volume}{A713}},
  \bibinfo{pages}{390} (\bibinfo{year}{2003}).

\bibitem[{\citenamefont{Duguet et~al.}(2004)\citenamefont{Duguet, Bender,
  Bonche, and Heenen}}]{duguet03c}
\bibinfo{author}{\bibfnamefont{T.}~\bibnamefont{Duguet}},
  \bibinfo{author}{\bibfnamefont{M.}~\bibnamefont{Bender}},
  \bibinfo{author}{\bibfnamefont{P.}~\bibnamefont{Bonche}}, \bibnamefont{and}
  \bibinfo{author}{\bibfnamefont{P.-H.} \bibnamefont{Heenen}},
  \bibinfo{journal}{Phys. Lett. B} \textbf{\bibinfo{volume}{559}},
  \bibinfo{pages}{201} (\bibinfo{year}{2004}).

\bibitem[{\citenamefont{Egido and Robledo}(2004)}]{egido04}
\bibinfo{author}{\bibfnamefont{J.~L.} \bibnamefont{Egido}} \bibnamefont{and}
  \bibinfo{author}{\bibfnamefont{L.~M.} \bibnamefont{Robledo}},
  \bibinfo{journal}{Lecture Notes in Physics} \textbf{\bibinfo{volume}{641}},
  \bibinfo{pages}{269} (\bibinfo{year}{2004}).

\bibitem[{\citenamefont{Rotival et~al.}()\citenamefont{Rotival, Bennaceur, and
  Duguet}}]{rotival08b}
\bibinfo{author}{\bibfnamefont{V.}~\bibnamefont{Rotival}},
  \bibinfo{author}{\bibfnamefont{K.}~\bibnamefont{Bennaceur}},
  \bibnamefont{and} \bibinfo{author}{\bibfnamefont{T.}~\bibnamefont{Duguet}},
  \eprint{arXiv:0711.1275}.

\bibitem[{\citenamefont{Mizutori et~al.}(2000)\citenamefont{Mizutori,
  Dobaczewski, Lalazissis, Nazarewicz, and Reinhard}}]{mizutori00}
\bibinfo{author}{\bibfnamefont{S.}~\bibnamefont{Mizutori}},
  \bibinfo{author}{\bibfnamefont{J.}~\bibnamefont{Dobaczewski}},
  \bibinfo{author}{\bibfnamefont{G.~A.} \bibnamefont{Lalazissis}},
  \bibinfo{author}{\bibfnamefont{W.}~\bibnamefont{Nazarewicz}},
  \bibnamefont{and} \bibinfo{author}{\bibfnamefont{P.-G.}
  \bibnamefont{Reinhard}}, \bibinfo{journal}{Phys. Rev. C}
  \textbf{\bibinfo{volume}{61}}, \bibinfo{pages}{044326}
  (\bibinfo{year}{2000}).

\bibitem[{\citenamefont{Tajima et~al.}(1992)\citenamefont{Tajima, Flocard,
  Bonche, Dobaczewski, and Heenen}}]{tajima92a}
\bibinfo{author}{\bibfnamefont{N.}~\bibnamefont{Tajima}},
  \bibinfo{author}{\bibfnamefont{H.}~\bibnamefont{Flocard}},
  \bibinfo{author}{\bibfnamefont{P.}~\bibnamefont{Bonche}},
  \bibinfo{author}{\bibfnamefont{J.}~\bibnamefont{Dobaczewski}},
  \bibnamefont{and} \bibinfo{author}{\bibfnamefont{P.-H.}
  \bibnamefont{Heenen}}, \bibinfo{journal}{Nucl. Phys.}
  \textbf{\bibinfo{volume}{A542}}, \bibinfo{pages}{409} (\bibinfo{year}{1992}).

\bibitem[{\citenamefont{Riisager et~al.}(1992)\citenamefont{Riisager, Jensen,
  and M{\o}ller}}]{riisager92}
\bibinfo{author}{\bibfnamefont{K.}~\bibnamefont{Riisager}},
  \bibinfo{author}{\bibfnamefont{A.~S.} \bibnamefont{Jensen}},
  \bibnamefont{and}
  \bibinfo{author}{\bibfnamefont{P.}~\bibnamefont{M{\o}ller}},
  \bibinfo{journal}{Nucl. Phys.} \textbf{\bibinfo{volume}{A548}},
  \bibinfo{pages}{393} (\bibinfo{year}{1992}).

\bibitem[{\citenamefont{Koopmans}(1934)}]{koopmans34}
\bibinfo{author}{\bibfnamefont{T.~H.} \bibnamefont{Koopmans}},
  \bibinfo{journal}{Physica} \textbf{\bibinfo{volume}{1}}, \bibinfo{pages}{104}
  (\bibinfo{year}{1934}).

\bibitem[{\citenamefont{Jensen et~al.}(2004)\citenamefont{Jensen, Riisager,
  Fedorov, and Garrido}}]{jensen04}
\bibinfo{author}{\bibfnamefont{A.~S.} \bibnamefont{Jensen}},
  \bibinfo{author}{\bibfnamefont{K.}~\bibnamefont{Riisager}},
  \bibinfo{author}{\bibfnamefont{D.~V.} \bibnamefont{Fedorov}},
  \bibnamefont{and} \bibinfo{author}{\bibfnamefont{E.}~\bibnamefont{Garrido}},
  \bibinfo{journal}{Rev. Mod. Phys.} \textbf{\bibinfo{volume}{76}},
  \bibinfo{pages}{215} (\bibinfo{year}{2004}).

\bibitem[{\citenamefont{Dobaczewski et~al.}(1984)\citenamefont{Dobaczewski,
  Flocard, and Treiner}}]{doba84a}
\bibinfo{author}{\bibfnamefont{J.}~\bibnamefont{Dobaczewski}},
  \bibinfo{author}{\bibfnamefont{H.}~\bibnamefont{Flocard}}, \bibnamefont{and}
  \bibinfo{author}{\bibfnamefont{J.}~\bibnamefont{Treiner}},
  \bibinfo{journal}{Nucl. Phys.} \textbf{\bibinfo{volume}{A422}},
  \bibinfo{pages}{103} (\bibinfo{year}{1984}).

\bibitem[{\citenamefont{Dobaczewski et~al.}(1996)\citenamefont{Dobaczewski,
  Nazarewicz, Werner, Berger, Chinn, and Decharg\'{e}}}]{doba96}
\bibinfo{author}{\bibfnamefont{J.}~\bibnamefont{Dobaczewski}},
  \bibinfo{author}{\bibfnamefont{W.}~\bibnamefont{Nazarewicz}},
  \bibinfo{author}{\bibfnamefont{T.~R.} \bibnamefont{Werner}},
  \bibinfo{author}{\bibfnamefont{J.-F.} \bibnamefont{Berger}},
  \bibinfo{author}{\bibfnamefont{C.~R.} \bibnamefont{Chinn}}, \bibnamefont{and}
  \bibinfo{author}{\bibfnamefont{J.}~\bibnamefont{Decharg\'{e}}},
  \bibinfo{journal}{Phys. Rev. C} \textbf{\bibinfo{volume}{53}},
  \bibinfo{pages}{2809} (\bibinfo{year}{1996}).

\bibitem[{\citenamefont{Terasaki et~al.}(1996)\citenamefont{Terasaki, Heenen,
  Flocard, and Bonche}}]{terasaki96}
\bibinfo{author}{\bibfnamefont{J.}~\bibnamefont{Terasaki}},
  \bibinfo{author}{\bibfnamefont{P.-H.} \bibnamefont{Heenen}},
  \bibinfo{author}{\bibfnamefont{H.}~\bibnamefont{Flocard}}, \bibnamefont{and}
  \bibinfo{author}{\bibfnamefont{P.}~\bibnamefont{Bonche}},
  \bibinfo{journal}{Nucl. Phys.} \textbf{\bibinfo{volume}{A600}},
  \bibinfo{pages}{371} (\bibinfo{year}{1996}).

\bibitem[{\citenamefont{Toki et~al.}(1991)\citenamefont{Toki, Sugahara, Hirata,
  Carlson, and Tanihata}}]{toki91}
\bibinfo{author}{\bibfnamefont{H.}~\bibnamefont{Toki}},
  \bibinfo{author}{\bibfnamefont{Y.}~\bibnamefont{Sugahara}},
  \bibinfo{author}{\bibfnamefont{D.}~\bibnamefont{Hirata}},
  \bibinfo{author}{\bibfnamefont{B.~V.} \bibnamefont{Carlson}},
  \bibnamefont{and} \bibinfo{author}{\bibfnamefont{I.}~\bibnamefont{Tanihata}},
  \bibinfo{journal}{Nucl. Phys.} \textbf{\bibinfo{volume}{A524}},
  \bibinfo{pages}{633} (\bibinfo{year}{1991}).

\bibitem[{\citenamefont{Sharma et~al.}(1993)\citenamefont{Sharma, Nagajaran,
  and Ring}}]{sharma93}
\bibinfo{author}{\bibfnamefont{M.~M.} \bibnamefont{Sharma}},
  \bibinfo{author}{\bibfnamefont{M.~A.} \bibnamefont{Nagajaran}},
  \bibnamefont{and} \bibinfo{author}{\bibfnamefont{P.}~\bibnamefont{Ring}},
  \bibinfo{journal}{Phys. Lett. B} \textbf{\bibinfo{volume}{312}},
  \bibinfo{pages}{377} (\bibinfo{year}{1993}).

\bibitem[{\citenamefont{Sugahara and Toki}(1994)}]{suagahara94}
\bibinfo{author}{\bibfnamefont{Y.}~\bibnamefont{Sugahara}} \bibnamefont{and}
  \bibinfo{author}{\bibfnamefont{H.}~\bibnamefont{Toki}},
  \bibinfo{journal}{Nucl. Phys.} \textbf{\bibinfo{volume}{A579}},
  \bibinfo{pages}{557} (\bibinfo{year}{1994}).

\bibitem[{\citenamefont{Bennaceur et~al.}(1999)\citenamefont{Bennaceur,
  Dobaczewski, and Ploszajczak}}]{bennaceur99}
\bibinfo{author}{\bibfnamefont{K.}~\bibnamefont{Bennaceur}},
  \bibinfo{author}{\bibfnamefont{J.}~\bibnamefont{Dobaczewski}},
  \bibnamefont{and}
  \bibinfo{author}{\bibfnamefont{M.}~\bibnamefont{Ploszajczak}},
  \bibinfo{journal}{Phys. Rev. C} \textbf{\bibinfo{volume}{60}},
  \bibinfo{pages}{034308} (\bibinfo{year}{1999}).

\bibitem[{\citenamefont{Bennaceur et~al.}(2000)\citenamefont{Bennaceur,
  Dobaczewski, and Ploszajczak}}]{bennaceur00}
\bibinfo{author}{\bibfnamefont{K.}~\bibnamefont{Bennaceur}},
  \bibinfo{author}{\bibfnamefont{J.}~\bibnamefont{Dobaczewski}},
  \bibnamefont{and}
  \bibinfo{author}{\bibfnamefont{M.}~\bibnamefont{Ploszajczak}},
  \bibinfo{journal}{Phys. Lett. B} \textbf{\bibinfo{volume}{496}},
  \bibinfo{pages}{154} (\bibinfo{year}{2000}).

\bibitem[{\citenamefont{Hamamoto and Mottelson}(2003)}]{hamamoto03}
\bibinfo{author}{\bibfnamefont{I.}~\bibnamefont{Hamamoto}} \bibnamefont{and}
  \bibinfo{author}{\bibfnamefont{B.~R.} \bibnamefont{Mottelson}},
  \bibinfo{journal}{Phys. Rev. C} \textbf{\bibinfo{volume}{68}},
  \bibinfo{pages}{034312} (\bibinfo{year}{2003}).

\bibitem[{\citenamefont{Hamamoto and Mottelson}(2004)}]{hamamoto04}
\bibinfo{author}{\bibfnamefont{I.}~\bibnamefont{Hamamoto}} \bibnamefont{and}
  \bibinfo{author}{\bibfnamefont{B.~R.} \bibnamefont{Mottelson}},
  \bibinfo{journal}{Phys. Rev. C} \textbf{\bibinfo{volume}{69}},
  \bibinfo{pages}{064302} (\bibinfo{year}{2004}).

\bibitem[{\citenamefont{Grasso et~al.}(2006)\citenamefont{Grasso, Yoshida,
  Sandulescu, and {Van Giai}}}]{grasso06}
\bibinfo{author}{\bibfnamefont{M.}~\bibnamefont{Grasso}},
  \bibinfo{author}{\bibfnamefont{S.}~\bibnamefont{Yoshida}},
  \bibinfo{author}{\bibfnamefont{N.}~\bibnamefont{Sandulescu}},
  \bibnamefont{and} \bibinfo{author}{\bibfnamefont{N.}~\bibnamefont{{Van
  Giai}}}, \bibinfo{journal}{Phys. Rev. C} \textbf{\bibinfo{volume}{74}},
  \bibinfo{pages}{064317} (\bibinfo{year}{2006}).

\bibitem[{\citenamefont{Meng and Ring}(1998)}]{meng98}
\bibinfo{author}{\bibfnamefont{J.}~\bibnamefont{Meng}} \bibnamefont{and}
  \bibinfo{author}{\bibfnamefont{P.}~\bibnamefont{Ring}},
  \bibinfo{journal}{Phys. Rev. Lett.} \textbf{\bibinfo{volume}{80}},
  \bibinfo{pages}{460} (\bibinfo{year}{1998}).

\bibitem[{\citenamefont{Nerlo-Pomorska
  et~al.}(2000)\citenamefont{Nerlo-Pomorska, Pomorski, Berger, and
  Decharg\'e}}]{nerlo00}
\bibinfo{author}{\bibfnamefont{B.}~\bibnamefont{Nerlo-Pomorska}},
  \bibinfo{author}{\bibfnamefont{K.}~\bibnamefont{Pomorski}},
  \bibinfo{author}{\bibfnamefont{J.-F.} \bibnamefont{Berger}},
  \bibnamefont{and}
  \bibinfo{author}{\bibfnamefont{J.}~\bibnamefont{Decharg\'e}},
  \bibinfo{journal}{Eur. Phys. J. A} \textbf{\bibinfo{volume}{8}},
  \bibinfo{pages}{19} (\bibinfo{year}{2000}).

\bibitem[{\citenamefont{Im and Meng}(2000)}]{im00}
\bibinfo{author}{\bibfnamefont{S.}~\bibnamefont{Im}} \bibnamefont{and}
  \bibinfo{author}{\bibfnamefont{J.}~\bibnamefont{Meng}},
  \bibinfo{journal}{Phys. Rev. C} \textbf{\bibinfo{volume}{61}},
  \bibinfo{pages}{047302} (\bibinfo{year}{2000}).

\bibitem[{\citenamefont{Sandulescu et~al.}(2003)\citenamefont{Sandulescu, Geng,
  Toki, and Hillhouse}}]{sandulescu03}
\bibinfo{author}{\bibfnamefont{N.}~\bibnamefont{Sandulescu}},
  \bibinfo{author}{\bibfnamefont{L.~S.} \bibnamefont{Geng}},
  \bibinfo{author}{\bibfnamefont{H.}~\bibnamefont{Toki}}, \bibnamefont{and}
  \bibinfo{author}{\bibfnamefont{G.~C.} \bibnamefont{Hillhouse}},
  \bibinfo{journal}{Phys. Rev. C} \textbf{\bibinfo{volume}{68}},
  \bibinfo{pages}{054323} (\bibinfo{year}{2003}).

\bibitem[{\citenamefont{Geng et~al.}(2004)\citenamefont{Geng, Toki, and
  Meng}}]{geng04}
\bibinfo{author}{\bibfnamefont{L.~S.} \bibnamefont{Geng}},
  \bibinfo{author}{\bibfnamefont{H.}~\bibnamefont{Toki}}, \bibnamefont{and}
  \bibinfo{author}{\bibfnamefont{J.}~\bibnamefont{Meng}},
  \bibinfo{journal}{Mod. Phys. Lett. A} \textbf{\bibinfo{volume}{19}},
  \bibinfo{pages}{2171} (\bibinfo{year}{2004}).

\bibitem[{\citenamefont{Kaushik et~al.}(2005)\citenamefont{Kaushik, Singh, and
  Yadav}}]{kaushik05}
\bibinfo{author}{\bibfnamefont{M.}~\bibnamefont{Kaushik}},
  \bibinfo{author}{\bibfnamefont{D.}~\bibnamefont{Singh}}, \bibnamefont{and}
  \bibinfo{author}{\bibfnamefont{H.~L.} \bibnamefont{Yadav}},
  \bibinfo{journal}{Acta Phys. Slov.} \textbf{\bibinfo{volume}{55}},
  \bibinfo{pages}{181} (\bibinfo{year}{2005}).

\bibitem[{\citenamefont{Terasaki et~al.}(2006)\citenamefont{Terasaki, Zhang,
  Zhou, and Meng}}]{terasaki06}
\bibinfo{author}{\bibfnamefont{J.}~\bibnamefont{Terasaki}},
  \bibinfo{author}{\bibfnamefont{S.~Q.} \bibnamefont{Zhang}},
  \bibinfo{author}{\bibfnamefont{S.~G.} \bibnamefont{Zhou}}, \bibnamefont{and}
  \bibinfo{author}{\bibfnamefont{J.}~\bibnamefont{Meng}},
  \bibinfo{journal}{Phys. Rev. C} \textbf{\bibinfo{volume}{74}},
  \bibinfo{pages}{054318} (\bibinfo{year}{2006}).

\bibitem[{\citenamefont{Bennaceur and Dobaczewski}(2005)}]{bennaceur05a}
\bibinfo{author}{\bibfnamefont{K.}~\bibnamefont{Bennaceur}} \bibnamefont{and}
  \bibinfo{author}{\bibfnamefont{J.}~\bibnamefont{Dobaczewski}},
  \bibinfo{journal}{Comp. Phys. Comm.} \textbf{\bibinfo{volume}{168}},
  \bibinfo{pages}{96} (\bibinfo{year}{2005}).

\bibitem[{\citenamefont{Chabanat et~al.}(1997)\citenamefont{Chabanat, Meyer,
  Bonche, Schaeffer, and Haensel}}]{chabanat97}
\bibinfo{author}{\bibfnamefont{E.}~\bibnamefont{Chabanat}},
  \bibinfo{author}{\bibfnamefont{J.}~\bibnamefont{Meyer}},
  \bibinfo{author}{\bibfnamefont{P.}~\bibnamefont{Bonche}},
  \bibinfo{author}{\bibfnamefont{R.}~\bibnamefont{Schaeffer}},
  \bibnamefont{and} \bibinfo{author}{\bibfnamefont{P.}~\bibnamefont{Haensel}},
  \bibinfo{journal}{Nucl. Phys.} \textbf{\bibinfo{volume}{A627}},
  \bibinfo{pages}{710} (\bibinfo{year}{1997}).

\bibitem[{\citenamefont{Chabanat et~al.}(1998)\citenamefont{Chabanat, Bonche,
  Haensel, Meyer, and Schaeffer}}]{chabanat98}
\bibinfo{author}{\bibfnamefont{E.}~\bibnamefont{Chabanat}},
  \bibinfo{author}{\bibfnamefont{P.}~\bibnamefont{Bonche}},
  \bibinfo{author}{\bibfnamefont{P.}~\bibnamefont{Haensel}},
  \bibinfo{author}{\bibfnamefont{J.}~\bibnamefont{Meyer}}, \bibnamefont{and}
  \bibinfo{author}{\bibfnamefont{R.}~\bibnamefont{Schaeffer}},
  \bibinfo{journal}{Nucl. Phys.} \textbf{\bibinfo{volume}{A635}},
  \bibinfo{pages}{231} (\bibinfo{year}{1998}).

\bibitem[{\citenamefont{Tondeur}(1979)}]{tondeur79}
\bibinfo{author}{\bibfnamefont{F.}~\bibnamefont{Tondeur}},
  \bibinfo{journal}{Nucl. Phys.} \textbf{\bibinfo{volume}{A315}},
  \bibinfo{pages}{353} (\bibinfo{year}{1979}).

\bibitem[{\citenamefont{Krieger et~al.}(1990)\citenamefont{Krieger, Bonche,
  Flocard, Quentin, and Weiss}}]{krieger90}
\bibinfo{author}{\bibfnamefont{S.~J.} \bibnamefont{Krieger}},
  \bibinfo{author}{\bibfnamefont{P.}~\bibnamefont{Bonche}},
  \bibinfo{author}{\bibfnamefont{H.}~\bibnamefont{Flocard}},
  \bibinfo{author}{\bibfnamefont{P.}~\bibnamefont{Quentin}}, \bibnamefont{and}
  \bibinfo{author}{\bibfnamefont{M.~S.} \bibnamefont{Weiss}},
  \bibinfo{journal}{Nucl. Phys.} \textbf{\bibinfo{volume}{A517}},
  \bibinfo{pages}{275} (\bibinfo{year}{1990}).

\bibitem[{\citenamefont{Bertsch and Esbensen}(1991)}]{bertsch}
\bibinfo{author}{\bibfnamefont{G.~F.} \bibnamefont{Bertsch}} \bibnamefont{and}
  \bibinfo{author}{\bibfnamefont{H.}~\bibnamefont{Esbensen}},
  \bibinfo{journal}{Ann. Phys. (NY)} \textbf{\bibinfo{volume}{209}},
  \bibinfo{pages}{327} (\bibinfo{year}{1991}).

\bibitem[{\citenamefont{Tajima et~al.}(1993)\citenamefont{Tajima, Bonche,
  Flocard, Heenen, and Weiss}}]{tajima93}
\bibinfo{author}{\bibfnamefont{N.}~\bibnamefont{Tajima}},
  \bibinfo{author}{\bibfnamefont{P.}~\bibnamefont{Bonche}},
  \bibinfo{author}{\bibfnamefont{H.}~\bibnamefont{Flocard}},
  \bibinfo{author}{\bibfnamefont{P.-H.} \bibnamefont{Heenen}},
  \bibnamefont{and} \bibinfo{author}{\bibfnamefont{M.~S.} \bibnamefont{Weiss}},
  \bibinfo{journal}{Nucl. Phys.} \textbf{\bibinfo{volume}{A551}},
  \bibinfo{pages}{434} (\bibinfo{year}{1993}).

\bibitem[{\citenamefont{Terasaki et~al.}(1995)\citenamefont{Terasaki, Heenen,
  Bonche, Dobaczewski, and Flocard}}]{terasaki95}
\bibinfo{author}{\bibfnamefont{J.}~\bibnamefont{Terasaki}},
  \bibinfo{author}{\bibfnamefont{P.-H.} \bibnamefont{Heenen}},
  \bibinfo{author}{\bibfnamefont{P.}~\bibnamefont{Bonche}},
  \bibinfo{author}{\bibfnamefont{J.}~\bibnamefont{Dobaczewski}},
  \bibnamefont{and} \bibinfo{author}{\bibfnamefont{H.}~\bibnamefont{Flocard}},
  \bibinfo{journal}{Nucl. Phys.} \textbf{\bibinfo{volume}{A593}},
  \bibinfo{pages}{1} (\bibinfo{year}{1995}).

\bibitem[{\citenamefont{Hilaire and Girod}(2007)}]{hilaire07}
\bibinfo{author}{\bibfnamefont{S.}~\bibnamefont{Hilaire}} \bibnamefont{and}
  \bibinfo{author}{\bibfnamefont{M.}~\bibnamefont{Girod}},
  \bibinfo{journal}{Eur. Phys. J. A} \textbf{\bibinfo{volume}{33}},
  \bibinfo{pages}{237} (\bibinfo{year}{2007}).

\bibitem[{\citenamefont{Dobaczewski et~al.}(2004)\citenamefont{Dobaczewski,
  Stoitsov, and Nazarewicz}}]{doba04b}
\bibinfo{author}{\bibfnamefont{J.}~\bibnamefont{Dobaczewski}},
  \bibinfo{author}{\bibfnamefont{M.~V.} \bibnamefont{Stoitsov}},
  \bibnamefont{and}
  \bibinfo{author}{\bibfnamefont{W.}~\bibnamefont{Nazarewicz}},
  \bibinfo{journal}{AIP Conf. Proc.} \textbf{\bibinfo{volume}{726}},
  \bibinfo{pages}{51} (\bibinfo{year}{2004}).

\bibitem[{\citenamefont{Angeli}(2004)}]{angeli04}
\bibinfo{author}{\bibfnamefont{I.}~\bibnamefont{Angeli}},
  \bibinfo{journal}{Atomic Data and Nuclear Data Tables}
  \textbf{\bibinfo{volume}{87}}, \bibinfo{pages}{185} (\bibinfo{year}{2004}).

\bibitem[{\citenamefont{Audi et~al.}(2003)\citenamefont{Audi, Wapstra, and
  Thibault}}]{audi2003}
\bibinfo{author}{\bibfnamefont{G.}~\bibnamefont{Audi}},
  \bibinfo{author}{\bibfnamefont{A.~H.} \bibnamefont{Wapstra}},
  \bibnamefont{and} \bibinfo{author}{\bibfnamefont{C.}~\bibnamefont{Thibault}},
  \bibinfo{journal}{Nucl. Phys.} \textbf{\bibinfo{volume}{A729}},
  \bibinfo{pages}{337} (\bibinfo{year}{2003}).

\bibitem[{\citenamefont{Lubi{\'n}ski et~al.}(1998)\citenamefont{Lubi{\'n}ski,
  Jastrz\c{e}bski, Trzci\'nska, Kurcewicz, Hartmann, Schmid, {von Egidy},
  Smola\'nczuk, and Wycech}}]{lubinski98}
\bibinfo{author}{\bibfnamefont{P.}~\bibnamefont{Lubi{\'n}ski}},
  \bibinfo{author}{\bibfnamefont{J.}~\bibnamefont{Jastrz\c{e}bski}},
  \bibinfo{author}{\bibfnamefont{A.}~\bibnamefont{Trzci\'nska}},
  \bibinfo{author}{\bibfnamefont{W.}~\bibnamefont{Kurcewicz}},
  \bibinfo{author}{\bibfnamefont{F.~J.} \bibnamefont{Hartmann}},
  \bibinfo{author}{\bibfnamefont{W.}~\bibnamefont{Schmid}},
  \bibinfo{author}{\bibfnamefont{T.}~\bibnamefont{{von Egidy}}},
  \bibinfo{author}{\bibfnamefont{R.}~\bibnamefont{Smola\'nczuk}},
  \bibnamefont{and} \bibinfo{author}{\bibfnamefont{S.}~\bibnamefont{Wycech}},
  \bibinfo{journal}{Phys. Rev. C} \textbf{\bibinfo{volume}{57}},
  \bibinfo{pages}{2962} (\bibinfo{year}{1998}).

\bibitem[{\citenamefont{Helm}(1956)}]{helm56}
\bibinfo{author}{\bibfnamefont{R.~H.} \bibnamefont{Helm}},
  \bibinfo{journal}{Phys. Rev.} \textbf{\bibinfo{volume}{104}},
  \bibinfo{pages}{1466} (\bibinfo{year}{1956}).

\bibitem[{\citenamefont{Rosen et~al.}(1957)\citenamefont{Rosen, Raphael, and
  {\"U}berall}}]{rosen57}
\bibinfo{author}{\bibfnamefont{M.}~\bibnamefont{Rosen}},
  \bibinfo{author}{\bibfnamefont{R.}~\bibnamefont{Raphael}}, \bibnamefont{and}
  \bibinfo{author}{\bibfnamefont{H.}~\bibnamefont{{\"U}berall}},
  \bibinfo{journal}{Phys. Rev.} \textbf{\bibinfo{volume}{163}},
  \bibinfo{pages}{927} (\bibinfo{year}{1957}).

\bibitem[{\citenamefont{Raphael and Rosen}(1970)}]{raphael70}
\bibinfo{author}{\bibfnamefont{R.}~\bibnamefont{Raphael}} \bibnamefont{and}
  \bibinfo{author}{\bibfnamefont{M.}~\bibnamefont{Rosen}},
  \bibinfo{journal}{Phys. Rev. C} \textbf{\bibinfo{volume}{1}},
  \bibinfo{pages}{547} (\bibinfo{year}{1970}).

\bibitem[{\citenamefont{Tajima}(2004)}]{tajima04}
\bibinfo{author}{\bibfnamefont{N.}~\bibnamefont{Tajima}},
  \bibinfo{journal}{Phys. Rev. C} \textbf{\bibinfo{volume}{69}},
  \bibinfo{pages}{034305} (\bibinfo{year}{2004}).

\bibitem[{\citenamefont{Friedrich and Voegler}(1982)}]{friedrich82}
\bibinfo{author}{\bibfnamefont{J.}~\bibnamefont{Friedrich}} \bibnamefont{and}
  \bibinfo{author}{\bibfnamefont{N.}~\bibnamefont{Voegler}},
  \bibinfo{journal}{Nucl. Phys.} \textbf{\bibinfo{volume}{A373}},
  \bibinfo{pages}{192} (\bibinfo{year}{1982}).

\bibitem[{\citenamefont{{Van Neck} et~al.}(1996)\citenamefont{{Van Neck},
  Dieperink, and Waroquier}}]{vanneck96}
\bibinfo{author}{\bibfnamefont{D.}~\bibnamefont{{Van Neck}}},
  \bibinfo{author}{\bibfnamefont{A.~E.~L.} \bibnamefont{Dieperink}},
  \bibnamefont{and}
  \bibinfo{author}{\bibfnamefont{M.}~\bibnamefont{Waroquier}},
  \bibinfo{journal}{Phys. Rev. C} \textbf{\bibinfo{volume}{53}},
  \bibinfo{pages}{2231} (\bibinfo{year}{1996}).

\bibitem[{\citenamefont{{Van Neck} et~al.}(1998)\citenamefont{{Van Neck},
  Waroquier, Dieperink, Pieper, and Pandharipande}}]{vanneck98a}
\bibinfo{author}{\bibfnamefont{D.}~\bibnamefont{{Van Neck}}},
  \bibinfo{author}{\bibfnamefont{M.}~\bibnamefont{Waroquier}},
  \bibinfo{author}{\bibfnamefont{A.~E.~L.} \bibnamefont{Dieperink}},
  \bibinfo{author}{\bibfnamefont{S.~C.} \bibnamefont{Pieper}},
  \bibnamefont{and} \bibinfo{author}{\bibfnamefont{V.~R.}
  \bibnamefont{Pandharipande}}, \bibinfo{journal}{Phys. Rev. C}
  \textbf{\bibinfo{volume}{57}}, \bibinfo{pages}{2308} (\bibinfo{year}{1998}).

\bibitem[{\citenamefont{{Van Neck} and Waroquier}(1998)}]{vanneck98b}
\bibinfo{author}{\bibfnamefont{D.}~\bibnamefont{{Van Neck}}} \bibnamefont{and}
  \bibinfo{author}{\bibfnamefont{M.}~\bibnamefont{Waroquier}},
  \bibinfo{journal}{Phys. Rev. C} \textbf{\bibinfo{volume}{58}},
  \bibinfo{pages}{3359} (\bibinfo{year}{1998}).

\bibitem[{\citenamefont{Escher et~al.}(2001)\citenamefont{Escher, Jennings, and
  Sherif}}]{escher01}
\bibinfo{author}{\bibfnamefont{J.}~\bibnamefont{Escher}},
  \bibinfo{author}{\bibfnamefont{B.~K.} \bibnamefont{Jennings}},
  \bibnamefont{and} \bibinfo{author}{\bibfnamefont{H.~S.}
  \bibnamefont{Sherif}}, \bibinfo{journal}{Phys. Rev. C}
  \textbf{\bibinfo{volume}{64}}, \bibinfo{pages}{065801}
  (\bibinfo{year}{2001}).

\bibitem[{\citenamefont{Cl{\'e}ment}(1973{\natexlab{a}})}]{clement73a}
\bibinfo{author}{\bibfnamefont{C.~F.} \bibnamefont{Cl{\'e}ment}},
  \bibinfo{journal}{Nucl. Phys.} \textbf{\bibinfo{volume}{A213}},
  \bibinfo{pages}{469} (\bibinfo{year}{1973}{\natexlab{a}}).

\bibitem[{\citenamefont{Cl{\'e}ment}(1973{\natexlab{b}})}]{clement73b}
\bibinfo{author}{\bibfnamefont{C.~F.} \bibnamefont{Cl{\'e}ment}},
  \bibinfo{journal}{Nucl. Phys.} \textbf{\bibinfo{volume}{A213}},
  \bibinfo{pages}{493} (\bibinfo{year}{1973}{\natexlab{b}}).

\bibitem[{\citenamefont{{Van Neck} et~al.}(1993)\citenamefont{{Van Neck},
  Waroquier, and Heyde}}]{vanneck93}
\bibinfo{author}{\bibfnamefont{D.}~\bibnamefont{{Van Neck}}},
  \bibinfo{author}{\bibfnamefont{M.}~\bibnamefont{Waroquier}},
  \bibnamefont{and} \bibinfo{author}{\bibfnamefont{K.}~\bibnamefont{Heyde}},
  \bibinfo{journal}{Phys. Lett. B} \textbf{\bibinfo{volume}{314}},
  \bibinfo{pages}{255} (\bibinfo{year}{1993}).

\bibitem[{\citenamefont{Shebeko et~al.}(2006)\citenamefont{Shebeko,
  Papakonstantinou, and Mavrommatis}}]{shebeko06}
\bibinfo{author}{\bibfnamefont{A.~V.} \bibnamefont{Shebeko}},
  \bibinfo{author}{\bibfnamefont{P.}~\bibnamefont{Papakonstantinou}},
  \bibnamefont{and}
  \bibinfo{author}{\bibfnamefont{E.}~\bibnamefont{Mavrommatis}},
  \bibinfo{journal}{Eur. Phys. J. A} \textbf{\bibinfo{volume}{27}},
  \bibinfo{pages}{143} (\bibinfo{year}{2006}).

\bibitem[{\citenamefont{Levy et~al.}(1984)\citenamefont{Levy, Perdew, and
  Sahni}}]{levy84}
\bibinfo{author}{\bibfnamefont{M.}~\bibnamefont{Levy}},
  \bibinfo{author}{\bibfnamefont{J.~P.} \bibnamefont{Perdew}},
  \bibnamefont{and} \bibinfo{author}{\bibfnamefont{V.}~\bibnamefont{Sahni}},
  \bibinfo{journal}{Phys. Rev. A} \textbf{\bibinfo{volume}{30}},
  \bibinfo{pages}{2745} (\bibinfo{year}{1984}).

\bibitem[{\citenamefont{Dreizler and Gross}(1990)}]{dreizler90}
\bibinfo{author}{\bibfnamefont{R.~M.} \bibnamefont{Dreizler}} \bibnamefont{and}
  \bibinfo{author}{\bibfnamefont{E.~K.~U.} \bibnamefont{Gross}},
  \emph{\bibinfo{title}{{Density Functional Theory}}}
  (\bibinfo{publisher}{Springer}, \bibinfo{address}{Berlin},
  \bibinfo{year}{1990}).

\bibitem[{\citenamefont{Abramowitz and Stegun}(1965)}]{abramowitz}
\bibinfo{author}{\bibfnamefont{M.}~\bibnamefont{Abramowitz}} \bibnamefont{and}
  \bibinfo{author}{\bibfnamefont{I.}~\bibnamefont{Stegun}},
  \emph{\bibinfo{title}{{Handbook of Mathematical Functions}}}
  (\bibinfo{publisher}{Dover}, \bibinfo{address}{New York},
  \bibinfo{year}{1965}).

\bibitem[{\citenamefont{Stoitsov et~al.}(2003)\citenamefont{Stoitsov,
  Dobaczewski, Nazarewicz, Pittel, and Dean}}]{stoitsov03a}
\bibinfo{author}{\bibfnamefont{M.~V.} \bibnamefont{Stoitsov}},
  \bibinfo{author}{\bibfnamefont{J.}~\bibnamefont{Dobaczewski}},
  \bibinfo{author}{\bibfnamefont{W.}~\bibnamefont{Nazarewicz}},
  \bibinfo{author}{\bibfnamefont{S.}~\bibnamefont{Pittel}}, \bibnamefont{and}
  \bibinfo{author}{\bibfnamefont{D.~J.} \bibnamefont{Dean}},
  \bibinfo{journal}{Phys. Rev.} \textbf{\bibinfo{volume}{C68}},
  \bibinfo{pages}{054312} (\bibinfo{year}{2003}).

\bibitem[{\citenamefont{Bertulani et~al.}(2002)\citenamefont{Bertulani, Hammer,
  and {Van Kolck}}}]{bertulani02}
\bibinfo{author}{\bibfnamefont{C.~A.} \bibnamefont{Bertulani}},
  \bibinfo{author}{\bibfnamefont{H.-W.} \bibnamefont{Hammer}},
  \bibnamefont{and} \bibinfo{author}{\bibfnamefont{U.}~\bibnamefont{{Van
  Kolck}}}, \bibinfo{journal}{Nucl. Phys.} \textbf{\bibinfo{volume}{A712}},
  \bibinfo{pages}{37} (\bibinfo{year}{2002}).

\bibitem[{\citenamefont{Engel}(2007)}]{engel07}
\bibinfo{author}{\bibfnamefont{J.}~\bibnamefont{Engel}},
  \bibinfo{journal}{Phys. Rev. C} \textbf{\bibinfo{volume}{75}},
  \bibinfo{pages}{014306} (\bibinfo{year}{2007}).

\bibitem[{\citenamefont{Dahlquist and Bj{\"o}rck}(1974)}]{dahlquist74}
\bibinfo{author}{\bibfnamefont{G.}~\bibnamefont{Dahlquist}} \bibnamefont{and}
  \bibinfo{author}{\bibfnamefont{{\AA}.}~\bibnamefont{Bj{\"o}rck}},
  \emph{\bibinfo{title}{{Numerical Methods}}}
  (\bibinfo{publisher}{Prentice-Hall}, \bibinfo{address}{Englewood Cliffs, NJ},
  \bibinfo{year}{1974}).

\bibitem[{\citenamefont{Nunes}()}]{nunespriv}
\bibinfo{author}{\bibfnamefont{F.}~\bibnamefont{Nunes}},
\bibinfo{title}{(private communication).}

\bibitem[{\citenamefont{Meng et~al.}(2002)\citenamefont{Meng, Toki, Zeng,
  Zhang, and Zhou}}]{meng04}
\bibinfo{author}{\bibfnamefont{J.}~\bibnamefont{Meng}},
  \bibinfo{author}{\bibfnamefont{H.}~\bibnamefont{Toki}},
  \bibinfo{author}{\bibfnamefont{J.~Y.} \bibnamefont{Zeng}},
  \bibinfo{author}{\bibfnamefont{S.~Q.} \bibnamefont{Zhang}}, \bibnamefont{and}
  \bibinfo{author}{\bibfnamefont{S.-G.} \bibnamefont{Zhou}},
  \bibinfo{journal}{Phys. Rev. C} \textbf{\bibinfo{volume}{65}},
  \bibinfo{pages}{041302(R)} (\bibinfo{year}{2002}).

\bibitem[{whi(2006)}]{whitepapernscl}
\emph{\bibinfo{title}{{Isotope Science Facility at Michigan State University;
  upgrade of the \uppercase{NSCL} rare isotope research capabilities}}}
  (\bibinfo{year}{2006}),
  \bibinfo{note}{\\\protect\url{http://www.nscl.msu.edu/future/isf}}.

\bibitem[{\citenamefont{Dieperink and {de Forest, Jr.}}(1974)}]{dieperink74}
\bibinfo{author}{\bibfnamefont{A.~E.~L.} \bibnamefont{Dieperink}}
  \bibnamefont{and} \bibinfo{author}{\bibfnamefont{T.}~\bibnamefont{{de Forest,
  Jr.}}}, \bibinfo{journal}{Phys. Rev. C} \textbf{\bibinfo{volume}{10}},
  \bibinfo{pages}{543} (\bibinfo{year}{1974}).

\bibitem[{\citenamefont{Giraud}(2008)}]{giraud77}
\bibinfo{author}{\bibfnamefont{B.~G.} \bibnamefont{Giraud}},
  \bibinfo{journal}{Phys. Rev. C} \textbf{\bibinfo{volume}{77}},
  \bibinfo{pages}{014311} (\bibinfo{year}{2008}).

\bibitem[{\citenamefont{Duguet et~al.}(2002{\natexlab{a}})\citenamefont{Duguet,
  Bonche, Heenen, and Meyer}}]{duguet02a}
\bibinfo{author}{\bibfnamefont{T.}~\bibnamefont{Duguet}},
  \bibinfo{author}{\bibfnamefont{P.}~\bibnamefont{Bonche}},
  \bibinfo{author}{\bibfnamefont{P.-H.} \bibnamefont{Heenen}},
  \bibnamefont{and} \bibinfo{author}{\bibfnamefont{J.}~\bibnamefont{Meyer}},
  \bibinfo{journal}{Phys. Rev. C} \textbf{\bibinfo{volume}{65}},
  \bibinfo{pages}{014310} (\bibinfo{year}{2002}{\natexlab{a}}).

\bibitem[{\citenamefont{Duguet et~al.}(2002{\natexlab{b}})\citenamefont{Duguet,
  Bonche, Heenen, and Meyer}}]{duguet02b}
\bibinfo{author}{\bibfnamefont{T.}~\bibnamefont{Duguet}},
  \bibinfo{author}{\bibfnamefont{P.}~\bibnamefont{Bonche}},
  \bibinfo{author}{\bibfnamefont{P.-H.} \bibnamefont{Heenen}},
  \bibnamefont{and} \bibinfo{author}{\bibfnamefont{J.}~\bibnamefont{Meyer}},
  \bibinfo{journal}{Phys. Rev. C} \textbf{\bibinfo{volume}{65}},
  \bibinfo{pages}{014311} (\bibinfo{year}{2002}{\natexlab{b}}).

\bibitem[{\citenamefont{{Van Neck} et~al.}(2006)\citenamefont{{Van Neck},
  Verdonck, Bonny, Ayers, and Waroquier}}]{vanneck06}
\bibinfo{author}{\bibfnamefont{D.}~\bibnamefont{{Van Neck}}},
  \bibinfo{author}{\bibfnamefont{S.}~\bibnamefont{Verdonck}},
  \bibinfo{author}{\bibfnamefont{G.}~\bibnamefont{Bonny}},
  \bibinfo{author}{\bibfnamefont{P.~W.} \bibnamefont{Ayers}}, \bibnamefont{and}
  \bibinfo{author}{\bibfnamefont{M.}~\bibnamefont{Waroquier}},
  \bibinfo{journal}{Phys. Rev. A} \textbf{\bibinfo{volume}{74}},
  \bibinfo{pages}{042501} (\bibinfo{year}{2006}).

\bibitem[{\citenamefont{Berdichevsky and Mosel}(1982)}]{berdich82}
\bibinfo{author}{\bibfnamefont{D.}~\bibnamefont{Berdichevsky}}
  \bibnamefont{and} \bibinfo{author}{\bibfnamefont{U.}~\bibnamefont{Mosel}},
  \bibinfo{journal}{Nucl. Phys.} \textbf{\bibinfo{volume}{A388}},
  \bibinfo{pages}{205} (\bibinfo{year}{1982}).

\bibitem[{\citenamefont{Gambhir and Patil}(1985)}]{gambhir85}
\bibinfo{author}{\bibfnamefont{Y.~K.} \bibnamefont{Gambhir}} \bibnamefont{and}
  \bibinfo{author}{\bibfnamefont{S.~H.} \bibnamefont{Patil}},
  \bibinfo{journal}{Z. Phys.} \textbf{\bibinfo{volume}{A321}},
  \bibinfo{pages}{161} (\bibinfo{year}{1985}).

\bibitem[{\citenamefont{Gambhir and Patil}(1986)}]{gambhir86}
\bibinfo{author}{\bibfnamefont{Y.~K.} \bibnamefont{Gambhir}} \bibnamefont{and}
  \bibinfo{author}{\bibfnamefont{S.~H.} \bibnamefont{Patil}},
  \bibinfo{journal}{Z. Phys.} \textbf{\bibinfo{volume}{A324}},
  \bibinfo{pages}{9} (\bibinfo{year}{1986}).

\bibitem[{\citenamefont{Bhagwat et~al.}(2000)\citenamefont{Bhagwat, Gambhir,
  and Patil}}]{bhagwat00}
\bibinfo{author}{\bibfnamefont{A.}~\bibnamefont{Bhagwat}},
  \bibinfo{author}{\bibfnamefont{Y.~K.} \bibnamefont{Gambhir}},
  \bibnamefont{and} \bibinfo{author}{\bibfnamefont{S.~H.} \bibnamefont{Patil}},
  \bibinfo{journal}{Eur. Phys. J. A} \textbf{\bibinfo{volume}{8}},
  \bibinfo{pages}{511} (\bibinfo{year}{2000}).

\end{thebibliography}
\end{document}